\documentclass[aps,10pt,pre,floatfix,twocolumn]{revtex4-1}
\usepackage{color}
\usepackage{graphicx}
\usepackage{comment}
\usepackage{siunitx}
\usepackage{subfigure,amsmath, amsthm, amssymb}
\usepackage{braket}
\usepackage{multirow}
\usepackage[section]{placeins}
\usepackage{tikz}
\usepackage{epstopdf}
\usepackage{amsfonts}
\usetikzlibrary{arrows}
\usetikzlibrary{external}
 \usepackage[T1]{fontenc}
\usepackage{mathpazo}
\usetikzlibrary{3d,calc}
\usepackage[normalem]{ulem}
\usepackage[ colorlinks=true,linkcolor=blue,filecolor=blue,citecolor = blue,urlcolor=cyan]{hyperref}

\begin{document}

\title{Active Random Walks in One and Two Dimensions} 

\author{Stephy Jose}
\email{stephyjose@tifrh.res.in}
\affiliation{TIFR Centre for Interdisciplinary Sciences, Tata Institute of Fundamental Research, Hyderabad 500046, India}
\author{Dipanjan Mandal}
\email{dipkar.308@gmail.com}
\affiliation{TIFR Centre for Interdisciplinary Sciences, Tata Institute of Fundamental Research, Hyderabad 500046, India}
\author{Mustansir Barma}
\email{barma@tifrh.res.in}
\affiliation{TIFR Centre for Interdisciplinary Sciences, Tata Institute of Fundamental Research, Hyderabad 500046, India}
\author{Kabir Ramola}
\email{kramola@tifrh.res.in}
\affiliation{TIFR Centre for Interdisciplinary Sciences, Tata Institute of Fundamental Research, Hyderabad 500046, India}
\date{\today}


\begin{abstract}
We investigate active lattice walks: biased continuous time random walks which perform orientational diffusion between lattice directions in one and two spatial dimensions. We study the occupation probability of an arbitrary site on the lattice in one and two dimensions, and derive exact results in the continuum limit.
Next, we compute the large deviation free energy function in both one and two dimensions, which we use to compute the moments and the cumulants of the displacements exactly at late times. Our exact results demonstrate that the cross-correlations between the motion in the $x$ and $y$ directions in two dimensions persist in the large deviation function. We also demonstrate that the large deviation function of an active particle with diffusion displays two regimes, with differing diffusive behaviors.
We verify our analytic results with kinetic Monte Carlo simulations of an active lattice walker in one and two dimensions.
\end{abstract}
\maketitle

\section{Introduction}
\label{intro}
Active matter, consisting of particles that 
perform directed motion using ambient or internal or ambient energy, is an important class of non-equilibrium systems that has recently been of broad interest. Examples range from granular particles, flocks of animals to bacterial suspensions~\cite{walsh2017noise,schnitzer1993theory,gautrais2009analyzing,cavagna2010scale,cates2012diffusive,ramaswamy2010mechanics}.
Fluctuation-dissipation relations do not generally apply to such cases, because these systems break detailed balance at the microscopic
scales. 
Interacting active particles can exhibit several non-equilibrium collective phenomena~\cite{tailleur2008statistical,slowman2017exact}, such
as domain-formation, swarming or flocking. Characteristics of global order are found in such interacting systems (bird flocks, fish schools, colloids or even human crowds)~\cite{gautrais2009analyzing,cavagna2010scale,enculescu2011active}, as well as strong correlation without any leadership or external force. Such persistent motion can even result in clustering and non-equilibrium phase transitions such as in the paradigmatic Vicsek and Toner-Tu models~\cite{vicsek1995novel,toner1995long,lam2015self,czirok1999collective}. In such systems, Motility Induced Phase Separation (MIPS) or confinement induced aggregation occurs even in the absence of attractive interactions between the individuals in the flock~\cite{cates2013active,kourbane2018exact,merrigan2020arrested,lee2013active}.~Active particles also exhibit interesting stationary distributions when trapped by external confining potentials~\cite{malakar2020steady,sevilla2019stationary,dhar2019run}.

Several interesting microscopic models for the dynamics of active particles have been extensively studied, including the
run and tumble particle (RTP) model \cite{malakar2018steady,evans2018run,mori2020universal,mori2020universalp,singh2019generalised,angelani2014first,martens2012probability} and the active Brownian particle (ABP) model \cite{basu2018active,lindner2008diffusion,kumar2020active,romanczuk2012active,romanczuk2010collective}. Other studies have also analyzed models of active particle motion that diffuse between fixed orientations in two dimensions~\cite{santra2020run}.
A RTP performs directed motion with a fixed velocity along a selected orientation associated with an internal direction of bias and this orientation flips stochastically at a fixed rate $\gamma$. run and tumble particle motion consists of a sequence of forward steps followed by a sudden reorientation or tumble. The duration  $\tau$, of the forward flight or run is random and is exponentially distributed via $P(\tau)=\gamma e^{-\gamma \tau}$. The ABP also self propels at a fixed speed, but changes its direction gradually by rotational diffusion (with rotational diffusivity $D_r$). Here, ${D_r}^{-1}$ sets the characteristic time for the rotational diffusion. An isolated ABP performs a random walk at large timescales and becomes indistinguishable from RTP dynamics when all the microscopic parameters are uniform and isotropic~\cite{cates2013active,solon2015active}. 

In this paper, we study biased Continuous Time Random Walks (CTRWs) \cite{montroll1965random,montroll1979enriched,kutner2017continuous,mainardi2020advent} with Poisson distributed jumps performing orientational diffusion between lattice directions in one and two dimensions. These are often called the two state RTP model in one dimension and the four state RTP model in two dimensions respectively. Multi-state persistent random walks arise in several situations and are useful theoretical tools~\cite{masoliver2017continuous}. They have also been used to model correlation effects on frequency dependent conductivity in superionic conductors \cite{shlesinger1979correlation}. A crucial motivation in studying active particles on lattices stems from the difficulty of solving the Fokker-Planck equations associated with such systems in the continuum. Since the differential equations associated with lattice walks admit a formal analytic solution, lattice systems can lead to a direct insight into the behavior of active particle dynamics even in higher dimensions. Additionally, the continuum limit can be obtained from such a lattice model by taking the appropriate limits. All the techniques introduced in this paper can be generalized to higher dimensions and to different lattice structures.

A central aspect of our study is the large deviation function associated with RTP motion. There have been many studies on the large deviation functions~\cite{touchette2009large} for different models of active particle motion in different dimensions~\cite{van2019central,mori2021condensation,proesmans2020phase,mori2021first,mallmin2019comparison,gradenigo2019first,dean2021position}. In this paper, we calculate the exact large deviation free energy function for a RTP with diffusion in one and two dimensions, which we use to analyze the large deviation rate functions associated with RTP motion. We show that the rate function corresponding to the occupation probability distribution in one dimension and the rate functions corresponding to the marginal occupation probability distributions in two dimensions exhibit two regions with differing diffusive behaviors.

This paper is organized as follows: In Section~\ref{sec_model}, we discuss the RTP model in one and two dimensions and define the parameters used in the study. In Section~\ref{cond_occ_prob}, we derive the evolution equations for the occupation probability of an arbitrary lattice site. We show that the continuum space limit emerges by taking appropriate limits of the occupation probability in Fourier space. We recover the position distribution in Ref.~\cite{malakar2018steady} for a RTP in one dimension and the marginal position distributions in Ref.~\cite{santra2020run} describing the evolution of the $x$ and $y$ components of the position of a RTP in two dimensions in the continuum limit. In Section~\ref{mc}, we compute the moments, cumulants and large deviation free energy functions for the one and two dimensional cases. We also compute the associated large deviation rate function for the one dimensional process, as well as the two dimensional process projected along an arbitrary angle. We verify our analytic results with kinetic Monte Carlo (kMC) simulation results of an active random walker on one and two dimensional lattices. 

\section{The Model}
\label{sec_model}

We consider the motion of an active random walker on one and two dimensional infinite lattices with lattice points labeled by integer variables $x$ and $(x,y)$ respectively.
The walker can be biased along any of the lattice orientations, labeled by an internal state $m$, and performs orientational diffusion between these states.

\begin{figure} [t!]
\begin{center}
\subfigure[]{\includegraphics[width=.45\linewidth]{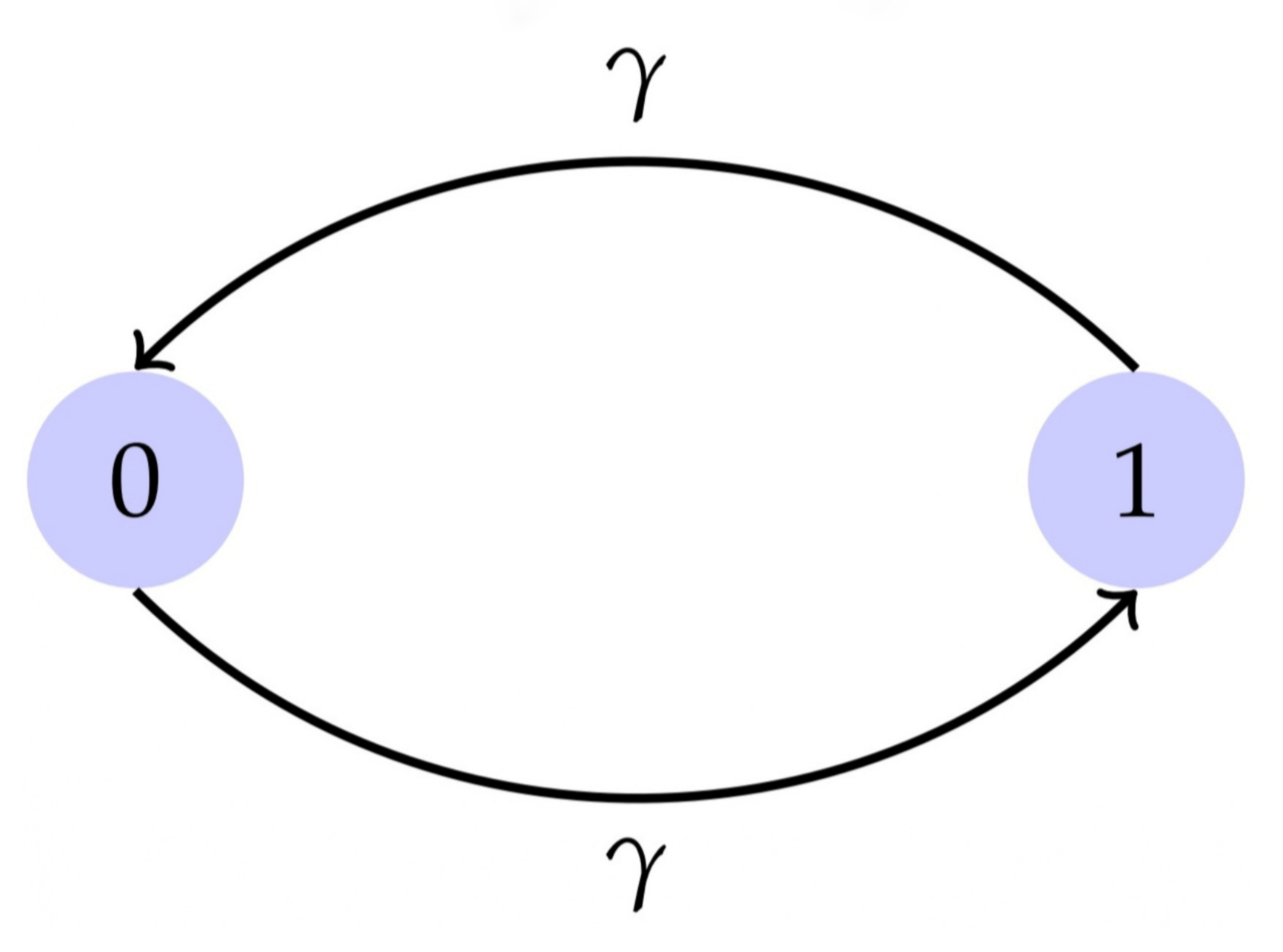} }
\subfigure[]{\includegraphics[width=.45\linewidth]{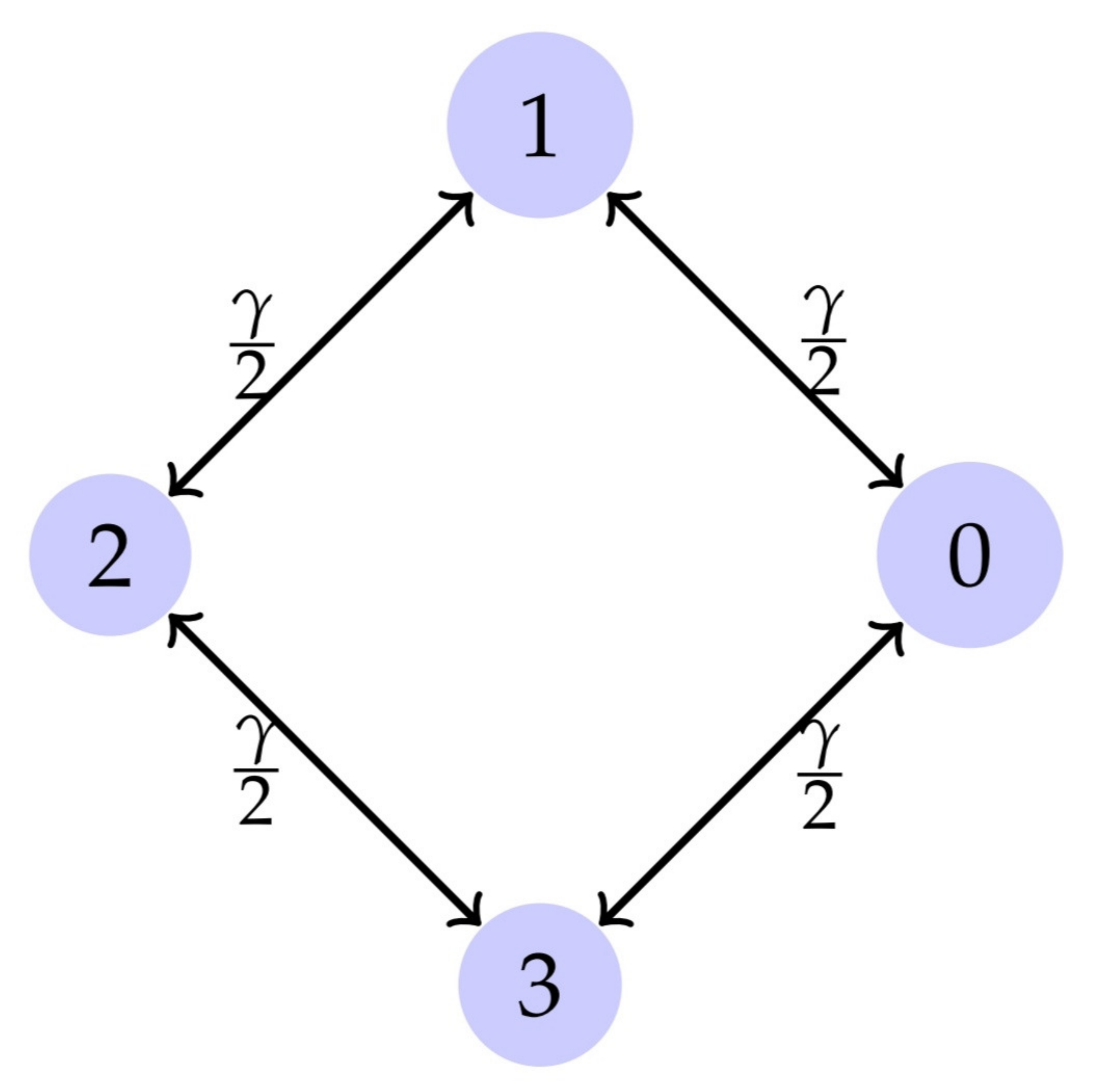} }
 \caption{(a)~An active particle in one dimension can flip its internal state from $0$ to $1$ or $1$ to $0$ with a rate $\gamma$ each.~(b)~An active particle in two dimensions can flip its internal state between $0$,~$1$,~$2$ or~$3$ with a rate $\frac{\gamma}{2}$ each.}\label{fig_states}
\end{center}
\end{figure} 
\begin{table}[ht]
\begin{center}
\parbox{0.96\linewidth}{
\begin{center}
\begin{tabular}{|c|c|c|}
 \hline
State & \multicolumn{2}{|c|}{Rate} \\
 \cline{2-3}
& $+ x$ & $-x$\\
 \hline
$0$   & $ D_{1d}+\epsilon$    & $ D_{1d}-\epsilon $\\
\hline
$1$&$D_{1d}-\epsilon$&   $D_{1d}+\epsilon $\\
 \hline
\end{tabular}
\end{center}
 \caption{The translational rates along different lattice directions for a RTP on a one dimensional lattice.}\label{table1}
}\hspace{0.3 cm}
\parbox{.96\linewidth}{
\begin{center}
\begin{tabular}{|c|c|c|c|c|}
 \hline
State & \multicolumn{4}{|c|}{Rate} \\
 \cline{2-5}
& $+ x$ & $-x$ & $+y$& $-y$\\
 \hline
$0$   & $ D_{2d}+\epsilon$    & $ D_{2d}-\epsilon $&  $D_{2d}$&   $D_{2d}$\\
\hline
$1$&   $D_{2d}$ & $D_{2d}$   &$D_{2d}+\epsilon$&   $D_{2d}-\epsilon $\\
\hline
$2$ &$ D_{2d}-\epsilon$ & $ D_{2d}+\epsilon$&  $D_{2d}$&   $D_{2d}$\\
\hline
$3$   &$D_{2d}$ & $D_{2d}$&  $ D_{2d}-\epsilon$&   $ D_{2d}+\epsilon$\\
 \hline
\end{tabular}
 \caption{The translational rates along different lattice directions for a RTP on a two dimensional square lattice.}
\label{table2}
\end{center}
}
\end{center}
\end{table}
\subsection{One Dimension}
In one dimension, the walker can be in any of the two possible internal states; $0$ or $1$.  Let $P_m(x,t)$ represent the probability of finding an active walker starting at the origin at time $t=0$, in the $m^{th}$ state at the lattice position $x$ at time $t$. The initial conditions are $P_m(x,t=0)=\frac{1}{2}\delta_{x,0},~m=0,1$. The walker is biased along the positive $x$ direction if it is in the state $0$ and is biased along the negative $x$ direction if it is in the state  $1$. The translational rates for the particle in different states along different directions are listed in Table~\ref{table1}. The intrinsic diffusion constant associated with the particle motion in one dimension is represented as $D_{1d}$. The value of the bias, $\epsilon$ is bounded between $0$ and $D_{1d}$. In addition to the translation, the particle may also flip its internal state with a finite rate $\gamma$ (refer to Fig.~\ref{fig_states}). 
A special case of the model considered in this paper appears in \cite{shlesinger1979correlation}, where a persistent random walk which can switch between two internal states in one dimension has been studied.
\subsection{Two Dimensions}
In two dimensions, the walker can be in any of the four possible internal states; $0,~1,~2$ or $3$. Let $P_m(x,y,t)$ represent the probability of finding an active walker starting at the origin at time $t=0$, in the $m^{th}$ state at the lattice position $(x,y)$ at time $t$. The initial conditions are $P_m(x,y,t=0)=\frac{1}{4}\delta_{x,0}\delta_{y,0},~m=0,1,2,3$. The walker is biased along the positive $x$, positive $y$, negative $x$ or negative $y$ direction if the walker is in the state $0,~1,~2$ or $3$ respectively. 
The translational rates for the particle in different states along different directions are listed in Table~\ref{table2}. The intrinsic diffusion constant associated with the particle motion in two dimensions is represented as $D_{2d}$. The value of the bias, $\epsilon$ is bounded between $0$ and $D_{2d}$. In addition to the translation, the particle may also flip its internal state to any of the other two possible states with a finite rate $\frac{\gamma}{2}$ each (refer to Fig.~\ref{fig_states}).

\subsection{Continuum Limit}

Throughout this study, we also analyze the continuum limit of these active random walk models.
All expressions in the continuum limit are referred to with a subscript  $\lq cont \rq$.
The continuum limit can be arrived at using standard scaling procedures where the microscopic rate constants scale with discretization parameters such as the lattice constant, $a$ \cite{redner2001guide}. In the Supplemental Material we display the convergence of our lattice results to the continuum theoretical predictions as the lattice spacing $a$ is reduced.

In this study, we compute the continuum limit of the occupation probabilities by analyzing their behavior in Fourier space, and keeping only the relevant terms as $|\vec{k}|a \to 0$~\cite{hilfer1995exact, hilfer2003fractional}.
As the lattice spacing $a$ is decreased, the bias and diffusion rates are rescaled as $\epsilon \to \frac{\epsilon}{a}$ and $D_{1d} \to \frac{D_{1d}}{a^2}$. This yields continuum Fokker Planck equations for RTP motion with discrete states analyzed in previous studies \cite{malakar2018steady,santra2020run}.
In Fourier space, this rescaling of the rates fixes the terms ${(ka)}^2 D_{1d}$ and $(ka) \epsilon$.
Throughout the calculations, we have set the lattice constant $a=1$ for simplicity, and therefore in the continuum limit expressions, we keep terms of order ${k}^2 D_{1d}$ and $k \epsilon$ in the $k \to 0$ limit of our lattice expressions.



\section{OCCUPATION PROBABILITY}
\label{cond_occ_prob}

In this Section, we study the occupation probability of lattice sites for a walker starting at the origin.
We derive evolution equations for the occupation probability of an arbitrary lattice site in one and two dimensions using which we compute the exact occupation probabilities both on the lattice as well as in the continuum limit.
\subsection{One Dimension}

Using the rates for translational motion in the two states given in Table~\ref{table1}, the equations obeyed by the occupation probabilities in the different states $P_{0,1}(x,t)$ can be expressed as
\begin{small}
\begin{eqnarray}
\label{eq:1}
 \frac{\partial P_0(x,t)}{\partial t}&=&\left(D_{1d}+\epsilon \right)P_0(x-1,t)+\left(D_{1d}-\epsilon \right)P_0(x+1,t)\nonumber\\
 &&-2D_{1d}P_0(x,t)+\gamma P_1(x,t)-\gamma P_0(x,t),\\
 \label{eq:2}
  \frac{\partial P_1(x,t)}{\partial t}&=&\left(D_{1d}+\epsilon \right)P_1(x+1,t)+\left(D_{1d}-\epsilon \right)P_1(x-1,t)\nonumber\\
  &&-2D_{1d}P_1(x,t)+\gamma P_0(x,t)-\gamma P_1(x,t),
\end{eqnarray}
\end{small}
where $D_{1d}$ refers to the intrinsic diffusion constant  associated with the active particle motion in one dimension.
These coupled equations can be written together in a compact form as follows:
\begin{small}
\begin{eqnarray}
\label{sq-1}
\frac{\partial P_{m}}{\partial t} = D_{1d} {\nabla}^2 P_{m} -  \epsilon ~ {\bf \hat{m}} . \vec{\nabla}P_{m} +\gamma \left( P_{m'}-P_{m}\right),
\end{eqnarray}
\end{small}
where $P_m=P_m(x,t)$ and the subscript $m$ denotes the internal state ($m=0,1$).~We define $m'=\mod\left(m+1,2\right)$. The modulo operator $\mod\left(m,n\right)$ gives the remainder when $m$ is divided by $n$. Here, $\vec{\nabla}$ is the discrete derivative operator and $\nabla^2$ is the discrete Laplacian operator on the one dimensional lattice defined as
\begin{small}
\begin{eqnarray}
\vec{\nabla}P_m(x,t)&=&\left(P_m(x+1,t)-P_m(x-1,t)\right){\bf \hat{x}},\nonumber\\{\nabla}^2 P_{m}(x,t)&=& P_m(x+1,t)+P_m(x-1,t)-2P_m(x,t).
\end{eqnarray}
\end{small}
Here, ${\bf \hat{m}}$ denotes the bias direction which is ${\bf \hat{x}}$ or~${\bf -\hat{x}}$ for states $0$ and~$1$ respectively.

We use the superscript tilde ($\sim$) to denote any transform (Fourier, Laplace or Fourier-Laplace) of the occupation probabilities. We use the Fourier transform for the space variables whereas the Laplace transform is used for the time variable.
We define the Fourier transform of the occupation probability $P_m(x,t)$, for a RTP on a one dimensional infinite lattice in the internal state $m$ as $\tilde P_m(k,t)=\sum_{x=-\infty}^{\infty}e^{ikx}P_m(x,t)$ and the Laplace transform of $\tilde P_m(k,t)$ as $\tilde P_m(k,s)=\int_0^{\infty}dt e^{-s t}\tilde P_m(k,t) $.
Taking a Fourier transform of Eq.~(\ref{sq-1}) yields
\begin{equation}
\label{matrix11}
\frac{\partial}{\partial t}
\ket{\tilde P_m (k,t)}
=
\mathcal{M}(k)\ket{\tilde P_m (k,t)},
\end{equation}
where the ket $\ket{\tilde P_m (k,t)}$ represents the column vector ${\begin{pmatrix}
\tilde P_0(k,t)\\
\tilde P_1(k,t)
\end{pmatrix}}$.
In the above equation, we have used the quantum mechanical bra-ket notation to represent the probabilities of the particle in different states.
The matrix $\mathcal{M}(k)$ is given as 
\begin{equation}
\label{matrix2}
\mathcal{M}(k)=
\begin{pmatrix}
 \mu & \gamma \\
 \gamma & {\mu}^*
 \end{pmatrix},
\end{equation}
with the coefficient, $\mu \equiv \mu(k)=2D_{1d}(\cos k-1)-\gamma + i 2 \epsilon \sin k$, and $\gamma$ represents the transition rate between the states. Next, taking the Laplace transform of Eq.~(\ref{matrix11}) yields
\begin{equation}
\label{matrix11a}
\left [s\mathbb{I}-\mathcal{M}(k)\right]~
\ket{\tilde P_m (k,s)}
= \ket{\tilde P_m (k,t=0)},
\end{equation}
where $\mathbb{I}$ is the two dimensional identity matrix.
As we start the process with a symmetric initial condition with equal probabilities of being in states $0$ and $1$, the initial condition in the Fourier domain is $\ket{\tilde P_m (k,t=0)}~=~{\begin{pmatrix}
\frac{1}{2} 
\frac{1}{2}
\end{pmatrix}}^T$.
Solving Eq.~(\ref{matrix11a}) along with the initial conditions, we obtain the Fourier-Laplace transform of the occupation probability as,
\begin{equation}
\label{fl}
\tilde P (k,s)=\frac{1}{(2D_{1d}(1-\cos k)+s)+\frac{4 \epsilon^2 \sin^2 k}{(2D_{1d}(1-\cos k)+s+2 \gamma)}},
\end{equation}
where  $\tilde P(k,s)~=~\tilde P_0(k,s)+\tilde P_1(k,s)$.
Next, performing a Laplace inversion of Eq.~(\ref{fl}) yields
\begin{small}
\begin{eqnarray}
\label{comp}
&&\tilde{P}(k,t)=\nonumber\\&&e^{-t(2D_{1d}(1-\cos{k})+{\gamma})}\Bigg[\cosh{\left( t R(k) \right)}
+\frac{\gamma}{R(k)}\sinh{\left( t R(k)\right)}\Bigg], 
\end{eqnarray} 
\end{small}
where $R(k)$ is defined as
\begin{equation}
\label{rk}
R(k)=\sqrt{\gamma^2-4 \epsilon^2 \sin^2 k}.
\end{equation}
Equation~(\ref{comp}) is the exact expression for the Fourier transform of the site occupation probability of a run and tumble particle on a one dimensional infinite lattice. This expression can also be alternatively derived by diagonalizing the matrix $\mathcal{M}(k)$ provided in Eq.~(5) for each $k$ and solving the resulting eigenvalue equation for symmetric initial conditions $P_0( x, t = 0 ) = P_1 ( x, t = 0 ) = (1/2) \delta_ {x,0}$. We have provided the details of this calculation in Appendix A. 
The occupation probability in real space $P(x,t)$, can be obtained by taking the inverse Fourier transform of $\tilde P (k,t)$. Since we consider an infinite lattice, this is given as
\begin{equation}
\label{inv_ft0}
P(x,t)=\frac{1}{2 \pi}\int_{-\pi}^{\pi}dk e^{-ikx}\tilde{P}(k,t).
\end{equation}

We next analyze the continuum limit of the lattice expressions for the site occupation probabilities. In the $k \rightarrow 0$ limit, the expression in Eq.~(\ref{comp}) converges to the result in Ref.~\cite{malakar2018steady} for a diffusive RTP in continuous space with diffusion coefficient $D_{1d}$ and velocity $v=2 \epsilon$. Since we study the process in discrete space and  continuous time, the non-trivial limit of the lattice walk we consider in order to derive the distribution in continuous space is the $k \to 0$ limit keeping $k^2 D_{1d}$ and $k \epsilon$ fixed.
Keeping relevant terms up to  $\mathcal{O}\left(k^2\right)$ in Eq.~(\ref{comp}) yields
\begin{small}
\begin{eqnarray}
\label{comp1}
&&\lim_{k \rightarrow 0}\tilde{P}(k,t)\nonumber\\&&=e^{-t\left(D_{1d} k^2+\gamma\right)}\Bigg[ \cosh{\left( t R_c(k) \right)}
+\frac{\gamma}{R_c(k)}\sinh{\left( t R_c(k)\right)}\Bigg],
\end{eqnarray} 
\end{small}
where 
$R_c(k)=\lim_{k \to 0} R{(k)}=\sqrt{\gamma^2-4\epsilon^2{k}^2}$ refers to the continuum limit of the expression in Eq.~(\ref{rk}).
An inverse Fourier transform of Eq.~(\ref{comp1}) yields the probability distribution in continuous space, given as
\begin{equation}
\label{inv_ft1}
{P(x,t)}_{cont}=\lim_{k \rightarrow 0}\frac{1}{2 \pi}\int_{-\infty}^{\infty}dk e^{-ikx}\tilde{P}(k,t).
\end{equation}
Evaluating the integral in Eq.~(\ref{inv_ft1}) along with the expression in Eq.~(\ref{comp1}) yields
\begin{small}
\begin{equation}
\label{comp3}
\begin{aligned}
{P(x,t)}_{cont}=\\
&\ \frac{\gamma   e^{- \gamma t}}{4\epsilon
    \sqrt{4 \pi D_{1d} t}} \int_{-\infty}^{\infty}dy e^{-\frac{ (x-y)^2}{4D_{1d} t}}\Theta (t-\frac{\left| y\right|}{2 \epsilon} ) \\
    & \left[I_0\left(\gamma 
   \sqrt{t^2-\frac{y^2}{4 \epsilon^2}}\right)+\frac{t I_1\left(\gamma  \sqrt{t^2-\frac{y^2}{4 \epsilon^2}}\right)}{\sqrt{t^2-\frac{y^2}{4 \epsilon^2}}}\right]\\
      & +\frac{ \cosh \left(\frac{ \epsilon}{D_{1d}} x\right) e^{-\frac{\epsilon^2}{D_{1d}} t-\frac{x^2}{4 D_{1d}t}-\gamma  t}}{ \sqrt{4 \pi D_{1d} t}}, \\
\end{aligned}
\end{equation}
\end{small}
where $I_n$ is the modified Bessel function of the first kind of order $n$ and $\Theta$ is the Heaviside step function. We note that this continuum limit answer matches the expression derived in Ref.~\cite{malakar2018steady} for the probability that a run and tumble particle is at position $x$ at time $t$.

To test the above theoretical predictions, we perform kinetic Monte Carlo simulations. Kinetic Monte Carlo simulations~\cite{prados1997dynamical,bortz1975new,voter2007introduction} unlike Monte Carlo simulations allow us to relate simulation steps to physical time. Usual Monte Carlo techniques require very small time discretization for accurate integration. But the processes we study take place at large timescales and the system essentially remains inactive at the short timescales. Kinetic Monte Carlo techniques overcome this limitation by performing direct jumps to events thus saving simulation time. The knowledge of the rates describing the motion of the particle helps to associate a Poisson distributed time interval between consecutive events. The length of the time intervals varies during the simulation. In the RTP model considered in this paper, the events could be either the hop to an adjacent lattice site or the sudden tumble (in other words; change of internal state) of the particle. The initial position of the particle is set at the origin. An update in the position or the state of the particle (which in turn is determined by the corresponding probability rates) occurs after every Poisson distributed time interval.

\begin{figure}[t]
\begin{center}
 \includegraphics[width=1.05\linewidth]{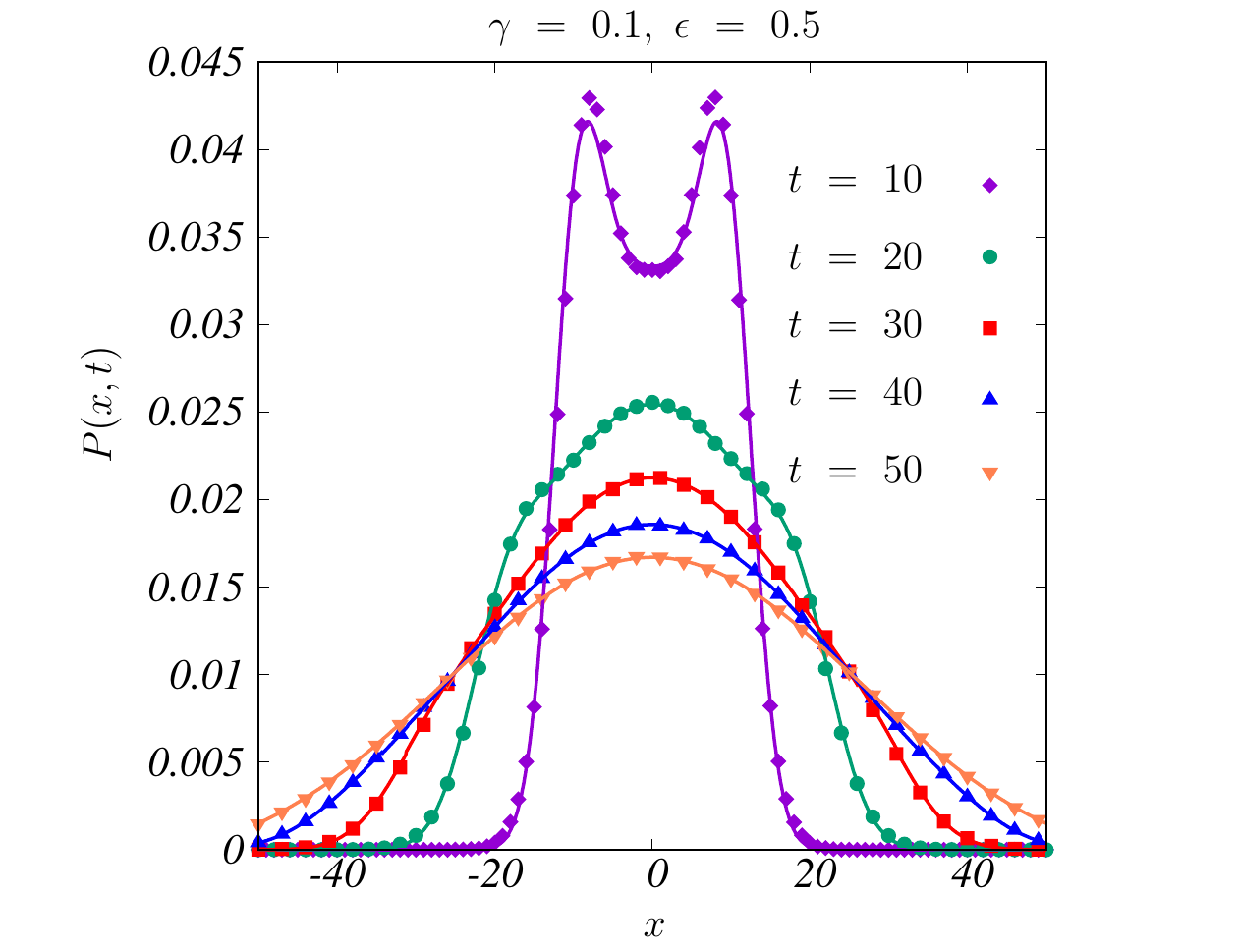}
 \caption{Occupation probability of a RTP on a one dimensional infinite lattice $P(x,t)$, obtained from performing continuous time kinetic Monte Carlo simulations (points) plotted against the theoretical result for the continuum limit of the occupation probability given in Eq.~(\ref{comp3}) (solid curves) at different times, $t$. In the long time limit ($t\gg\frac{1}{\gamma}$), both the results agree well. The fixed parameter values used are $\gamma=0.1$, $\epsilon=0.5$ and $D_{1d}=0.5$. The simulation data is averaged over ${10}^7$ realizations.}\label{fig_pxt_1d}
 \end{center}
\end{figure}

In Fig.~\ref{fig_pxt_1d}, we compare the expression in Eq.~(\ref{comp3}) with kinetic Monte Carlo simulation results for the occupation probability $P(x,t)$, of a RTP on a one dimensional infinite lattice with lattice spacing $a=1$. In the long time limit ($t\gg\frac{1}{\gamma}$), active particle motion in discrete space converges to the motion in continuous space. 

\subsection{Two Dimensions}

Analogous to the one dimensional case discussed above, the occupation probabilities for the four states in two dimensions are governed by the following coupled differential equations
\begin{small}
\begin{eqnarray}
\label{sq0}
\hspace{-0.2 cm}
\frac{\partial P_{m}}{\partial t} = D_{2d} {\nabla}^2 P_{m} -  \epsilon {\bf \hat{m}} . \vec{\nabla}P_{m} +\frac{\gamma}{2}\left( P_{m_{+1}} + P_{m_{-1}}-2 P_{m}\right),
\end{eqnarray}
\end{small}
where $D_{2d}$ refers to the intrinsic diffusion constant  associated with the active particle motion in two dimensions. Here, $P_m=P_m(x,y,t)$ and the subscript $m$ denotes the internal state ($m=0,~1,~2,~3$). 
The modulo operator $\mod\left(m,n\right)$ represents the remainder when $m$ is divided by $n$.
We also define $m_{+1}=\mod\left(m+5,4\right)$ and $m_{-1}=\mod\left(m+3,4\right)$.
$\vec{\nabla}$ is the discrete gradient operator and $\nabla^2$ is the discrete Laplacian operator on the square lattice defined as
\begin{eqnarray} 
{\bf \hat{x}} . \vec{\nabla} P_m(x,y,t) &=&  P_m(x+1,y,t)-P_m(x-1,y,t) ,\nonumber\\
{\bf \hat{y}} . \vec{\nabla} P_m(x,y,t) &=&   P_m(x,y+1,t)-P_m(x,y-1,t) ,\nonumber\\
{\nabla}^2 P_{m}(x,y,t)&=& P_m(x+1,y,t)+P_m(x-1,y,t)\nonumber\\&&+P_m(x,y+1,t)+P_m(x,y-1,t)\nonumber\\&&-4 P_m(x,y,t).
\end{eqnarray}
Here, ${\bf \hat{m}}$ denotes the bias direction which is ${\bf \hat{x}},~{\bf \hat{y}},~{\bf -\hat{x}}$ or ${\bf -\hat{y}}$ for states $0,~1,~2$ and $3$ respectively. The intrinsic diffusion constant associated with run and tumble particle motion in two dimensions is denoted as $D_{2d}$.

We define the Fourier transform of the occupation probability of a RTP in the $m^{th}$ state on a two dimensional infinite square lattice as $\tilde P_m(k_x,k_y,t)=\sum_{x=-\infty}^{\infty}\sum_{y=-\infty}^{\infty}e^{i(k_xx+k_yy)}P_m(x,y,t)$ and the Laplace transform of $\tilde P_m(k_x,k_y,t)$ as $\tilde P_m(k_x,k_y,s)=\int_0^{\infty}dte^{-s t}\tilde P_m(k_x,k_y,t) $.
Taking the Fourier transform of Eq.~(\ref{sq0}) yields
\begin{equation}
\label{big}
\frac{\partial}{\partial t}
\ket{\tilde P_m (k_x,k_y,t)}
=
\mathcal{Q}(k_x,k_y)
\ket{\tilde P_m (k_x,k_y,t)},
\end{equation}
where the ket $\ket{\tilde P_m (k_x,k_y,t)}$ represents the column vector given as 
$
\ket{\tilde P_m (k_x,k_y,t)}
=
 \begin{pmatrix}
\tilde P_0(k_x,k_y,t) \\
\tilde P_1(k_x,k_y,t)\\
\tilde P_2(k_x,k_y,t)\\
\tilde P_3(k_x,k_y,t)
\end{pmatrix},
$
and the matrix $\mathcal{Q}(k_x,k_y)$ is defined as 
\begin{small}
\begin{equation} 
\label{qmatrix}
\mathcal{Q}(k_x,k_y)= \begin{pmatrix}
\mu & \gamma/2 & 0 & \gamma/2 \\
\gamma/2 & \nu & \gamma/2 & 0\\
0 & \gamma/2 & {\mu}^* & \gamma/2\\
\gamma/2 & 0 & \gamma/2 & {\nu}^*
\end{pmatrix}.
\end{equation}
\end{small}
The coefficients $\mu \equiv \mu(k_x,k_y)$ and $\nu \equiv \nu(k_x,k_y)$ are given as
\begin{eqnarray}
\mu(k_x,k_y)&=&2 D_{2d}[\cos k_x+\cos k_y-2]-\gamma + i 2 \epsilon \sin k_x,\nonumber\\
\nu(k_x,k_y)&=&2 D_{2d}[\cos k_x+\cos k_y-2]-\gamma + i 2 \epsilon \sin k_y.\nonumber\\
\end{eqnarray}
Next, taking the Laplace transform of Eq.~(\ref{big}) yields
\begin{equation}
\label{matrix101a}
\left [s\mathbb{I}-\mathcal{Q}(k_x,k_y) \right ]~
\ket{\tilde P_m (k_x,k_y,s)}
=
\ket{\tilde P_m (k_x,k_y,t=0)}.
\end{equation}
Here, $\mathbb{I}$ is the  four dimensional identity matrix. We assume symmetric initial conditions with equal probabilities of being in any of the four possible internal states. Thus the initial conditions in the Fourier space reduce to $ {\ket{\tilde P_m (k_x,k_y,t=0)}
=
{\begin{pmatrix}
\frac{1}{4} 
\frac{1}{4} 
\frac{1}{4} 
\frac{1}{4}
\end{pmatrix}}}^T$.
We next solve Eq.~(\ref{matrix101a}) along with the initial conditions to obtain the occupation probability in Fourier-Laplace domain, $\tilde P(k_x,k_y,s)=\tilde P_0(k_x,k_y,s)+\tilde P_1(k_x,k_y,s)+\tilde P_2(k_x,k_y,s)+\tilde P_3(k_x,k_y,s)$. This expression is quite large, and we provide the exact expression for $\tilde P(k_x,k_y,s)$ in Eq.~(\ref{pks2dn}).

Unfortunately, it is hard to invert the Fourier-Laplace transform exactly and obtain the full two dimensional occupation probability, $P(x,y,t)$. Hence we study the $x$ and $y$ motion separately.
We define the marginal occupation probabilities in $x$ and $y$ as 
$
P(x,t)=\sum_{y=- \infty}^{\infty}P(x,y,t)
$
and
$
P(y,t)=\sum_{x=- \infty}^{\infty}P(x,y,t).
$
We use the same symbol for the occupation probability in one dimension and the marginal occupation probability in two dimensions to avoid a proliferation of symbols.
We focus on the marginal function in $x$, which is the same as the marginal function in $y$ because of the symmetric initial conditions. The Fourier-Laplace transform of the marginal occupation probability in $x$ defined as $\tilde P (k_x,s)=\sum_{x=-\infty}^{\infty}\int_0^{\infty}dte^{ik_x x-s t}P(x,t)$ can be obtained from the two dimensional occupation probability given in Eq.~(\ref{pks2dn}) by setting $k_y=0$. This expression is provided in Eq.~(\ref{pks2dnprojected}). Performing the Laplace inversion of Eq.~(\ref{pks2dnprojected}) yields the marginal occupation probability in Fourier space, $
\tilde P(k_x,t)=L^{-1}\left[\tilde P(k_x,s) \right]$. We find
\begin{small}
\begin{eqnarray}
\label{comp2d}
\tilde{P}(k_x,t)&=&\Bigg[\frac{\gamma^2-2\epsilon^2\sin^2(k_x)}{{R(k_x)}^2}\cosh{\left( t R(k_x) \right)-\frac{2\epsilon^2\sin^2(k_x)}{{R(k_x)}^2}}\nonumber\\ 
&&+\frac{\gamma}{R(k_x)}\sinh{\left( t R(k_x)\right)}\Bigg]e^{-t(2D_{2d}(1-\cos{k_x})+ \gamma)}, \nonumber\\&&
\end{eqnarray} 
\end{small}
where the function $R(k_x)$ is defined in Eq.~(\ref{rk}). The occupation probability in real space $P(x,t)$, can be obtained by performing the inverse Fourier transform of $\tilde P (k_x,t)$.

We next take the continuum limit of Eq.~(\ref{comp2d}) and analyze the marginal distributions in the continuum limit.
In the limit $k_x \rightarrow 0$, this expression converges to the result in Ref.~\cite{santra2020run} for the Fourier transform of the marginal occupation probability distribution in two dimensions for a four state RTP model in continuous space with velocity $v=2 \epsilon$ (except for the diffusion term which is absent in their model). Similar to the case in one dimension, the terms ${k_x}^2D_{2d}$ and $k_x \epsilon$ are held fixed. Keeping the relevant terms up to $O\left({k_x}^2\right)$ in Eq.~(\ref{comp2d}) yields
\begin{eqnarray}
\label{comp12d}
\lim_{k_x \rightarrow 0}\tilde{P}(k_x,t)&=&\Bigg[\frac{\gamma^2-2\epsilon^2{k_x}^2}{{R_c(k_x)}^2}\cosh{\left( t R_c(k_x) \right)}-\frac{2\epsilon^2{k_x}^2}{{R_c(k_x)}^2}\nonumber\\
&&+\frac{\gamma}{R_c(k_x)}\sinh{\left( t R_c(k_x)\right)}\Bigg]e^{-t( D_{2d}{k_x}^2+\gamma)}. \nonumber\\&&
\end{eqnarray}
where $R_c(k_x)$ is the continuum limit of $R(k_x)$ defined as $R_c(k_x)=\lim_{k_x \to 0}R{(k_x)}=\sqrt{\gamma^2-4\epsilon^2{k_x}^2}$.
The Fourier transform in Eq.~(\ref{comp12d}) can be inverted exactly as in Ref.~\cite{santra2020run} which gives the marginal distribution in continuous space, ${P(x,t)}_{cont}=\int_{- \infty}^{\infty}dy{P(x,y,t)}_{cont}$. 

\section{MOMENTS, CUMULANTS AND LARGE DEVIATION FUNCTIONS}
\label{mc}
We next analyze the moments and cumulants associated with RTP motion on the lattice and in the continuum, and show that they follow a large deviation principle. Using the exact forms of the occupation probabilities in Fourier and Fourier-Laplace domain, we analyze the large deviation free energy and rate functions in one and two dimensions.
\subsection{One Dimension}
\begin{figure} [t]
\begin{center}
 \includegraphics[width=1.05\linewidth]{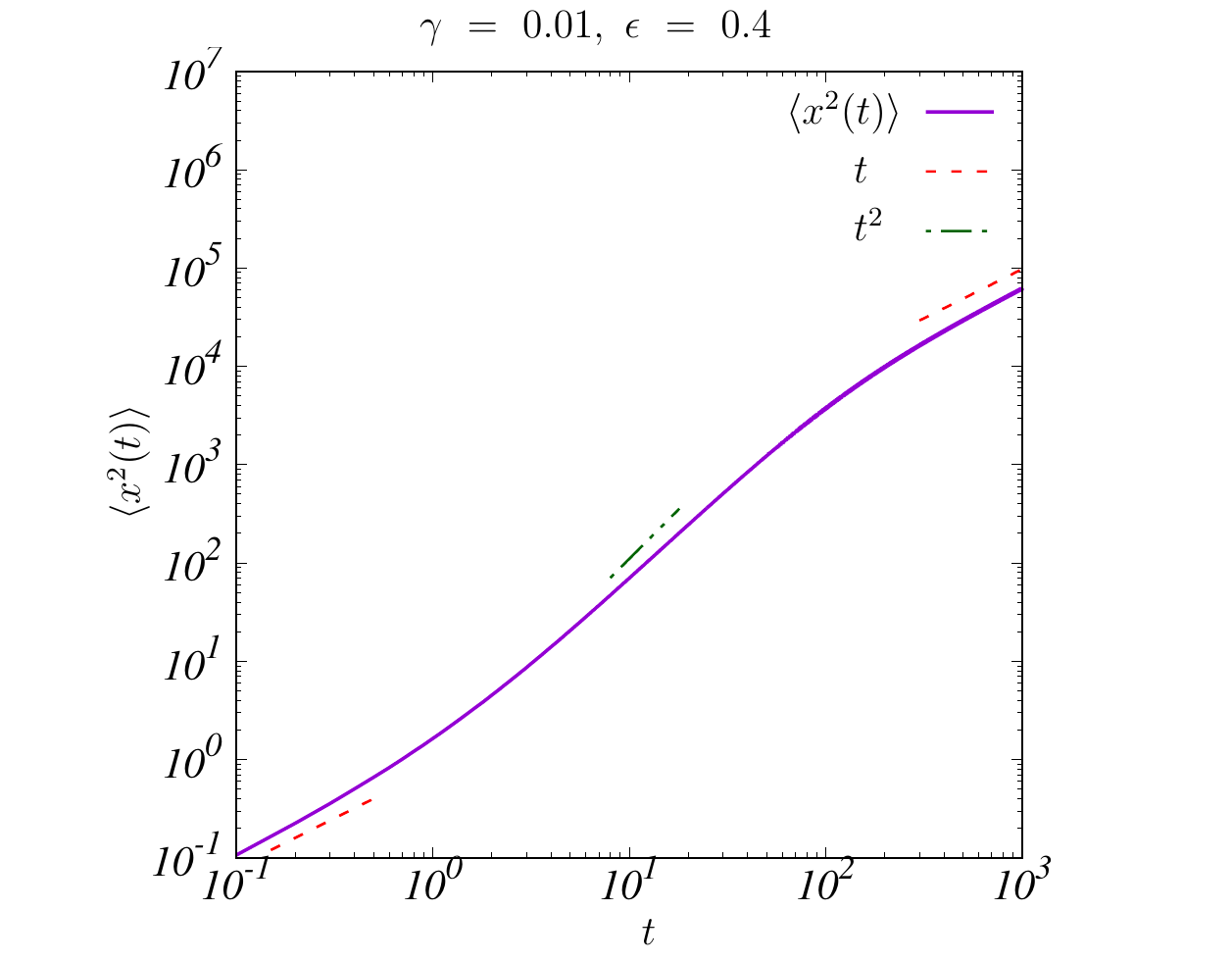}
 \caption{Mean square displacement of a RTP in one dimension $\langle {x^2(t)} \rangle$, given in Eq.~(\ref{xsq1d}), as a function of time. The values in the $x$ and $y$ axes are given in ${\log}_{10}$ scale. The figure displays three scaling regimes where the motion is diffusive at very short timescales ($\langle {x^2(t)} \rangle \propto t$,
till $t \approx \frac{D_{1d}}{2\epsilon^2}$), ballistic at intermediate timescales ($\langle {x^2(t)} \rangle \propto t^2$,
for $t \approx \frac{1}{\gamma}$) and again diffusive but with an enhanced diffusion
coefficient at large timescales ($\langle {x^2(t)} \rangle \propto t$,
for $t \gg\frac{1}{\gamma}$). The fixed parameters used in the plot are $\gamma=0.01$ and $\epsilon=0.4$.  The intrinsic diffusion constant $D_{1d}$ is set equal to $0.5$. }\label{fig_sec_moment_1d}
 \end{center}
\end{figure}

In one dimension, the Fourier transform of the occupation probability in Eq.~(\ref{comp}) is equivalent to the moment generating function. The moments can be computed as
\begin{eqnarray}
\langle {x^n(t)} \rangle &=& \frac{1}{i^n}\frac{\partial^n \tilde P(k,t)}{\partial k ^n} \bigg\rvert_{k=0}.
\label{eq_moments_1d}
\end{eqnarray}
From symmetry, the odd moments are zero. We have
\begin{equation} 
\langle x^{2n+1}(t) \rangle =0,~n=0,1,2,3,...~.
\end{equation}
The second moment has the explicit form
\begin{eqnarray}
\label{xsq1d}
\langle {x^2(t)} \rangle &=& -\frac{\partial^2 \tilde P(k,t)}{\partial k ^2} \bigg\rvert_{k=0}\nonumber\\&&= 2 {\mathcal{D}}_{1d}t- \frac{2\epsilon^2}{\gamma^2}(1-e^{-2 \gamma t}),
\end{eqnarray}
where ${\mathcal{D}}_{1d}$ is the modified diffusion constant due to activity in one dimension given as
\begin{equation} 
\label {d1dn}
{\mathcal{D}}_{1d}=  D_{1d}+\frac{2\epsilon^2}{\gamma},
 \end{equation}
and $D_{1d}$ is the intrinsic diffusion constant associated with the particle motion in one dimension.
We note that ${\mathcal{D}}_{1d}$ is the same effective diffusion constant that appears in the continuum version of the model in Ref.~\cite{malakar2018steady}. The Supplemental Material~\cite{SI} lists the first few non zero moments in one dimension for both the discrete and continuum cases. We observe that the moments for the discrete and continuum cases are the same up to the third order.
In the $t \rightarrow 0$ limit, Eq. (\ref{xsq1d}) reduces to
\begin{equation} 
\langle {x^2(t)} \rangle  \xrightarrow[t \rightarrow 0]{} 2 D_{1d} t+4  \epsilon ^2 t^2-\frac{8}{3}\epsilon ^2 \gamma  t^3 +O\left(t^4\right).
\end{equation}
Therefore, RTP motion is diffusive at very short timescales ($\langle {x^2(t)} \rangle \propto t$,
till $t \approx \frac{D_{1d}}{2\epsilon^2}$) and ballistic at intermediate timescales ($\langle {x^2(t)} \rangle \propto t^2$,
for $t \approx \frac{1}{\gamma}$).
In the $t \rightarrow \infty$ limit, the variance of the position of an active lattice walk in one dimension converges to that of a one dimensional Brownian motion with the modified diffusion constant ${\mathcal{D}}_{1d}$ and
$
 \langle {x^2(t)} \rangle \xrightarrow[t \rightarrow \infty]{} 2 {\mathcal{D}}_{1d}t.
$
Therefore, at large times, the behavior is once again diffusive but with an enhanced diffusion
coefficient ($\langle {x^2(t)} \rangle \propto t$,
for $t \gg\frac{1}{\gamma}$).
In Fig.~\ref{fig_sec_moment_1d}, we display the plot of the mean square displacement $\langle {x^2(t)} \rangle$ given in Eq.~(\ref{xsq1d}) as a function of time for fixed parameter values $\gamma$ and $\epsilon$.
 
\begin{figure} [t]
\begin{center}
 \includegraphics[width=1.05\linewidth]{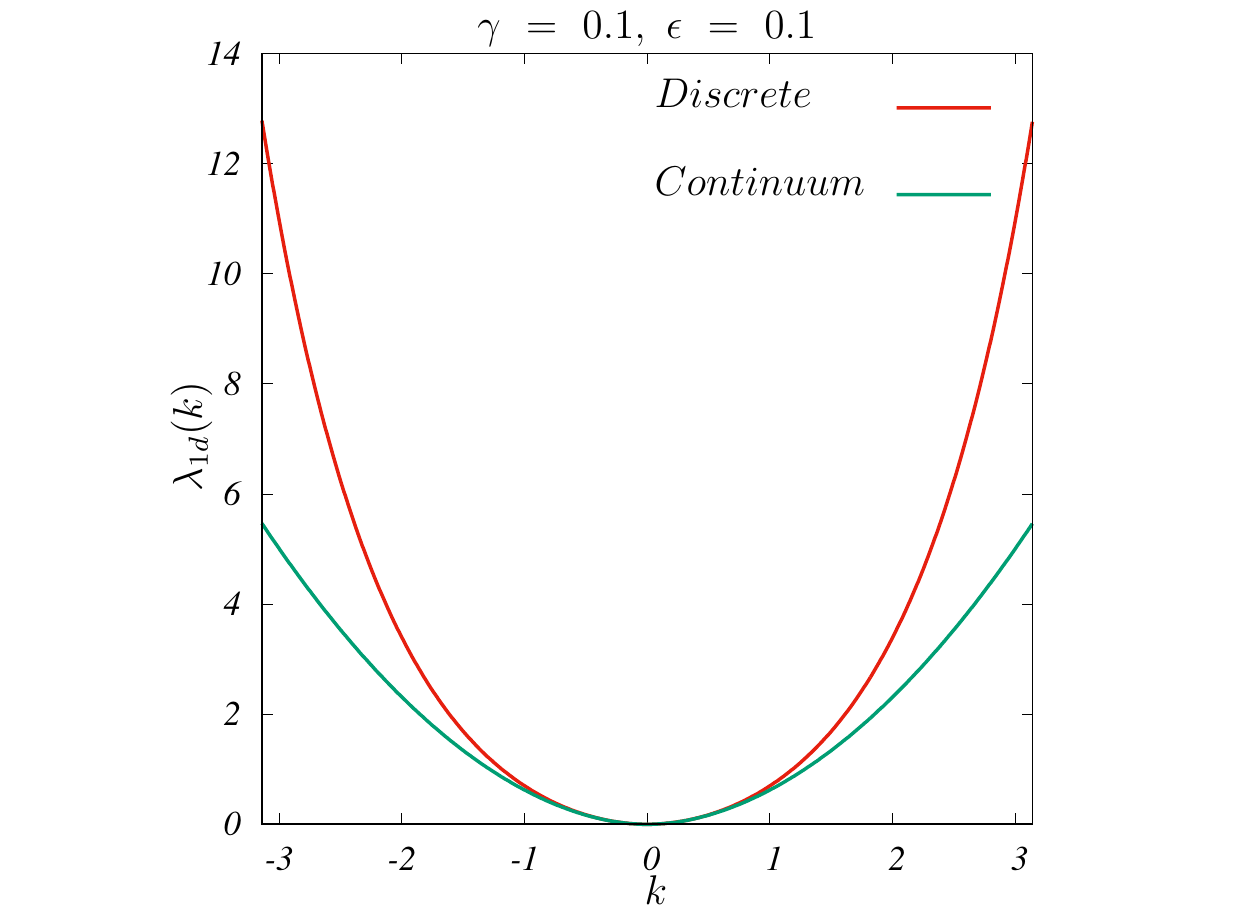}
 \caption{The large deviation free energy function ${\lambda_{1d}(k)}$ computed using Eq.~(\ref{psi1d}) and Eq.~(\ref{cpsi1d}) plotted as a function of $k$ for a RTP in one dimension for the discrete and continuum cases respectively. The fixed parameter values used are $\gamma=0.1,~\epsilon=0.1$ and $D_{1d}=0.5$.}\label{fig_free_energy_1d_comp}
 \end{center}
\end{figure}

We next study the cumulant generating function $\tilde{G}(k,t)$, which is the logarithm of the moment generating function. We have
\begin{small}
\begin{eqnarray}
\tilde{G}(k,t)&=&\log(\tilde{P}(k,t))=-t(2D_{1d}(1-\cos{k})+\gamma) \nonumber\\&&+\log \Big(\cosh{\left( t R(k) \right)}+\frac{\gamma}{R(k)}\sinh{\left( t R(k)\right)}\Big),
\label{eq_gk_1d}
\end{eqnarray} 
\end{small}
where $R(k)$ is defined in Eq.~(\ref{rk}).
The cumulants associated with the displacements at any time $t$ can be computed as
\begin{eqnarray}
{\langle {x^n(t)} \rangle}_c &=& \frac{1}{i^n}\frac{\partial^n \tilde G(k,t)}{\partial k ^n} \bigg\rvert_{k=0}.
\label{eq_cumulants_1d}
\end{eqnarray}
Once again from symmetry, the odd cumulants are zero.
We provide the list of a first few non-zero cumulants in the Supplemental Material \cite{SI}. It is easy to show, using the expression in Eqs.~(\ref{eq_gk_1d}) and (\ref{eq_cumulants_1d}) that the magnitude of all cumulants grow as $t$ in the large time limit. This observation points to a large deviation principle. We investigate the large deviation free energy function $\lambda_{1d}(k)$, which is the scaled cumulant generating function. 
To be consistent with the standard notations, let us replace $k$ with $-ik$. We define
\begin{equation} 
\label{largedevf}
\lambda_{1d}(k)=\lim_{t \longrightarrow \infty}\frac{1}{t}\tilde G(-ik,t).
\end{equation}
The large deviation free energy function $\lambda_{1d}(k)$ is differentiable and the subsequent derivatives with respect to $k$ give the cumulants of the distribution $P(x,t)$ in the long time limit. We have 
\begin{equation}
\lim_{t \longrightarrow \infty}{\langle x^{n}(t) \rangle}_c = t   \frac{\partial^n \lambda_{1d}(k)}{\partial k ^n} \bigg\rvert_{k=0}.
\label{eq_cumulants_free_energy_1d}
\end{equation}
The large deviation free energy function
$\lambda_{1d}(k)$ can be identified to be the largest eigenvalue of the Markov matrix $\mathcal{M}(-ik)$ given in Eq.~(\ref{matrix2}). This can be seen from Eq.~(\ref{large_eigenvalue}) where $\tilde P(k,t)$ can be expressed in terms of the eigenvalues of the evolution matrix. The asymptotic behavior of $\tilde P(k,t)$ is determined by the largest eigenvalue of $\mathcal{M}(-ik)$. As the large time limit involves the exponential of the eigenvalue, another route to arrive at $ \lambda_{1d}(k)$ is by computing the poles of $\tilde P(-ik,s)$ in the variable $s$. Determining the relevant pole of the expression in Eq.~(\ref{fl}), we arrive at the following form for the large deviation free energy function for a RTP on a one dimensional lattice:
\begin{equation} 
\label{psi1d}
\lambda_{1d}(k)=2D_{1d}(-1+\cosh k)-\gamma +\sqrt{\gamma ^2+4 \epsilon ^2 \sinh ^2 k},
\end{equation}
which is the same form that appears in Ref.~\cite{van2019central} for a generalized one dimensional model. The asymptotic limits of the first few cumulants computed using the expression in Eq.~(\ref{eq_cumulants_free_energy_1d}) are matched with the asymptotic limits of the exact expressions for the cumulants valid at all times computed using Eq.~(\ref{eq_cumulants_1d}) in the Supplemental Material \cite{SI}. We next analyze the continuum limit of the large deviation free energy function by computing the small $k$ limit (keeping the relevant terms up to $O(k^2)$) of the above equation with $k^2D_{1d}$ and $k \epsilon$ held fixed. This yields the large deviation free energy function in the continuum limit,
\begin{equation} 
\label{cpsi1d}
{\lambda_{1d}(k)}_{cont}=D_{1d}k^2- \gamma +\sqrt{\gamma ^2+4 k^2 \epsilon ^2}.
\end{equation}
The above expression has the following asymptotic limits:
\begin{equation}
{\lambda_{1d}(k)}_{cont}\xrightarrow[k \rightarrow 0]{}{\mathcal{D}}_{1d}k^2,
\end{equation}
and
\begin{equation}
{\lambda_{1d}(k)}_{cont}\xrightarrow[k \rightarrow \infty]{}D_{1d}k^2+2 \epsilon k,
\end{equation}
where the expression for ${\mathcal{D}}_{1d}$ is provided in Eq.~(\ref{d1dn}). The large deviation free energy function for the discrete and continuum cases for fixed parameter values $\gamma$ and $\epsilon$ is plotted in Fig.~\ref{fig_free_energy_1d_comp}.
 A fundamental quantity of interest is the large deviation function or rate function $I_{1d}(X)$, which describes the asymptotic behavior of the occupation probability $P(X=\frac{x}{t},t)$. We have
\begin{equation}
\label{ldform}
    \lim_{t \rightarrow \infty} P(X,t)=e^{-tI_{1d}(X)}
\end{equation}
where $X=\frac{x}{t}$.
In the rest of this Section, we work with the scaled coordinate $X$ to represent the position.
As the exact expressions for the probability distributions are not always available, another route to deriving the large deviation function is through the large deviation free energy function that is easier to compute. This is accomplished using the Gartner-Ellis Theorem \cite{ellis1984large,touchette2009large} which states that if $\lambda_{1d}(k)$ exists and is differentiable, then $P(X,t)$ obeys a large deviation principle, with the rate function $I_{1d}(X)$ given by the Legendre-Fenchel transform of $\lambda_{1d}(k)$;
\begin{equation}
\label{ldf_exact}
I_{1d}(X)=\underset{k}{\max}\{ kX-\lambda_{1d}(k) \}.
\end{equation}
In the continuum limit, the above equation reduces to
\begin{equation}
\label{cont}
    {I_{1d}(X)}_{cont}=\underset{k}{\max}\{ kX-{\lambda_{1d}(k)}_{cont} \}.
\end{equation}

\begin{figure} [t]
\begin{center}
\includegraphics[width=1.05\linewidth]{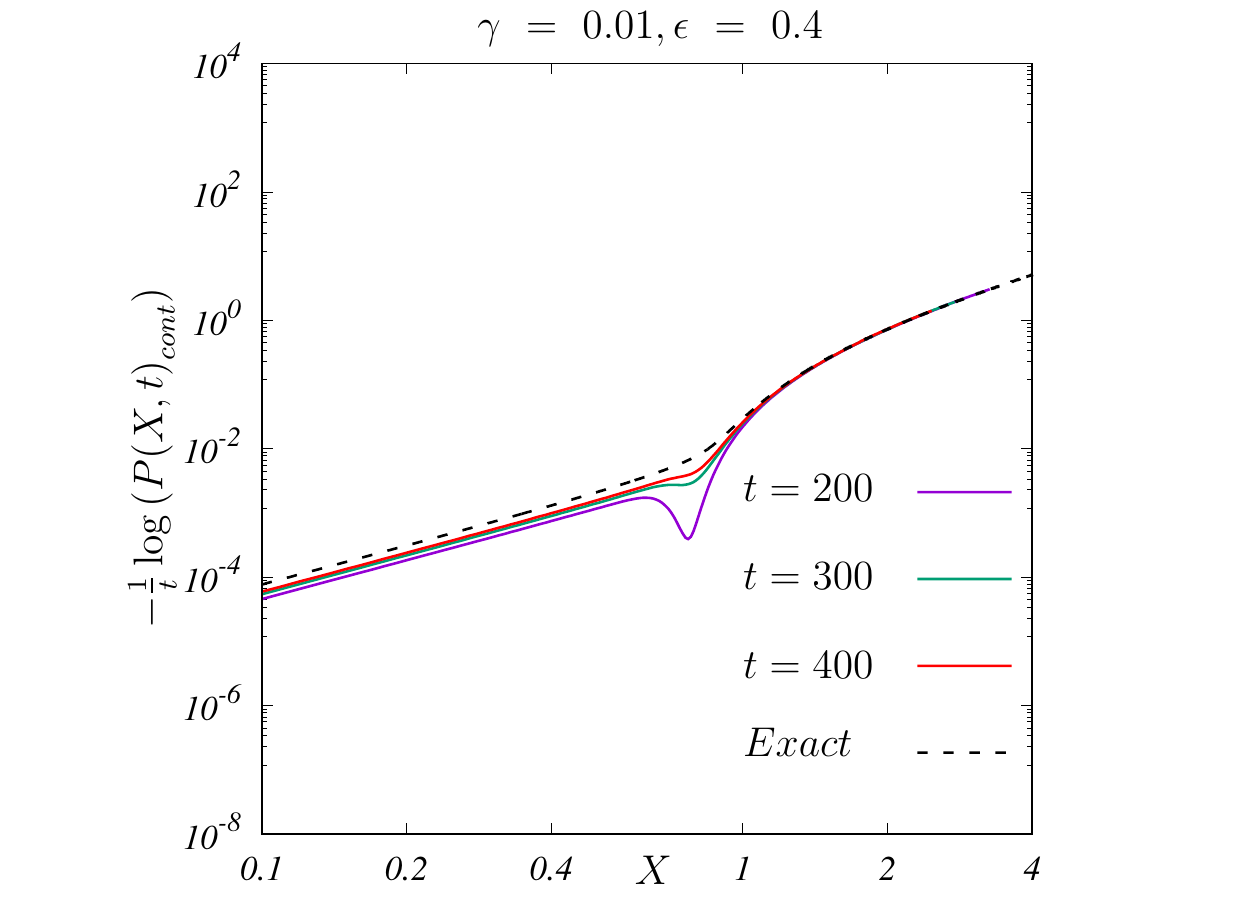}
\caption{The exact large deviation function ${I_{1d}(X)}_{cont}$ computed using Eq.~(\ref{cont}) for a RTP in one dimension in the continuum limit plotted as a function of $X$ for fixed parameter values; $\gamma=0.01,~\epsilon=0.4$ and $D_{1d}=0.5$ (the dashed line). The values in the $y$ axis are given in ${\log}_{10}$ scale. The function displays two regimes with differing diffusive behaviors. The plot is symmetric in $X$ and hence follows the same pattern for negative $X$ values. The solid curves are obtained by performing the numerical integration of Eq.~(\ref{l_t}) using the explicit expression for ${P(X,t)}_{cont}$ given in Eq.~(\ref{comp3}) at different times. At large times, the numerical integration results converge to the theoretical prediction in Eq.~(\ref{cont}). The inverted peaks in the solid curves near the crossover point between the two regimes at finite times reflect the peaks in the occupation probability at short times (refer to Fig.~\ref{fig_pxt_1d}).}
\label{fig_large_dev_1d_comp}
 \end{center}
\end{figure}

As the exact occupation probability distribution in the continuum limit is known (given in Eq.~(\ref{comp3})), we can alternatively use this result to compute the large deviation function in the long time limit as
\begin{equation}
\label{l_t}
{I_{1d}(X)}_{cont}=\lim_{t \rightarrow \infty}-\frac{1}{t}\log \left ({P(X,t)}_{cont} \right ).
\end{equation}
In Fig.~\ref{fig_large_dev_1d_comp}, we plot $-\frac{1}{t}\log \left ({P(X,t)}_{cont} \right )$ at different times computed using the expression for ${P(X,t)}_{cont}$ given in Eq.~(\ref{comp3}). We find that they converge to the right large deviation function computed using Eq.~(\ref{cont}) in the long time limit. 

\begin{figure} [t]
\begin{center}
 \includegraphics[width=1.05\linewidth]{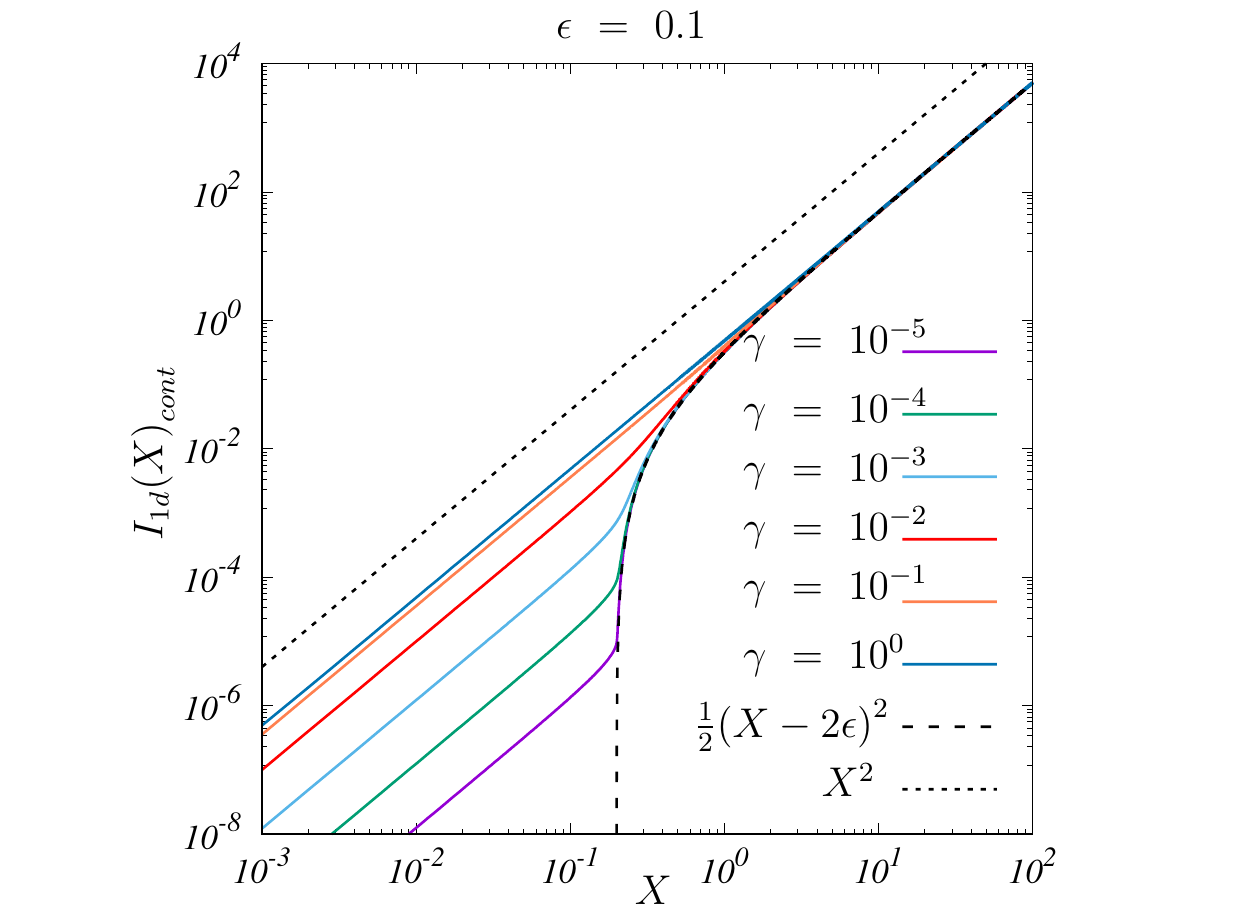}
 \caption{The large deviation function ${I_{1d}(X)}_{cont}$ for a RTP in one dimension in the continuum limit computed using Eq.~(\ref{cont}) plotted as a function of $X$ for different flipping rates $\gamma$. The values in the $x$ and $y$ axes are given in ${\log}_{10}$ scale. The bias $\epsilon$ is fixed to be $0.1$ and the intrinsic diffusion constant $D_{1d}$ is set equal to $0.5$. The rate function grows as $\frac{X^2}{4{\mathcal{D}}_{1d}}$ in the small $X$ limit (up to a scale of  $X \approx \frac{2 \epsilon}{1-\sqrt\frac{D_{1d}}{{{\mathcal{D}}_{1d}}}}$) and then it grows as $\frac{{\left( X-2 \epsilon \right) }^2}{4 D_{1d}}$. The plot is symmetric in $X$ and hence follows the same pattern for negative $X$ values.}\label{fig_large_dev_1d}
 \end{center}
\end{figure}

In Fig.~\ref{fig_large_dev_1d}, we display the plot of the continuum limit large deviation function ${I(X)}_{cont}$, as a function of $X$ which by definition is the Legendre-Fenchel transform of ${\lambda_{1d}(k)}_{cont}$ given in Eq.~(\ref{cpsi1d}) for different parameter values. For small $X$, the rate function ${I_{1d}(X)}_{cont}$ scales as $\frac{X^2}{4{\mathcal{D}}_{1d}}$ and for large $X$, it scales as $\frac{{\left( X-2 \epsilon \right) }^2}{4D_{1d}}$. We have
\begin{equation}
\label{l_theory}
    {I_{1d}(X)}_{cont}= 
\begin{cases}
   \frac{X^2}{4{\mathcal{D}}_{1d}},&\text{for } X \ll \frac{2 \epsilon}{1-\sqrt{\frac{D_{1d}}{{\mathcal{D}}_{1d}}}},\\
     \frac{{\left( X-2 \epsilon \right) }^2}{4D_{1d}},&\text{for } X \gg \frac{2 \epsilon}{1-\sqrt{\frac{D_{1d}}{{\mathcal{D}}_{1d}}}}.
\end{cases}
\end{equation}
Using Eq.~(\ref{l_theory}) along with the continuum limit of Eq.~(\ref{ldform}), we arrive at the large deviation form:

\begin{equation}
\label{pxt_ldf_theory}
  \lim_{t \rightarrow \infty}  {P(X,t)}_{cont}= 
\begin{cases}
    e^{-t\frac{X^2}{4{\mathcal{D}}_{1d}}},&\text{for } X \ll \frac{2 \epsilon}{1-\sqrt{\frac{D_{1d}}{{\mathcal{D}}_{1d}}}},\\
     e^{-t\frac{{\left( X-2 \epsilon \right) }^2}{4D_{1d}}},&\text{for } X \gg \frac{2\epsilon}{1-\sqrt{\frac{D_{1d}}{{\mathcal{D}}_{1d}}}}.
\end{cases}
\end{equation}

Although the exact closed form expression for the large deviation function can be derived using Eq.~(\ref{cont}), this expression is rather long, and we have not quoted this result here. Instead, it is instructive to examine the behavior of the large deviation function in the limit of zero diffusion.
We first analyze the zero diffusion limit of the free energy function in one dimension. 
Setting the intrinsic diffusion constant, $D_{1d}$ to zero yields the free energy function for a purely active process as
\begin{equation} 
\label{cpsi1do}
{\lambda_{1d}(k)}_{cont,0}=- \gamma +\sqrt{\gamma ^2+4 k^2 \epsilon ^2}.
\end{equation}
Here, the subscript $\lq cont,0 \rq$ indicates the continuum limit for the zero diffusion case.
Expanding Eq.~(\ref{cpsi1do}) about $k=0$ yields
\begin{equation}
{\lambda_{1d}(k)}_{cont,0}\xrightarrow[k \rightarrow 0]{} {\mathcal{D}}_{1d,0}k^2 ,
\end{equation}
where 
\begin{equation}
{\mathcal{D}}_{1d,0}=\frac{2 \epsilon^2}{\gamma}, 
\label{d1dn0}
\end{equation}
is the effective diffusion constant for a purely active lattice walk without diffusion in one dimension. In the limit of large $k$, we find
\begin{equation}
{\lambda_{1d}(k)}_{cont,0}\xrightarrow[k \rightarrow \infty]{} 2 \epsilon k .
\end{equation}
The Legendre-Fenchel transform of Eq.~(\ref{cpsi1do}) yields the rate function
\begin{equation}
{I_{1d}(X)}_{cont,0}=\gamma \left ( 1-\sqrt{1-\frac{X^2}{4 \epsilon^2}} \right ),
\label{ldf1d}
\end{equation}
which is non-zero only in the bounded interval $X \in [-2 \epsilon,2 \epsilon]$. This can also be seen from the linear behavior of the large deviation free energy function in the large $k$ limit implying a bound on the maximum scaled displacement. The linear behaviour sets a cutoff distance $X_{\text{max}} = \lim_{k\to \infty}{\lambda_{1d}(k)}_{cont,0}/k$ beyond which the function $kX-{\lambda_{1d}(k)}_{cont,0}$ in Eq.~(\ref{cont}) is unbounded and the maximum does not exist for $X \not\in [-2 \epsilon,2 \epsilon]$. The expression for the large deviation function provided in Eq.~(\ref{ldf1d}) has been previously derived in Refs.~\cite{proesmans2020phase,banerjee2020current, mori2021condensation} for a purely active run and tumble particle without diffusion in one dimension.

An expansion of ${I_{1d}(X)}_{cont,0}$ around $X=0$ yields
\begin{equation}
{I_{1d}(X)}_{cont,0}\xrightarrow[X \rightarrow 0]{}\frac{ X^2}{4 {\mathcal{D}}_{1d,0}}+\frac{X^4}{64 \epsilon ^2 {\mathcal{D}}_{1d,0}}+\frac{ X^6}{512 \epsilon ^4 {\mathcal{D}}_{1d,0}}+...
\end{equation}
where the expression for ${\mathcal{D}}_{1d,0}$ is provided in Eq.~(\ref{d1dn0}). Therefore at late times, the distribution $P(X,t) \sim \exp(-t I_{1d}(X))$ is Gaussian near the origin whereas the tails of the distribution are highly non-Gaussian.

We next present an alternative method to arrive at the limiting forms of the large deviation function in one dimension given in Eq.~(\ref{l_theory}). The run and tumble particle motion we consider is the sum of two independent random processes; active motion without diffusion and a completely diffusive process~\cite{malakar2018steady}. The total probability density ${P(X,t)}_{cont}$ can therefore be written as the convolution of the probability densities corresponding to the two separate processes;
\begin{small}
\begin{equation}
\label{convolution1}
{P(X,t)}_{cont}= \int_{-\infty}^{\infty}{f_A(Y,t)}_{cont}{f_B(X-Y,t)}_{cont}dY,
\end{equation}
\end{small}
where ${f_A(Y,t)}_{cont}$ and ${f_B(X-Y,t)}_{cont}$ represent the position distributions arising from RTP motion without diffusion and diffusion without activity respectively.
 We next consider the limiting distributions for both the processes separately. From the exact expression for the rate function given in Eq.~(\ref{ldf1d}) for a purely active process, we arrive at the following limiting form for the occupation probability in the long time limit:
\begin{eqnarray}
\label{f1}
\lim_{t \rightarrow \infty}{f_A(X,t)}_{cont}=e^{-t \gamma \left ( 1- \sqrt{1-\frac{X^2}{4 \epsilon^2}} \right ) }\Theta \left (2 \epsilon- |X|\right) ,\nonumber\\
\end{eqnarray}
where the Heaviside step function $\Theta$ represents a bound on the spatial extent of the particle, as the distribution ${f_A(X,t)}_{cont}$ is non zero only in the bounded interval $X \in \left [-2 \epsilon, 2 \epsilon \right ]$.
In the small $X$ limit, Eq.~(\ref{f1}) can be approximated as
\begin{eqnarray}
\label{f1_0}
\lim_{t \rightarrow \infty, x \rightarrow 0}{f_A(X,t)}_{cont}=e^{-t \frac{X^2}{4 {\mathcal{D}}_{1d,0}}}\Theta \left (2 \epsilon- |X|\right ).
\end{eqnarray}
For a completely diffusive process with diffusion constant $D_{1d}$, the probability density, ${f_B(X,t)}_{cont}$ in the long time limit assumes the following form:
\begin{eqnarray}
\label{f2}
\lim_{t \rightarrow \infty}{f_B(X,t)}_{cont}=e^{-t \frac{X^2}{4 D_{1d}}}.
\end{eqnarray}

Using Eq.~(\ref{f1_0}) and Eq.~(\ref{f2}) in Eq.~(\ref{convolution1}), we obtain

\begin{figure} [t]
\begin{center}
 \includegraphics[width=1.05\linewidth]{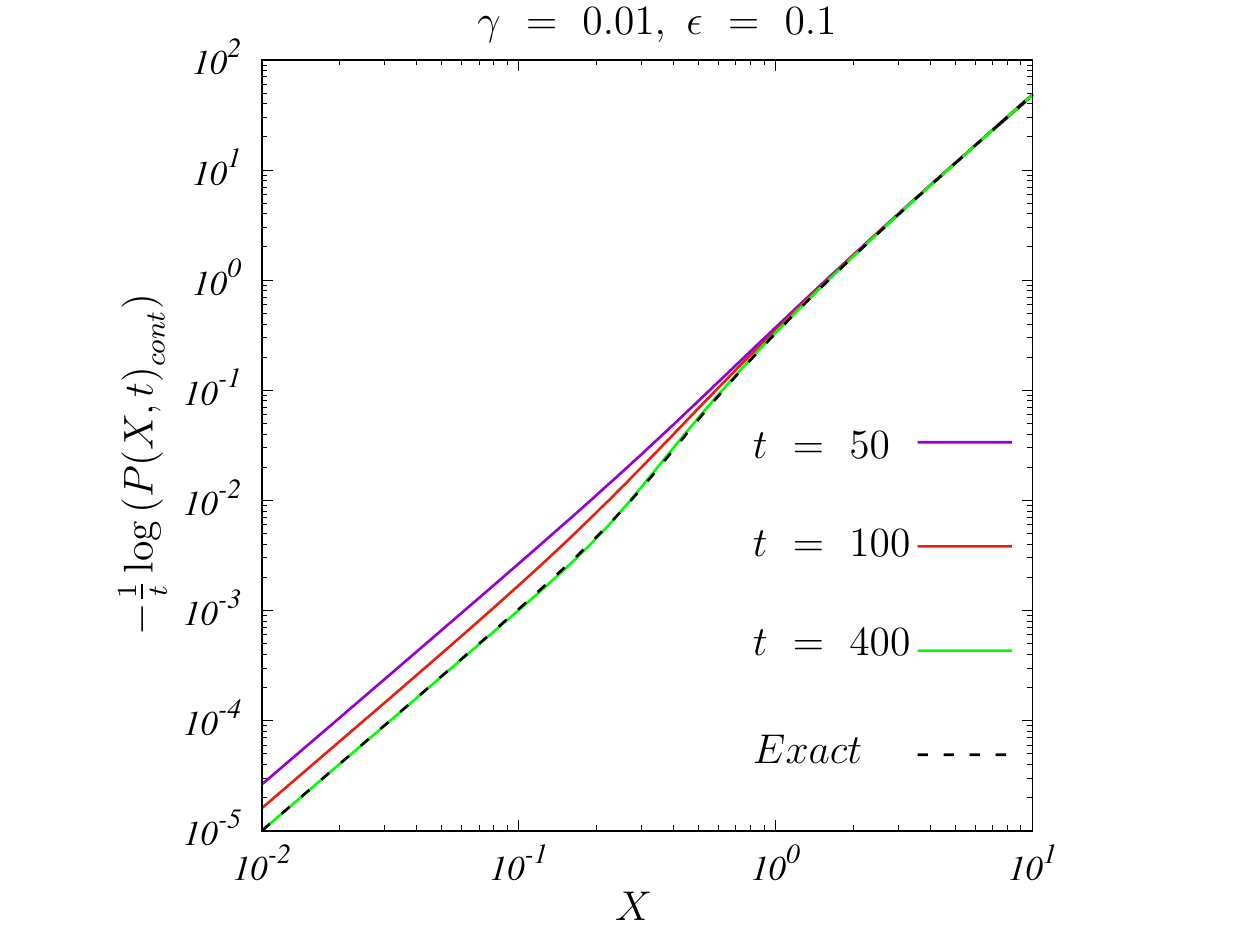}
 \caption{
 The large deviation function ${I_{1d}(X)}_{cont}$ computed using Eq.~(\ref{cont}) for a RTP in one dimension in the continuum limit plotted as a function of $X$ for fixed parameter values $\gamma=0.01$, $\epsilon=0.1$ and $D_{1d}=0.5$ (the dashed curve). The values in the $x$ and $y$ axes are given in ${\log}_{10}$ scale. The solid curves correspond to the function $-\frac{1}{t}\log \left ( {P (X,t)}_{cont}\right )$ at different times, where the explicit form of the function ${P(X,t)}_{cont}$ is given in Eq.~(\ref{erf1dmain0}). At large times, the solid curves converge to the dashed curve. }\label{fig_convolution_1d}
 \end{center}
\end{figure}

\begin{small} 
\begin{equation}
 \label{erf1dmain0}
\lim_{t \rightarrow \infty}{P(X,t)}_{cont}=
  \frac{\text{exp}\left( {-\frac{t X^2 \gamma }{4 D_{1d} \gamma +8 \epsilon ^2}} \right ) \sqrt{2\pi D_{1d} } \epsilon }{\sqrt{t \left(D_{1d}{\gamma }+2 \epsilon ^2\right)}}f(X,t),
\end{equation}
 \end{small}
 where
 \begin{small}
 \begin{eqnarray}
  \label{erf1dmain}
     &&f(X,t)=\nonumber\\&&
         \text{erf}\left(\frac{D_{1d} \gamma+\epsilon  (-X+2 \epsilon )}{\sqrt{\frac{(2 D_{1d}(D_{1d}\gamma+2 \epsilon ^2)}{t}}}\right)+\text{erf}\left(\frac{D_{1d} \gamma+\epsilon  (X+2 \epsilon
   )}{\sqrt{\frac{(2 D_{1d}(D_{1d} \gamma+2 \epsilon ^2)}{t}}}\right),\nonumber\\
 \end{eqnarray}
 \end{small}
and $ \text{erf}$ is the error function. The expression in Eq.~(\ref{erf1dmain0}) along with Eq.~(\ref{erf1dmain}) gives a good approximation to the large deviation function in one dimension (refer to Fig.~\ref{fig_convolution_1d}).

The distribution in Eq.~(\ref{erf1dmain0}) naturally accounts for the two regimes in the rate function of an active particle with diffusion, suggesting an interplay between activity and diffusion in the long time limit. In the first regime, the behavior is diffusive with the diffusion constant having an explicit dependence on the flip rate. In the second regime, the behavior is once again diffusive with a bias-induced shift with no explicit dependence on the flip rate. This can be interpreted as follows: pure active motion can take the particle from the origin to a maximum scaled distance of $X= 2\epsilon$ (which corresponds to a scenario where the particle has not flipped its state up to the time under consideration). The particle can explore regions beyond this scaled distance only through a combination of diffusion and active motion. As we are considering the process in continuous time, the number of steps in a time interval is determined by the microscopic rates, and the individual processes can be considered independently. We can decompose the motion of the particle into a series of active and diffusive steps, where the order of the steps does not affect the final position of the particle. We can therefore analyze the process first with the active steps, which take the particle to a location $Y < 2 \epsilon$, and then perform the diffusive steps from this location. Each of these locations $Y$ therefore represents the source of diffusion, which is precisely the form of the convolution in Eq.~(\ref{convolution1}).
Summing the contribution from each of the sources at $Y$, leads to an effective shift in the origin of the diffusion by the typical scale of $Y$. 

The large deviation function becomes sharper around $X= 2\epsilon$ as the flipping rate $\gamma$ is reduced.
This occurs because for small $\gamma$, the distribution of $Y$ is sharply peaked around $2 \epsilon$, becoming a delta function as $\gamma \to 0$. Therefore in the $\gamma \to 0$ limit, the large deviation function of the full process converges to a Gaussian shifted by $2 \epsilon$. This represents the lower envelope in Fig.~\ref{fig_large_dev_1d}, with the large deviation functions for larger $\gamma$ displaying a crossover between these two regimes. 
Such transitions in the different regimes of the total displacement $R$ in the large deviation function have also been observed in recent studies of different models of active particles~\cite{mori2021condensation,proesmans2020phase,mori2021first}

\subsection{Two Dimensions}

\begin{figure} [t]
\begin{center}
\hspace{-1cm}
 \includegraphics[width=1.05\linewidth]{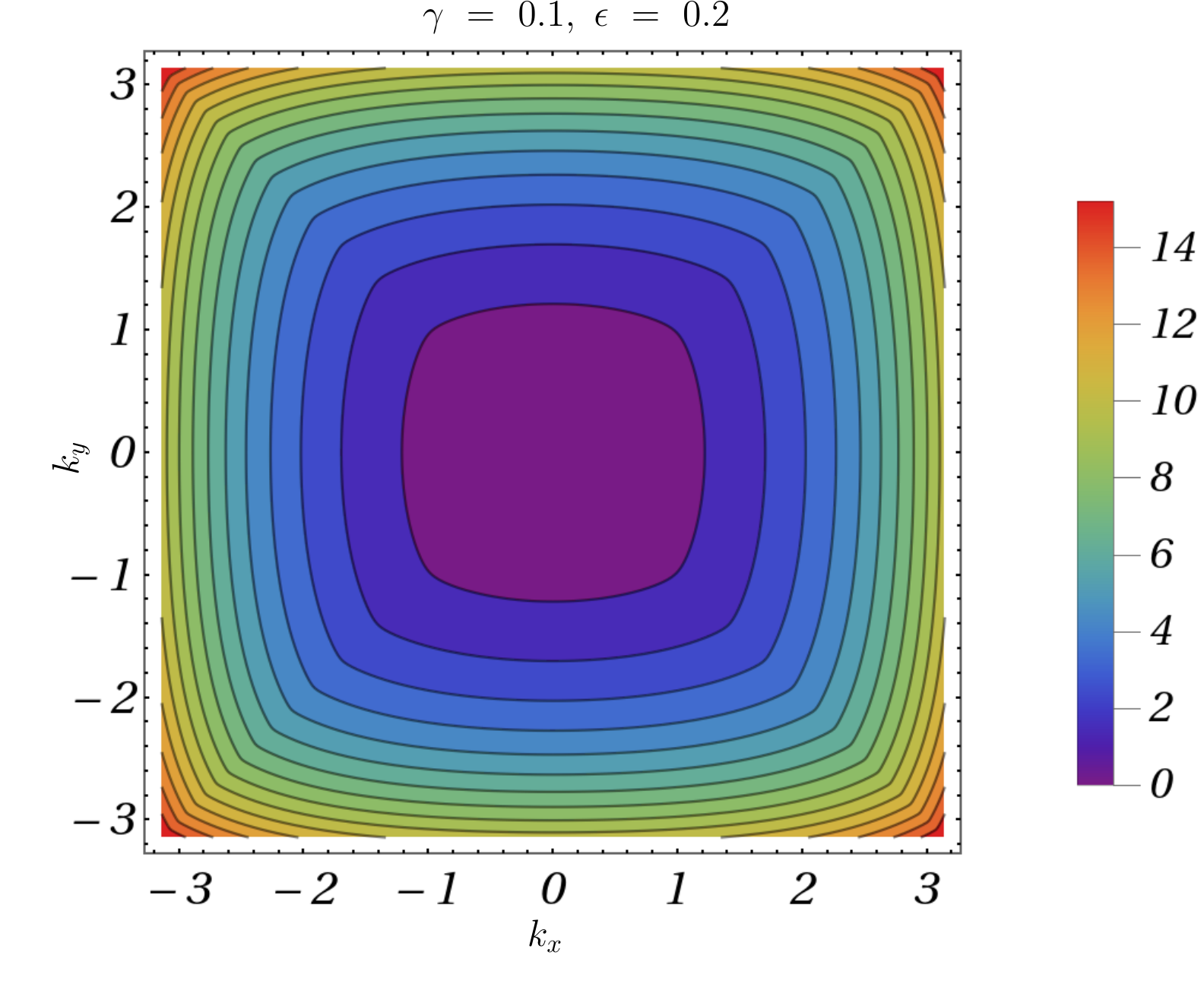}
 \caption{The large deviation free energy function, ${\lambda_{2d}(k_x,k_y)}$ computed using Eq.~(\ref{cpsi2d}) for a RTP on a two dimensional square lattice plotted as a function of $k_x$ and $k_y$. The fixed parameter values used are $\gamma=0.1,~\epsilon=0.2$ and $D_{2d}=0.25$.}\label{fig_free_energy_2d_full}
 \end{center}
\end{figure}

We next analyze the moments, cumulants and the large deviation functions associated with active random walks in two dimensions.
In this case, the Fourier transform of the occupation probability $\tilde P(k_x,k_y,t)$, is not readily available in a simple form. Therefore, in order to compute the moments of the displacements, we utilize the exact expression for the Fourier-Laplace transform of the occupation probability given in Eq.~(\ref{pks2dn}). The moments and the cross correlations between the displacements along $x$ and $y$ directions can then be computed as
  \begin{equation} 
 \langle {{x^n(t)}{y^m(t)}} \rangle = L^{-1}\left[\frac{1}{{(i)}^{n+m}} \frac{\partial^{n}\partial^{m} \tilde P(k_x,k_y,s)}{\partial {k_x} ^{n}\partial {k_y} ^{m}} \bigg\rvert_{k_x=k_y=0}\right].
 \label{eq_moments_2d}
 \end{equation} 
The four-fold rotational symmetry of the distribution of the occupation probability leads to
\begin{equation} 
\langle x^{2n+1}(t)y^m(t) \rangle =\langle x^m(t)y^{2n+1}(t) \rangle=0,
\end{equation}
where $m,~n=0,~1,~2,~3,...~.$
The second moment has the explicit form
\begin{equation} 
\langle {x^2(t)} \rangle=\langle {y^2(t)} \rangle=2 {\mathcal{D}}_{2d}t- \frac{4\epsilon^2}{\gamma^2}(1-e^{- \gamma t}),
\end{equation}
where ${\mathcal{D}}_{2d}$ is the modified diffusion constant due to activity in two dimensions given as
\begin{equation} 
\label{d2d}
{\mathcal{D}}_{2d}=  D_{2d}+\frac{2\epsilon^2}{\gamma}.
 \end{equation}
In the $t \rightarrow 0$ limit, the variance of the position along the $x$-direction has the form
  \begin{equation} 
\langle {x^2(t)} \rangle \xrightarrow[t \rightarrow 0]{}2D_{2d}t+2 \epsilon ^2 t^2 -\frac{2}{3} \gamma  \epsilon ^2 t^3 +O\left(t^4\right).
 \end{equation}
 In the $t \rightarrow \infty$ limit, the variance in the position of an active lattice walk in two dimensions reduces to that of a two dimensional Brownian motion with the modified diffusion constant ${\mathcal{D}}_{2d}$ and $\langle {r^2(t)} \rangle~=~\langle {x^2(t)+y^2(t)} \rangle \xrightarrow[t \rightarrow \infty]{} 4 {\mathcal{D}}_{2d}t$.
We provide the list of a first few non zero moments and cumulants (computed using the expressions for the moments) in two dimensions, as well as a match with direct numerical simulations in the Supplemental Material \cite{SI}.
As in one dimension, the cumulants in two dimensions vary linearly with $t$ in the large time limit suggesting a large deviation form. 
Following the same procedure as in one-dimension, the large deviation free energy function in two dimensions $\lambda_{2d}(k_x,k_y)$ can be determined from the largest eigenvalue of the Markov matrix $\mathcal{Q}(-ik_x,-ik_y)$ in Eq.~(\ref{qmatrix}). We obtain
\begin{small}
\begin{eqnarray} 
\label{cpsi2d}
\lambda_{2d}(k_x,k_y)&=& \frac{1}{2} \Bigg[ \sqrt{2 \left (\gamma^2-4 \epsilon ^2+2 \epsilon ^2 g(2k_x,2k_y)+ f(k_x,k_y) \right )}\nonumber\\&&-2 \gamma -8D_{2d}+4 D_{2d}g(k_x,k_y)\Bigg],
\end{eqnarray}
\end{small}
where the functions $g(k_x,k_y)$ and $f(k_x,k_y)$ are defined as
\begin{equation}
 g(k_x,k_y)=\cosh k_x+\cosh k_y,
\end{equation}
\begin{small}
\begin{eqnarray}
 &&f(k_x,k_y)=\biggl(
\gamma ^4-8 \gamma ^2 \epsilon ^2+4 \epsilon ^4+4 \gamma ^2 \epsilon ^2 g(2k_x,2k_y)\nonumber\\
&&+2 \epsilon ^4 g(4k_x,4k_y)-8\epsilon^4 \cosh (2 k_x) \cosh (2 k_y)
  \biggr)^{\!1/2}.
\end{eqnarray}
\end{small}
The above exact form for the large deviation free energy function demonstrates that the cross-correlations between the $x$ and $y$ motion in two dimensions persist at large times, as the above function does not reduce to a simple product form.
In Fig.~\ref{fig_free_energy_2d_full}, we display a plot of the two dimensional free energy function ${\lambda_{2d}(k_x,k_y)}$, for fixed parameter values $\gamma$ and $\epsilon$.

We next turn our attention to the cumulants of the displacements in $x$ and $y$, which carry the signatures of the correlations between $x$ and $y$ motion. Formally these can be obtained as
\begin{eqnarray}
{\langle x^{n}(t)y^{m}(t) \rangle}_c = \frac{1}{{(i)}^{n+m}} \frac{\partial^{n}\partial^{m} \log \left [ \tilde P(k_x,k_y,t) \right ]}{\partial {k_x} ^{n}\partial {k_y} ^{m}} \bigg\rvert_{k_x=k_y=0}. \nonumber\\
\end{eqnarray}
As the exact expression for $\tilde P(k_x, k_y, t)$ is known only in terms of its Laplace transform, we use the expressions for the moments given in Eq.~(\ref{eq_moments_2d}) to compute the first few cumulants at all times.
An alternative way of obtaining the asymptotic behavior of the cumulants is by using the exact form of the large deviation free energy function. We can compute the asymptotic limit of the cumulants as
\begin{small}
\begin{equation}
\lim_{t \longrightarrow \infty}{\langle x^{n}(t)y^{m}(t) \rangle}_c = t   \frac{\partial^n \partial^m  \lambda_{2d}(k_x,k_y)}{\partial k_x ^n \partial k_y ^m} \bigg\rvert_{k_x=0,~k_y=0} .
\label{eq_cumulants_free_energy_2d}
\end{equation}
\end{small}
We show that the asymptotic limits of the cumulants computed using the above expression match with the asymptotic limits of the exact expressions for the first few cumulants in the Supplemental Material \cite{SI}.

We next analyze the continuum limit of the large deviation free energy function in Eq.~(\ref{cpsi2d}), by taking the $k \to 0$ limit and keeping terms up to $\mathcal{O}(k^2)$. We have
\begin{small}
\begin{eqnarray} 
\label{cpsi2dcont}
{\lambda_{2d}(k_x,k_y)}_{cont}&=&  \Bigg[\frac{1}{\sqrt{2}} \sqrt{\gamma ^2+4
   \left({k_x}^2+{k_y}^2\right) \epsilon ^2+{ f(k_x,k_y)}_{cont}}\nonumber\\&&+D_{2d}({k_x}^2+{k_y}^2)- \gamma \Bigg],
\end{eqnarray}
\end{small}
where
\begin{small}
\begin{equation}
{f(k_x,k_y)}_{cont}=\sqrt{\gamma ^4+8 \left({k_x}^2+{k_y}^2\right) \gamma ^2 \epsilon ^2+16 {\left( {k_x}^2-{k_y}^2 \right)}^2\epsilon ^4}.
\end{equation}
\end{small}
While taking the continuum limit with $k_x \to 0, k_y \to 0$, the terms ${k_x}^2D_{2d}, {k_y}^2D_{2d}, k_x \epsilon$ and $k_y \epsilon$ are held fixed.

\begin{figure} [t]
\begin{center}
\includegraphics[width=1.05\linewidth]{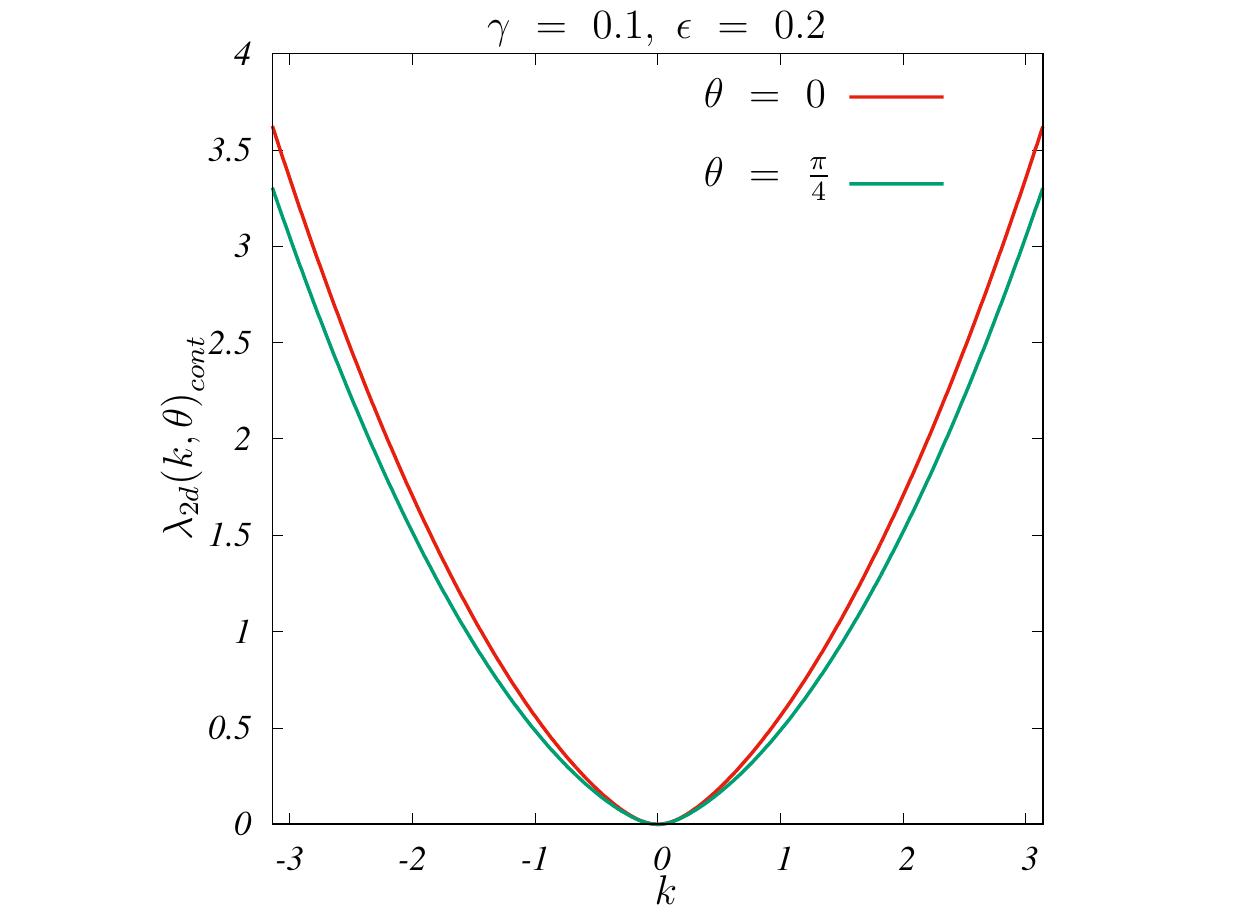}
\caption{The large deviation free energy function projected along the $\theta$ direction, ${\lambda_{2d}(k,\theta)}_{cont}$ plotted as a function of $k$ for two different directions $\theta=0,~\frac{\pi}{4}$. These have been computed using Eq.~(\ref{cpsi2dcontpolar}) for a RTP in two dimensions in the continuum limit. The fixed parameter values used are $\gamma=0.1,~\epsilon=0.2$ and $D_{2d}=0.25$.}
\label{fig_free_energy_2d_mar}
 \end{center}
\end{figure}

This large deviation free energy function has a much simpler form in polar coordinates, given as
\begin{small}
\begin{eqnarray} 
\label{cpsi2dcontpolar}
{\lambda_{2d}(k,\theta)}_{cont}&=& \frac{\gamma  \sqrt{1+k^2 \delta +\sqrt{1+2 k^2 \delta +k^4 \delta ^2 \cos ^2(2 \theta
   )}}}{\sqrt{2}}\nonumber\\&&-\gamma +D_{2d}k^2,\nonumber\\
\end{eqnarray}
\end{small}
where $k=\sqrt{{k_x}^2+{k_y}^2}$, $\theta={\tan}^{-1}\frac{k_y}{k_x}$ and $\delta=4 \frac{\epsilon^2}{\gamma^2}$.
The above expression has the following asymptotic limits:\begin{equation}
{\lambda_{2d}(k,\theta)}_{cont}\xrightarrow[k \rightarrow 0]{}{\mathcal{D}}_{2d}k^2,
\end{equation}
where ${\mathcal{D}}_{2d}$ is given in Eq.~(\ref{d2d}) and
\begin{equation}
{\lambda_{2d}(k,\theta)}_{cont}\xrightarrow[k \rightarrow \infty]{}D_{2d}k^2+k g(\theta),
\end{equation}
where
\begin{equation} 
g(\theta)=\sqrt{2} \epsilon  \sqrt{1+\left| \cos (2 \theta )\right| }.
\label{gtheta}
\end{equation}
It is clear from Eq.~(\ref{cpsi2dcontpolar}) that the projections of the free energy function along various angles are different (as shown in Fig.~\ref{fig_free_energy_2d_mar}), pointing to the fact that at large times, the large deviation functions associated with the process are not rotationally invariant.

Although the Gartner-Ellis theorem can be extended to higher dimensions, we follow the procedure used by us in the one dimensional case to analyze the two dimensional case, using the marginal distributions. We focus on the marginal probability distributions which arise by an integration over one of the spatial coordinates. We use the exact large deviation free energy function to derive 
the rate functions associated with the marginal occupation probability distributions in two dimensions.
For example, the marginal distribution along the $\theta$ direction can be computed as
\begin{small}
\begin{eqnarray}
\nonumber
P_{2d,\theta}(r,t) =  \int_{-\infty}^{\infty} dx \int_{-\infty}^{\infty} dy {P(x,y,t)}_{cont} \delta(r - x \cos \theta - y \sin \theta).\\
\end{eqnarray}
\end{small}
\begin{figure} [t!]
\begin{center}
\vspace{-0.25 cm}
 \includegraphics[width=1.15\linewidth]{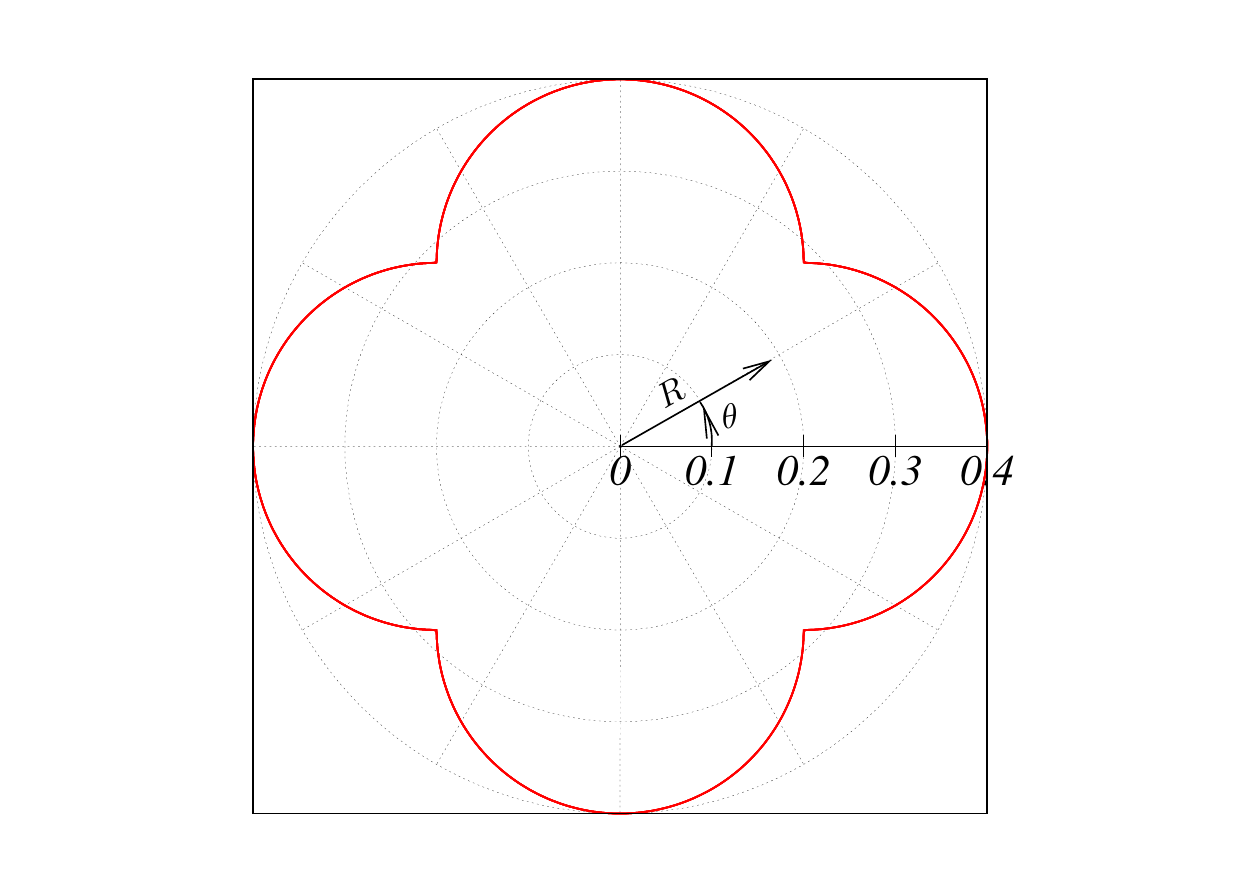}
 \caption{Polar plot of the angular dependence of the bound on the marginal rate function in the zero diffusion limit, $g (\theta)$ given in Eq.~(\ref{gtheta}). Here, $\epsilon$ is fixed to be $0.2$. The shape of the function $g(\theta)$ is a consequence of the inherent square symmetry associated with the two dimensional process. The function $g (\theta)$ represents the arc traced out by the corners of a square as it is rotated by an angle $\theta$, which yields the limits of the projected occupation probability distribution along the $\theta$ direction. } \label{fig_lightcone}
 \end{center}
\end{figure}

It is easy to show that the large deviation free energy function associated with the projected distribution is related to the free energy function in polar coordinates given in Eq.~({\ref{cpsi2dcontpolar}}). The large deviation free energy function at a particular $\theta$ yields the free energy function corresponding to the probability distribution projected along the $\theta$ direction in real space, ${\lambda_{2d}(k,\theta)}_{cont}$. 
This can be seen by substituting $k_x=k_x'\cos \theta-k_y'\sin \theta$ and $k_y=k_x'\sin \theta+k_y'\cos \theta$ in Eq.~(\ref{cpsi2dcont}) and setting one coordinate (say $k_y'$) to zero

As in one dimension, we analyze the zero diffusion limit of the free energy function in Eq.~(\ref{cpsi2dcontpolar}) where we could derive exact closed form expressions for the rate functions associated with the marginal occupation probability distributions in two dimensions. 
If the diffusion constant is zero in Eq.~(\ref{cpsi2dcontpolar}), we have
\begin{small}
\begin{eqnarray} 
\label{cpsi2dcontpolar0}
{\lambda_{2d}(k,\theta)}_{cont,0}&=&\frac{\gamma  \sqrt{1+k^2 \delta +\sqrt{1+2 k^2 \delta +k^4 \delta ^2 \cos ^2(2 \theta
   )}}}{\sqrt{2}}-\gamma.\nonumber\\
\end{eqnarray}
\end{small}
The above expression follows a quadratic dispersion relation in the small $k$ limit
\begin{equation}
{\lambda_{2d}(k,\theta)}_{cont,0}  \xrightarrow[k \rightarrow 0]{}{\mathcal{D}}_{2d,0}k^2,
\end{equation}
where 
\begin{equation}
 {\mathcal{D}}_{2d,0}=\frac{2 \epsilon^2}{\gamma},   
\end{equation}
is the modified diffusion constant due to a purely active process in two dimensions. In the large $k$ limit, we find a linear dispersion relation
\begin{eqnarray} 
\label{cpsi2dcontpolar0k}
{\lambda_{2d}(k,\theta)}_{cont,0}\xrightarrow[k \rightarrow \infty]{}k g(\theta).
\end{eqnarray}
Here, $g(\theta)$ sets a bound on the spatial extent of the marginal rate function without diffusion in real space. In Fig.~\ref{fig_lightcone}, we display the polar plot of this function $g(\theta)$ for the fixed parameter value $\epsilon=0.2$. 
This function displays a four-fold symmetry, which is a consequence of the inherent square symmetry associated with the two dimensional process. This can be understood as follows: the position distribution of the two-dimensional RTP motion without diffusion is bounded within a square shaped region given by $\Theta(2 \epsilon t -|x|)\Theta(2 \epsilon t - |y|)$. The function $g (\theta)$ therefore represents the arc traced out by the corners of this square as it is rotated by an angle $\theta$, which yields the limits of the projected occupation probability distribution along the $\theta$ direction.
\begin{figure} [t]
\begin{center}
\includegraphics[width=1.05\linewidth]{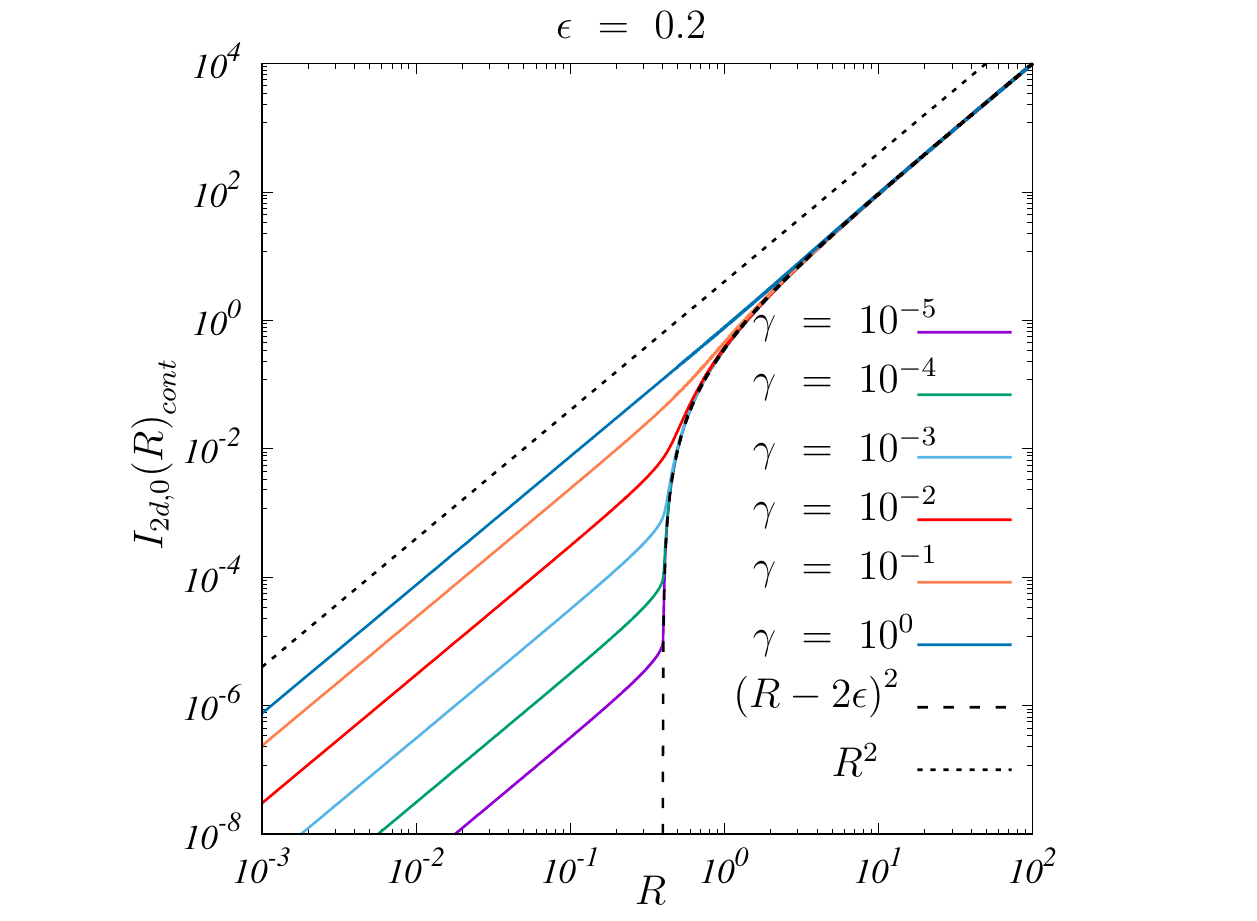}
\caption{The large deviation function projected along the $\theta=0$ direction, ${I_{2d,0}(R)}_{cont}$  plotted as a function of $R$ for different flipping rates $\gamma$ (solid curves). The values in the $x$ and $y$ axes are given in ${\log}_{10}$ scale. These have been computed using Eq.~(\ref{cont2dmarginal0}) for a RTP in two dimensions in the continuum limit. The bias $\epsilon$ is fixed to be $0.2$ and the intrinsic diffusion constant $D_{2d}$ is equal to $0.25$. The rate function grows as $\frac{R^2}{4{\mathcal{D}}_{2d}}$ in the small $R$ limit (upto a scale of  $R \approx \frac{2 \epsilon}{1-\sqrt{\frac{D_{2d}}{{\mathcal{D}}_{2d}}}}$) and then it grows as $\frac{{\left( R-2 \epsilon \right) }^2}{4 D_{2d}}$. The plot is symmetric in $R$ and hence follows the same pattern for negative $R$ values.}\label{fig_large_dev_2d}
 \end{center}
\end{figure}
We next focus on the marginal large deviation function
\begin{equation}
\label{l_t_2d}
{I_{2d,\theta}(R)}_{cont}=\lim_{t \rightarrow \infty}-\frac{1}{t}\log \left ({P_{2d,\theta}(R,t)}_{cont} \right ),
\end{equation}
where $R=\frac{r}{t}$ is the scaled displacement along the $\theta$ direction.
Similar to the procedure in one dimension, we can utilize the Gartner-Ellis theorem to derive the large deviation rate functions associated with the marginal distributions ${I_{2d,\theta}(R)}_{cont}$ in two dimensions, as the marginal distributions under consideration involve a single variable.
The rate functions corresponding to the marginal distribution along the $\theta$ direction can be computed as
\begin{equation}
\label{cont2dmarginal0}
    {I_{2d,\theta}(R)}_{cont}=\underset{k}{\max}\{ kR-{\lambda_{2d,\theta}(k)}_{cont} \}.
\end{equation}

Next, we derive the large deviation function for active motion without diffusion in two dimensions, projected along an arbitrary direction $\theta$.
In the large $k$ regime, the large deviation free energy function displays linear behaviour in $k$ as given in Eq.~(\ref{cpsi2dcontpolar0k}). Similar to the one dimensional case, this implies a bound on the maximum scaled displacement in the large deviation function, when projected along the direction $\theta$. This can be seen through the inversion of the free energy function using the Gartner-Ellis theorem in Eq.~(\ref{cont2dmarginal0}). The linear behaviour therefore sets a cutoff distance $R_{\text{max}} = \lim_{k\to \infty}{\lambda_{2d,\theta}(k)}_{cont,0}/k$. At distances greater than $R_{\text{max}}$, the function $kR-{\lambda_{2d,\theta}(k)}_{cont,0}$ is unbounded, and therefore the maximum does not exist, implying that the large deviation function does not exist for distances $R > R_{\text{max}}$.
Therefore, for arbitrary $\theta$ and for small $R$, Eq.~(\ref{cont2dmarginal0}) predicts
\begin{equation}
\label{ldfmarginal0beta}
{I_{2d,\theta}(R)}_{cont,0}\approx \frac{R^2}{4 {\mathcal{D}}_{2d,0}}\Theta \left( g(\theta)-R \right),
\end{equation}
where ${\mathcal{D}}_{2d,0}=\frac{2 \epsilon^2}{\gamma}$.
The expressions for the exact large deviation functions are greatly simplified along the directions $\theta=0$ and $\frac{\pi}{4}$ in the zero diffusion limit. We have the following closed form expressions:
\begin{small}
\begin{eqnarray}
\label{ldfmarginal00}
    {I_{2d,0}(R)}_{cont,0}&=&\gamma \left ( 1-\sqrt{1-\frac{R^2}{4 \epsilon^2}} \right )\Theta \left( 2 \epsilon-R \right), \nonumber\\
    {I_{2d,\frac{\pi}{4}}(R)}_{cont,0}&=&\frac{\gamma}{2} \left ( 1-\sqrt{1-\frac{R^2}{2 \epsilon^2}} \right )\Theta \left( \sqrt{2} \epsilon-R \right).
\end{eqnarray}
\end{small}
The details of the above calculations are given in Appendix~\ref{Amarginalrate}. Clearly, the large deviation function in two dimensions is not isotropic. This is because the underlying process we consider remains anisotropic even at large times. This is different from the two dimensional cases studied in Refs.~\cite{proesmans2020phase,mori2021condensation} where the direction of movement of the particle is chosen isotropically after each tumble.

Analogous to the one-dimensional case, the corresponding rate functions with diffusion have the form
\begin{equation}
\label{l_theory2d}
    {I_{2d,\theta}(R)}_{cont}= 
\begin{cases}
   \frac{R^2}{4{\mathcal{D}}_{2d}},&\text{for } R \ll \frac{g(\theta)}{1-\sqrt{\frac{D_{2d}}{{\mathcal{D}}_{2d}}}},\\
     \frac{{\left( R-g(\theta) \right) }^2}{4 D_{2d}},&\text{for } R \gg \frac{g(\theta)}{1-\sqrt{\frac{D_{2d}}{{\mathcal{D}}_{2d}}}}.
\end{cases}
\end{equation}
We therefore conclude that the marginal distributions display the large deviation form
\begin{equation}
    \lim_{t \rightarrow \infty}{P_{2d,\theta}(R,t)}_{cont}= 
\begin{cases}
    e^{-t\frac{R^2}{4{\mathcal{D}}_{2d}}},&\text{for } R \ll \frac{g(\theta)}{1-\sqrt{\frac{D_{2d}}{{\mathcal{D}}_{2d}}}},\\
     e^{-t\frac{{\left( R-g(\theta) \right) }^2}{4 D_{2d}}},&\text{for } R \gg \frac{g(\theta)}{1-\sqrt{\frac{D_{2d}}{{\mathcal{D}}_{2d}}}}.
\end{cases}
\end{equation}

In Fig.~\ref{fig_large_dev_2d}, the large deviation function in the continuum limit ${I_{2d,0}(R)}_{cont}$ projected along the $\theta=0$ direction is plotted for fixed parameter values. For small $R$, the rate function ${I_{2d,0}(R)}_{cont}$ scales as $\frac{R^2}{4{\mathcal{D}}_{2d}}$ and for large $R$, it scales as $\frac{{\left( R-2 \epsilon \right) }^2}{4 D_{2d}}$. This is similar to the behavior of the large deviation function of the one-dimensional case, where two regions with differing diffusive behaviors are observed. The scale of $R$ at which the crossover between the two regimes occur along different directions $\theta$ is determined by the function $g(\theta)$ plotted in Fig.~\ref{fig_lightcone}. The symmetry associated with the function $g(\theta)$ points to the inherent square symmetry associated with the two dimensional process. 



\section{Discussion}
In this paper, we have investigated active lattice walks in one and two spatial dimensions. 
We analyzed the occupation probability of an arbitrary site on the lattice and showed that the lattice version of active particle motion can also be used to derive the distributions in continuous space by taking appropriate limits.  We computed the moments, cumulants and cross correlations in the position of the active random walker. Next, we computed the exact large deviation free energy function in both one and two dimensions, which we used to compute the moments and the cumulants of the displacements exactly at late times.  We also demonstrated that the large deviation rate function associated with an active particle with diffusion displays two regimes, with differing diffusive behaviors. Specifically in two dimensions, we analyzed the large deviation rate functions projected along a given direction $\theta$, and demonstrated the emergence of these two regions. The two regimes can naturally be explained as a consequence of the two underlying processes in play: diffusion and active motion. At small values of the scaled length, the process is described by effective diffusion arising from both processes, while at large length scales the process may be interpreted as receiving contributions primarily from active motion {\it followed} by diffusion. This leads to the second regime in the large deviation function described by shifted diffusion. We also demonstrated that in the two dimensional process, the cross-correlations between the $x$ and $y$ motion persist in the large deviation function. Finally, we also verified our analytic results with kinetic Monte Carlo simulations of an active lattice walker in one and two dimensions.

It would be interesting to extend our results for active random walks to higher dimensions as well as to different types of underlying lattices. As we have shown that non-trivial corrections to diffusion can emerge at large lengthscales, it would also be of interest to study the first passage statistics of active random walks to better understand the nature of these regimes, as has been recently investigated in Refs.~\cite{lacroix2020universal,basu2018active,malakar2018steady,de2021survival}.
Lattice models of interacting particle systems have also been used as paradigmatic models to study the non-trivial collective effects associated with dense particulate matter \cite{de1985rigorous,de1986reaction}. It has recently been shown that phase transitions resembling Motility Induced Phase Separation can be realized in lattice models with activity~\cite{kourbane2018exact,agranov2021exact}, however with microscopic rules different from our case. It would therefore be interesting to generalize our study to interacting active lattice walks to better understand the nature of the non-equilibrium phases that appear in collections of active particles.

\section*{Acknowledgments}
We thank Roshan Maharana, Vishnu V.~Krishnan, Pappu Acharya, Debankur Das, Soham Mukhopadhyay and Jishnu Nampoothiri for useful discussions. 
This project was funded by intramural funds at TIFR Hyderabad from the Department of Atomic Energy (DAE). M.B. acknowledges support under the DAE Homi Bhabha Chair Professorship of the Department of Atomic Energy. 

\begin{widetext}

\begin{appendix}
\section{Fourier transform of the occupation probability of a general site on a  one dimensional lattice}
\label{A0}
Equation~(\ref{matrix11}) can also be solved by diagonalizing the matrix $\mathcal{M}(k)$ given in Eq.~(\ref{matrix2}).
Let $\lambda_+(k)$ and $\lambda_-(k)$ be the two eigenvalues of the matrix $\mathcal{M}(k)$. The eigenvalues can be expressed as
\begin{equation}
 \lambda_{\pm}(k)=2D_{1d}(\cos k-1)-\gamma \pm R(k), 
\end{equation}
where $R(k)$ is defined in Eq.~(\ref{rk}).
Solution of Eq.~(\ref{matrix11}) can then be written as
\begin{equation}
\ket{\tilde P_m (k,t)}=c_+(k)\vec X_+(k) e^{\lambda_+(k) t}+c_-(k)\vec X_- (k)e^{\lambda_-(k) t},
\end{equation}
where $\vec X_+(k)$ and $\vec X_-(k)$ are the eigenvectors of the matrix $\mathcal{M}(k)$ corresponding to the eigenvalues $\lambda_+(k)$ and $\lambda_-(k)$ respectively. These are given as 
\begin{equation}
\vec X_+(k)=
\begin{pmatrix}
\frac{i 2 \epsilon \sin k+ R(k)}{\gamma}\\
1
\end{pmatrix}
,~
\vec X_-(k)=
\begin{pmatrix}
\frac{i 2 \epsilon \sin k- R(k)}{\gamma}\\
1
\end{pmatrix}.
\end{equation}

Here, $c_+(k)$ and $c_-(k)$ are the prefactors to be determined from the initial conditions.
We use symmetric initial conditions; $P_0(x,t=0)=P_1(x,t=0)=\frac{1}{2}\delta_{x,0}$. That is, the particle has equal initial probabilities to be in the state $0$ or the state $1$. The total probability  can then be written as
\begin{eqnarray}
\label{large_eigenvalue}
\tilde{P}(k,t)&=&\tilde{P}_0(k,t)+\tilde{P}_1(k,t)
=a_+(k)e^{\lambda_+(k) t}+a_-(k)e^{\lambda_-(k) t},
\end{eqnarray}
where
\begin{eqnarray}
a_+(k)=\frac{1}{2}\left(1+\frac{\gamma}{R(k)} \right),~
a_-(k)=\frac{1}{2}\left(1-\frac{\gamma}{R(k)} \right).
\end{eqnarray}
After simplification, we obtain the expression in Eq.~(\ref{comp}).

\section{Moment generating function in two dimensions}
\label{Amoment}
The Laplace transform of the moment generating function in two dimensions is given as
\begin{eqnarray}
\label{pks2dn}
&&\tilde P(k_x,k_y,s)=\nonumber\\&&\frac{(-2 f+s+\gamma ) \left(4 f^2+(s+\gamma ) (s+2 \gamma )-2 f (2 s+3 \gamma )+(2-g) \epsilon ^2\right)}{(2 f-s) (2 f-s-2 \gamma ) (-2 f+s+\gamma )^2+2 (2-g) (-2 f+s+\gamma )^2 \epsilon ^2+4 (1-g) \epsilon ^4+4
   \epsilon ^4 \cos \left(2 k_x\right) \cos \left(2 k_y\right)},
\end{eqnarray}
where $f \equiv f(k_x,k_y)=D_{2d}(-2+\cos k_x +\cos k_y)$ and $g \equiv g(k_x,k_y)=\cos (2 k_x) +\cos (2 k_y)$.
Setting $k_y=0$ yields $\tilde P(k_x,s)=\tilde P(k_x,k_y=0,s)$. We obtain,
\begin{small}
\begin{eqnarray}
\label{pks2dnprojected}
\tilde P(k_x,s)=\frac{(-2 f+s+\gamma ) \left(4 f^2+(s+\gamma ) (s+2 \gamma )-2 f (2 s+3 \gamma )+(2-g) \epsilon ^2\right)}{(2 f-s) (2 f-s-2 \gamma ) (-2 f+s+\gamma )^2+2 (2-g) (-2 f+s+\gamma )^2 \epsilon ^2},
\end{eqnarray}
\end{small}
where $f \equiv f(k_x,k_y)=D_{2d}(-1+\cos k_x)$ and $g \equiv g(k_x,k_y)=1+\cos (2 k_x)$.
\section{Marginal rate function in two dimensions}
\label{Amarginalrate}
Let us take the free energy function in Eq.~(\ref{cpsi2dcontpolar0}) and substitute $\frac{\delta}{4} k^2=z$ and $\beta=1+\cos{4\theta}$. Thus we obtain
\begin{small}
\begin{eqnarray} 
\label{cpsi2dcontpolar1}
{\lambda_{2d,\beta}(z)}_{cont,0}&=& \frac{\gamma  \sqrt{1+4z +\sqrt{1+8 z +8 {z}^2 \beta}}}{\sqrt{2}}-\gamma .\nonumber\\
\end{eqnarray}
\end{small}
Using the transformation $1+8 z +8 {z}^2 \beta={g}^2$ gives
\begin{small}
\begin{eqnarray} 
\label{cpsi2dcontpolar2}
{\lambda_{2d,\beta}(g)}_{cont,0}&=& \frac{\gamma  \sqrt{1+4 z(g) +g}}{\sqrt{2}}-\gamma ,
\end{eqnarray}
\end{small}
in which $z(g)=\frac{-2+\sqrt{4+2 \left(-1+g^2\right) \beta }}{4 \beta }$.
Gartner-Ellis theorem in the transformed coordinates yields,
\begin{equation}
\label{ldfmarginal}
    {I_{2d,\beta}(R)}_{cont,0}=\underset{g}{\max}\{2 \sqrt{\frac{z(g)}{\delta}}R-{\lambda_{2d,\beta}(g)}_{cont,0} \}.
\end{equation}
For $\beta=2,0$, we get the closed form expressions in Eq.~(\ref{ldfmarginal00}).

\end{appendix}

\end{widetext}

\bibliographystyle{apsrev4-2}
\bibliography{Active_random_walks_bibliography.bib} 

\begin{thebibliography}{62}%
\makeatletter
\providecommand \@ifxundefined [1]{%
 \@ifx{#1\undefined}
}%
\providecommand \@ifnum [1]{%
 \ifnum #1\expandafter \@firstoftwo
 \else \expandafter \@secondoftwo
 \fi
}%
\providecommand \@ifx [1]{%
 \ifx #1\expandafter \@firstoftwo
 \else \expandafter \@secondoftwo
 \fi
}%
\providecommand \natexlab [1]{#1}%
\providecommand \enquote  [1]{``#1''}%
\providecommand \bibnamefont  [1]{#1}%
\providecommand \bibfnamefont [1]{#1}%
\providecommand \citenamefont [1]{#1}%
\providecommand \href@noop [0]{\@secondoftwo}%
\providecommand \href [0]{\begingroup \@sanitize@url \@href}%
\providecommand \@href[1]{\@@startlink{#1}\@@href}%
\providecommand \@@href[1]{\endgroup#1\@@endlink}%
\providecommand \@sanitize@url [0]{\catcode `\\12\catcode `\$12\catcode
  `\&12\catcode `\#12\catcode `\^12\catcode `\_12\catcode `\%12\relax}%
\providecommand \@@startlink[1]{}%
\providecommand \@@endlink[0]{}%
\providecommand \url  [0]{\begingroup\@sanitize@url \@url }%
\providecommand \@url [1]{\endgroup\@href {#1}{\urlprefix }}%
\providecommand \urlprefix  [0]{URL }%
\providecommand \Eprint [0]{\href }%
\providecommand \doibase [0]{https://doi.org/}%
\providecommand \selectlanguage [0]{\@gobble}%
\providecommand \bibinfo  [0]{\@secondoftwo}%
\providecommand \bibfield  [0]{\@secondoftwo}%
\providecommand \translation [1]{[#1]}%
\providecommand \BibitemOpen [0]{}%
\providecommand \bibitemStop [0]{}%
\providecommand \bibitemNoStop [0]{.\EOS\space}%
\providecommand \EOS [0]{\spacefactor3000\relax}%
\providecommand \BibitemShut  [1]{\csname bibitem#1\endcsname}%
\let\auto@bib@innerbib\@empty
\bibitem [{\citenamefont {Walsh}\ \emph {et~al.}(2017)\citenamefont {Walsh},
  \citenamefont {Wagner}, \citenamefont {Schlossberg}, \citenamefont {Olson},
  \citenamefont {Baskaran},\ and\ \citenamefont {Menon}}]{walsh2017noise}%
  \BibitemOpen
  \bibfield  {author} {\bibinfo {author} {\bibfnamefont {L.}~\bibnamefont
  {Walsh}}, \bibinfo {author} {\bibfnamefont {C.~G.}\ \bibnamefont {Wagner}},
  \bibinfo {author} {\bibfnamefont {S.}~\bibnamefont {Schlossberg}}, \bibinfo
  {author} {\bibfnamefont {C.}~\bibnamefont {Olson}}, \bibinfo {author}
  {\bibfnamefont {A.}~\bibnamefont {Baskaran}},\ and\ \bibinfo {author}
  {\bibfnamefont {N.}~\bibnamefont {Menon}},\ }\href@noop {} {\bibfield
  {journal} {\bibinfo  {journal} {Soft Matter}\ }\textbf {\bibinfo {volume}
  {13}},\ \bibinfo {pages} {8964} (\bibinfo {year} {2017})}\BibitemShut
  {NoStop}%
\bibitem [{\citenamefont {Schnitzer}(1993)}]{schnitzer1993theory}%
  \BibitemOpen
  \bibfield  {author} {\bibinfo {author} {\bibfnamefont {M.~J.}\ \bibnamefont
  {Schnitzer}},\ }\href@noop {} {\bibfield  {journal} {\bibinfo  {journal}
  {Physical Review E}\ }\textbf {\bibinfo {volume} {48}},\ \bibinfo {pages}
  {2553} (\bibinfo {year} {1993})}\BibitemShut {NoStop}%
\bibitem [{\citenamefont {Gautrais}\ \emph {et~al.}(2009)\citenamefont
  {Gautrais}, \citenamefont {Jost}, \citenamefont {Soria}, \citenamefont
  {Campo}, \citenamefont {Motsch}, \citenamefont {Fournier}, \citenamefont
  {Blanco},\ and\ \citenamefont {Theraulaz}}]{gautrais2009analyzing}%
  \BibitemOpen
  \bibfield  {author} {\bibinfo {author} {\bibfnamefont {J.}~\bibnamefont
  {Gautrais}}, \bibinfo {author} {\bibfnamefont {C.}~\bibnamefont {Jost}},
  \bibinfo {author} {\bibfnamefont {M.}~\bibnamefont {Soria}}, \bibinfo
  {author} {\bibfnamefont {A.}~\bibnamefont {Campo}}, \bibinfo {author}
  {\bibfnamefont {S.}~\bibnamefont {Motsch}}, \bibinfo {author} {\bibfnamefont
  {R.}~\bibnamefont {Fournier}}, \bibinfo {author} {\bibfnamefont
  {S.}~\bibnamefont {Blanco}},\ and\ \bibinfo {author} {\bibfnamefont
  {G.}~\bibnamefont {Theraulaz}},\ }\href@noop {} {\bibfield  {journal}
  {\bibinfo  {journal} {Journal of Mathematical Biology}\ }\textbf {\bibinfo
  {volume} {58}},\ \bibinfo {pages} {429} (\bibinfo {year} {2009})}\BibitemShut
  {NoStop}%
\bibitem [{\citenamefont {Cavagna}\ \emph {et~al.}(2010)\citenamefont
  {Cavagna}, \citenamefont {Cimarelli}, \citenamefont {Giardina}, \citenamefont
  {Parisi}, \citenamefont {Santagati}, \citenamefont {Stefanini},\ and\
  \citenamefont {Viale}}]{cavagna2010scale}%
  \BibitemOpen
  \bibfield  {author} {\bibinfo {author} {\bibfnamefont {A.}~\bibnamefont
  {Cavagna}}, \bibinfo {author} {\bibfnamefont {A.}~\bibnamefont {Cimarelli}},
  \bibinfo {author} {\bibfnamefont {I.}~\bibnamefont {Giardina}}, \bibinfo
  {author} {\bibfnamefont {G.}~\bibnamefont {Parisi}}, \bibinfo {author}
  {\bibfnamefont {R.}~\bibnamefont {Santagati}}, \bibinfo {author}
  {\bibfnamefont {F.}~\bibnamefont {Stefanini}},\ and\ \bibinfo {author}
  {\bibfnamefont {M.}~\bibnamefont {Viale}},\ }\href@noop {} {\bibfield
  {journal} {\bibinfo  {journal} {Proceedings of the National Academy of
  Sciences}\ }\textbf {\bibinfo {volume} {107}},\ \bibinfo {pages} {11865}
  (\bibinfo {year} {2010})}\BibitemShut {NoStop}%
\bibitem [{\citenamefont {Cates}(2012)}]{cates2012diffusive}%
  \BibitemOpen
  \bibfield  {author} {\bibinfo {author} {\bibfnamefont {M.~E.}\ \bibnamefont
  {Cates}},\ }\href@noop {} {\bibfield  {journal} {\bibinfo  {journal} {Reports
  on Progress in Physics}\ }\textbf {\bibinfo {volume} {75}},\ \bibinfo {pages}
  {042601} (\bibinfo {year} {2012})}\BibitemShut {NoStop}%
\bibitem [{\citenamefont {Ramaswamy}(2010)}]{ramaswamy2010mechanics}%
  \BibitemOpen
  \bibfield  {author} {\bibinfo {author} {\bibfnamefont {S.}~\bibnamefont
  {Ramaswamy}},\ }\href@noop {} {\bibfield  {journal} {\bibinfo  {journal}
  {Annual Review of Condensed Matter Physics}\ }\textbf {\bibinfo {volume}
  {1}},\ \bibinfo {pages} {323} (\bibinfo {year} {2010})}\BibitemShut {NoStop}%
\bibitem [{\citenamefont {Tailleur}\ and\ \citenamefont
  {Cates}(2008)}]{tailleur2008statistical}%
  \BibitemOpen
  \bibfield  {author} {\bibinfo {author} {\bibfnamefont {J.}~\bibnamefont
  {Tailleur}}\ and\ \bibinfo {author} {\bibfnamefont {M.}~\bibnamefont
  {Cates}},\ }\href@noop {} {\bibfield  {journal} {\bibinfo  {journal}
  {Physical Review Letters}\ }\textbf {\bibinfo {volume} {100}},\ \bibinfo
  {pages} {218103} (\bibinfo {year} {2008})}\BibitemShut {NoStop}%
\bibitem [{\citenamefont {Slowman}\ \emph {et~al.}(2017)\citenamefont
  {Slowman}, \citenamefont {Evans},\ and\ \citenamefont
  {Blythe}}]{slowman2017exact}%
  \BibitemOpen
  \bibfield  {author} {\bibinfo {author} {\bibfnamefont {A.}~\bibnamefont
  {Slowman}}, \bibinfo {author} {\bibfnamefont {M.}~\bibnamefont {Evans}},\
  and\ \bibinfo {author} {\bibfnamefont {R.}~\bibnamefont {Blythe}},\
  }\href@noop {} {\bibfield  {journal} {\bibinfo  {journal} {Journal of Physics
  A: Mathematical and Theoretical}\ }\textbf {\bibinfo {volume} {50}},\
  \bibinfo {pages} {375601} (\bibinfo {year} {2017})}\BibitemShut {NoStop}%
\bibitem [{\citenamefont {Enculescu}\ and\ \citenamefont
  {Stark}(2011)}]{enculescu2011active}%
  \BibitemOpen
  \bibfield  {author} {\bibinfo {author} {\bibfnamefont {M.}~\bibnamefont
  {Enculescu}}\ and\ \bibinfo {author} {\bibfnamefont {H.}~\bibnamefont
  {Stark}},\ }\href@noop {} {\bibfield  {journal} {\bibinfo  {journal}
  {Physical Review Letters}\ }\textbf {\bibinfo {volume} {107}},\ \bibinfo
  {pages} {058301} (\bibinfo {year} {2011})}\BibitemShut {NoStop}%
\bibitem [{\citenamefont {Vicsek}\ \emph {et~al.}(1995)\citenamefont {Vicsek},
  \citenamefont {Czir{\'o}k}, \citenamefont {Ben-Jacob}, \citenamefont
  {Cohen},\ and\ \citenamefont {Shochet}}]{vicsek1995novel}%
  \BibitemOpen
  \bibfield  {author} {\bibinfo {author} {\bibfnamefont {T.}~\bibnamefont
  {Vicsek}}, \bibinfo {author} {\bibfnamefont {A.}~\bibnamefont {Czir{\'o}k}},
  \bibinfo {author} {\bibfnamefont {E.}~\bibnamefont {Ben-Jacob}}, \bibinfo
  {author} {\bibfnamefont {I.}~\bibnamefont {Cohen}},\ and\ \bibinfo {author}
  {\bibfnamefont {O.}~\bibnamefont {Shochet}},\ }\href@noop {} {\bibfield
  {journal} {\bibinfo  {journal} {Physical Review Letters}\ }\textbf {\bibinfo
  {volume} {75}},\ \bibinfo {pages} {1226} (\bibinfo {year}
  {1995})}\BibitemShut {NoStop}%
\bibitem [{\citenamefont {Toner}\ and\ \citenamefont
  {Tu}(1995)}]{toner1995long}%
  \BibitemOpen
  \bibfield  {author} {\bibinfo {author} {\bibfnamefont {J.}~\bibnamefont
  {Toner}}\ and\ \bibinfo {author} {\bibfnamefont {Y.}~\bibnamefont {Tu}},\
  }\href@noop {} {\bibfield  {journal} {\bibinfo  {journal} {Physical Review
  Letters}\ }\textbf {\bibinfo {volume} {75}},\ \bibinfo {pages} {4326}
  (\bibinfo {year} {1995})}\BibitemShut {NoStop}%
\bibitem [{\citenamefont {Lam}\ \emph {et~al.}(2015)\citenamefont {Lam},
  \citenamefont {Schindler},\ and\ \citenamefont {Dauchot}}]{lam2015self}%
  \BibitemOpen
  \bibfield  {author} {\bibinfo {author} {\bibfnamefont {K.-D. N.~T.}\
  \bibnamefont {Lam}}, \bibinfo {author} {\bibfnamefont {M.}~\bibnamefont
  {Schindler}},\ and\ \bibinfo {author} {\bibfnamefont {O.}~\bibnamefont
  {Dauchot}},\ }\href@noop {} {\bibfield  {journal} {\bibinfo  {journal} {New
  Journal of Physics}\ }\textbf {\bibinfo {volume} {17}},\ \bibinfo {pages}
  {113056} (\bibinfo {year} {2015})}\BibitemShut {NoStop}%
\bibitem [{\citenamefont {Czir{\'o}k}\ \emph {et~al.}(1999)\citenamefont
  {Czir{\'o}k}, \citenamefont {Barab{\'a}si},\ and\ \citenamefont
  {Vicsek}}]{czirok1999collective}%
  \BibitemOpen
  \bibfield  {author} {\bibinfo {author} {\bibfnamefont {A.}~\bibnamefont
  {Czir{\'o}k}}, \bibinfo {author} {\bibfnamefont {A.-L.}\ \bibnamefont
  {Barab{\'a}si}},\ and\ \bibinfo {author} {\bibfnamefont {T.}~\bibnamefont
  {Vicsek}},\ }\href@noop {} {\bibfield  {journal} {\bibinfo  {journal}
  {Physical Review Letters}\ }\textbf {\bibinfo {volume} {82}},\ \bibinfo
  {pages} {209} (\bibinfo {year} {1999})}\BibitemShut {NoStop}%
\bibitem [{\citenamefont {Cates}\ and\ \citenamefont
  {Tailleur}(2013)}]{cates2013active}%
  \BibitemOpen
  \bibfield  {author} {\bibinfo {author} {\bibfnamefont {M.~E.}\ \bibnamefont
  {Cates}}\ and\ \bibinfo {author} {\bibfnamefont {J.}~\bibnamefont
  {Tailleur}},\ }\href@noop {} {\bibfield  {journal} {\bibinfo  {journal}
  {Europhysics Letters}\ }\textbf {\bibinfo {volume} {101}},\ \bibinfo {pages}
  {20010} (\bibinfo {year} {2013})}\BibitemShut {NoStop}%
\bibitem [{\citenamefont {Kourbane-Houssene}\ \emph {et~al.}(2018)\citenamefont
  {Kourbane-Houssene}, \citenamefont {Erignoux}, \citenamefont {Bodineau},\
  and\ \citenamefont {Tailleur}}]{kourbane2018exact}%
  \BibitemOpen
  \bibfield  {author} {\bibinfo {author} {\bibfnamefont {M.}~\bibnamefont
  {Kourbane-Houssene}}, \bibinfo {author} {\bibfnamefont {C.}~\bibnamefont
  {Erignoux}}, \bibinfo {author} {\bibfnamefont {T.}~\bibnamefont {Bodineau}},\
  and\ \bibinfo {author} {\bibfnamefont {J.}~\bibnamefont {Tailleur}},\
  }\href@noop {} {\bibfield  {journal} {\bibinfo  {journal} {Physical Review
  Letters}\ }\textbf {\bibinfo {volume} {120}},\ \bibinfo {pages} {268003}
  (\bibinfo {year} {2018})}\BibitemShut {NoStop}%
\bibitem [{\citenamefont {Merrigan}\ \emph {et~al.}(2020)\citenamefont
  {Merrigan}, \citenamefont {Ramola}, \citenamefont {Chatterjee}, \citenamefont
  {Segall}, \citenamefont {Shokef},\ and\ \citenamefont
  {Chakraborty}}]{merrigan2020arrested}%
  \BibitemOpen
  \bibfield  {author} {\bibinfo {author} {\bibfnamefont {C.}~\bibnamefont
  {Merrigan}}, \bibinfo {author} {\bibfnamefont {K.}~\bibnamefont {Ramola}},
  \bibinfo {author} {\bibfnamefont {R.}~\bibnamefont {Chatterjee}}, \bibinfo
  {author} {\bibfnamefont {N.}~\bibnamefont {Segall}}, \bibinfo {author}
  {\bibfnamefont {Y.}~\bibnamefont {Shokef}},\ and\ \bibinfo {author}
  {\bibfnamefont {B.}~\bibnamefont {Chakraborty}},\ }\href@noop {} {\bibfield
  {journal} {\bibinfo  {journal} {Physical Review Research}\ }\textbf {\bibinfo
  {volume} {2}},\ \bibinfo {pages} {013260} (\bibinfo {year}
  {2020})}\BibitemShut {NoStop}%
\bibitem [{\citenamefont {Lee}(2013)}]{lee2013active}%
  \BibitemOpen
  \bibfield  {author} {\bibinfo {author} {\bibfnamefont {C.~F.}\ \bibnamefont
  {Lee}},\ }\href@noop {} {\bibfield  {journal} {\bibinfo  {journal} {New
  Journal of Physics}\ }\textbf {\bibinfo {volume} {15}},\ \bibinfo {pages}
  {055007} (\bibinfo {year} {2013})}\BibitemShut {NoStop}%
\bibitem [{\citenamefont {Malakar}\ \emph {et~al.}(2020)\citenamefont
  {Malakar}, \citenamefont {Das}, \citenamefont {Kundu}, \citenamefont
  {Kumar},\ and\ \citenamefont {Dhar}}]{malakar2020steady}%
  \BibitemOpen
  \bibfield  {author} {\bibinfo {author} {\bibfnamefont {K.}~\bibnamefont
  {Malakar}}, \bibinfo {author} {\bibfnamefont {A.}~\bibnamefont {Das}},
  \bibinfo {author} {\bibfnamefont {A.}~\bibnamefont {Kundu}}, \bibinfo
  {author} {\bibfnamefont {K.~V.}\ \bibnamefont {Kumar}},\ and\ \bibinfo
  {author} {\bibfnamefont {A.}~\bibnamefont {Dhar}},\ }\href@noop {} {\bibfield
   {journal} {\bibinfo  {journal} {Physical Review E}\ }\textbf {\bibinfo
  {volume} {101}},\ \bibinfo {pages} {022610} (\bibinfo {year}
  {2020})}\BibitemShut {NoStop}%
\bibitem [{\citenamefont {Sevilla}\ \emph {et~al.}(2019)\citenamefont
  {Sevilla}, \citenamefont {Arzola},\ and\ \citenamefont
  {Cital}}]{sevilla2019stationary}%
  \BibitemOpen
  \bibfield  {author} {\bibinfo {author} {\bibfnamefont {F.~J.}\ \bibnamefont
  {Sevilla}}, \bibinfo {author} {\bibfnamefont {A.~V.}\ \bibnamefont
  {Arzola}},\ and\ \bibinfo {author} {\bibfnamefont {E.~P.}\ \bibnamefont
  {Cital}},\ }\href@noop {} {\bibfield  {journal} {\bibinfo  {journal}
  {Physical Review E}\ }\textbf {\bibinfo {volume} {99}},\ \bibinfo {pages}
  {012145} (\bibinfo {year} {2019})}\BibitemShut {NoStop}%
\bibitem [{\citenamefont {Dhar}\ \emph {et~al.}(2019)\citenamefont {Dhar},
  \citenamefont {Kundu}, \citenamefont {Majumdar}, \citenamefont
  {Sabhapandit},\ and\ \citenamefont {Schehr}}]{dhar2019run}%
  \BibitemOpen
  \bibfield  {author} {\bibinfo {author} {\bibfnamefont {A.}~\bibnamefont
  {Dhar}}, \bibinfo {author} {\bibfnamefont {A.}~\bibnamefont {Kundu}},
  \bibinfo {author} {\bibfnamefont {S.~N.}\ \bibnamefont {Majumdar}}, \bibinfo
  {author} {\bibfnamefont {S.}~\bibnamefont {Sabhapandit}},\ and\ \bibinfo
  {author} {\bibfnamefont {G.}~\bibnamefont {Schehr}},\ }\href@noop {}
  {\bibfield  {journal} {\bibinfo  {journal} {Physical Review E}\ }\textbf
  {\bibinfo {volume} {99}},\ \bibinfo {pages} {032132} (\bibinfo {year}
  {2019})}\BibitemShut {NoStop}%
\bibitem [{\citenamefont {Malakar}\ \emph {et~al.}(2018)\citenamefont
  {Malakar}, \citenamefont {Jemseena}, \citenamefont {Kundu}, \citenamefont
  {Kumar}, \citenamefont {Sabhapandit}, \citenamefont {Majumdar}, \citenamefont
  {Redner},\ and\ \citenamefont {Dhar}}]{malakar2018steady}%
  \BibitemOpen
  \bibfield  {author} {\bibinfo {author} {\bibfnamefont {K.}~\bibnamefont
  {Malakar}}, \bibinfo {author} {\bibfnamefont {V.}~\bibnamefont {Jemseena}},
  \bibinfo {author} {\bibfnamefont {A.}~\bibnamefont {Kundu}}, \bibinfo
  {author} {\bibfnamefont {K.~V.}\ \bibnamefont {Kumar}}, \bibinfo {author}
  {\bibfnamefont {S.}~\bibnamefont {Sabhapandit}}, \bibinfo {author}
  {\bibfnamefont {S.~N.}\ \bibnamefont {Majumdar}}, \bibinfo {author}
  {\bibfnamefont {S.}~\bibnamefont {Redner}},\ and\ \bibinfo {author}
  {\bibfnamefont {A.}~\bibnamefont {Dhar}},\ }\href@noop {} {\bibfield
  {journal} {\bibinfo  {journal} {Journal of Statistical Mechanics: Theory and
  Experiment}\ }\textbf {\bibinfo {volume} {2018}},\ \bibinfo {pages} {043215}
  (\bibinfo {year} {2018})}\BibitemShut {NoStop}%
\bibitem [{\citenamefont {Evans}\ and\ \citenamefont
  {Majumdar}(2018)}]{evans2018run}%
  \BibitemOpen
  \bibfield  {author} {\bibinfo {author} {\bibfnamefont {M.~R.}\ \bibnamefont
  {Evans}}\ and\ \bibinfo {author} {\bibfnamefont {S.~N.}\ \bibnamefont
  {Majumdar}},\ }\href@noop {} {\bibfield  {journal} {\bibinfo  {journal}
  {Journal of Physics A: Mathematical and Theoretical}\ }\textbf {\bibinfo
  {volume} {51}},\ \bibinfo {pages} {475003} (\bibinfo {year}
  {2018})}\BibitemShut {NoStop}%
\bibitem [{\citenamefont {Mori}\ \emph
  {et~al.}(2020{\natexlab{a}})\citenamefont {Mori}, \citenamefont {Le~Doussal},
  \citenamefont {Majumdar},\ and\ \citenamefont {Schehr}}]{mori2020universal}%
  \BibitemOpen
  \bibfield  {author} {\bibinfo {author} {\bibfnamefont {F.}~\bibnamefont
  {Mori}}, \bibinfo {author} {\bibfnamefont {P.}~\bibnamefont {Le~Doussal}},
  \bibinfo {author} {\bibfnamefont {S.~N.}\ \bibnamefont {Majumdar}},\ and\
  \bibinfo {author} {\bibfnamefont {G.}~\bibnamefont {Schehr}},\ }\href@noop {}
  {\bibfield  {journal} {\bibinfo  {journal} {Physical Review Letters}\
  }\textbf {\bibinfo {volume} {124}},\ \bibinfo {pages} {090603} (\bibinfo
  {year} {2020}{\natexlab{a}})}\BibitemShut {NoStop}%
\bibitem [{\citenamefont {Mori}\ \emph
  {et~al.}(2020{\natexlab{b}})\citenamefont {Mori}, \citenamefont {Le~Doussal},
  \citenamefont {Majumdar},\ and\ \citenamefont {Schehr}}]{mori2020universalp}%
  \BibitemOpen
  \bibfield  {author} {\bibinfo {author} {\bibfnamefont {F.}~\bibnamefont
  {Mori}}, \bibinfo {author} {\bibfnamefont {P.}~\bibnamefont {Le~Doussal}},
  \bibinfo {author} {\bibfnamefont {S.~N.}\ \bibnamefont {Majumdar}},\ and\
  \bibinfo {author} {\bibfnamefont {G.}~\bibnamefont {Schehr}},\ }\href@noop {}
  {\bibfield  {journal} {\bibinfo  {journal} {Physical Review E}\ }\textbf
  {\bibinfo {volume} {102}},\ \bibinfo {pages} {042133} (\bibinfo {year}
  {2020}{\natexlab{b}})}\BibitemShut {NoStop}%
\bibitem [{\citenamefont {Singh}\ and\ \citenamefont
  {Kundu}(2019)}]{singh2019generalised}%
  \BibitemOpen
  \bibfield  {author} {\bibinfo {author} {\bibfnamefont {P.}~\bibnamefont
  {Singh}}\ and\ \bibinfo {author} {\bibfnamefont {A.}~\bibnamefont {Kundu}},\
  }\href@noop {} {\bibfield  {journal} {\bibinfo  {journal} {Journal of
  Statistical Mechanics: Theory and Experiment}\ }\textbf {\bibinfo {volume}
  {2019}},\ \bibinfo {pages} {083205} (\bibinfo {year} {2019})}\BibitemShut
  {NoStop}%
\bibitem [{\citenamefont {Angelani}\ \emph {et~al.}(2014)\citenamefont
  {Angelani}, \citenamefont {Di~Leonardo},\ and\ \citenamefont
  {Paoluzzi}}]{angelani2014first}%
  \BibitemOpen
  \bibfield  {author} {\bibinfo {author} {\bibfnamefont {L.}~\bibnamefont
  {Angelani}}, \bibinfo {author} {\bibfnamefont {R.}~\bibnamefont
  {Di~Leonardo}},\ and\ \bibinfo {author} {\bibfnamefont {M.}~\bibnamefont
  {Paoluzzi}},\ }\href@noop {} {\bibfield  {journal} {\bibinfo  {journal} {The
  European Physical Journal E}\ }\textbf {\bibinfo {volume} {37}},\ \bibinfo
  {pages} {1} (\bibinfo {year} {2014})}\BibitemShut {NoStop}%
\bibitem [{\citenamefont {Martens}\ \emph {et~al.}(2012)\citenamefont
  {Martens}, \citenamefont {Angelani}, \citenamefont {Di~Leonardo},\ and\
  \citenamefont {Bocquet}}]{martens2012probability}%
  \BibitemOpen
  \bibfield  {author} {\bibinfo {author} {\bibfnamefont {K.}~\bibnamefont
  {Martens}}, \bibinfo {author} {\bibfnamefont {L.}~\bibnamefont {Angelani}},
  \bibinfo {author} {\bibfnamefont {R.}~\bibnamefont {Di~Leonardo}},\ and\
  \bibinfo {author} {\bibfnamefont {L.}~\bibnamefont {Bocquet}},\ }\href@noop
  {} {\bibfield  {journal} {\bibinfo  {journal} {The European Physical Journal
  E}\ }\textbf {\bibinfo {volume} {35}},\ \bibinfo {pages} {1} (\bibinfo {year}
  {2012})}\BibitemShut {NoStop}%
\bibitem [{\citenamefont {Basu}\ \emph {et~al.}(2018)\citenamefont {Basu},
  \citenamefont {Majumdar}, \citenamefont {Rosso},\ and\ \citenamefont
  {Schehr}}]{basu2018active}%
  \BibitemOpen
  \bibfield  {author} {\bibinfo {author} {\bibfnamefont {U.}~\bibnamefont
  {Basu}}, \bibinfo {author} {\bibfnamefont {S.~N.}\ \bibnamefont {Majumdar}},
  \bibinfo {author} {\bibfnamefont {A.}~\bibnamefont {Rosso}},\ and\ \bibinfo
  {author} {\bibfnamefont {G.}~\bibnamefont {Schehr}},\ }\href@noop {}
  {\bibfield  {journal} {\bibinfo  {journal} {Physical Review E}\ }\textbf
  {\bibinfo {volume} {98}},\ \bibinfo {pages} {062121} (\bibinfo {year}
  {2018})}\BibitemShut {NoStop}%
\bibitem [{\citenamefont {Lindner}\ and\ \citenamefont
  {Nicola}(2008)}]{lindner2008diffusion}%
  \BibitemOpen
  \bibfield  {author} {\bibinfo {author} {\bibfnamefont {B.}~\bibnamefont
  {Lindner}}\ and\ \bibinfo {author} {\bibfnamefont {E.}~\bibnamefont
  {Nicola}},\ }\href@noop {} {\bibfield  {journal} {\bibinfo  {journal} {The
  European Physical Journal Special Topics}\ }\textbf {\bibinfo {volume}
  {157}},\ \bibinfo {pages} {43} (\bibinfo {year} {2008})}\BibitemShut
  {NoStop}%
\bibitem [{\citenamefont {Kumar}\ \emph {et~al.}(2020)\citenamefont {Kumar},
  \citenamefont {Sadekar},\ and\ \citenamefont {Basu}}]{kumar2020active}%
  \BibitemOpen
  \bibfield  {author} {\bibinfo {author} {\bibfnamefont {V.}~\bibnamefont
  {Kumar}}, \bibinfo {author} {\bibfnamefont {O.}~\bibnamefont {Sadekar}},\
  and\ \bibinfo {author} {\bibfnamefont {U.}~\bibnamefont {Basu}},\ }\href@noop
  {} {\bibfield  {journal} {\bibinfo  {journal} {Physical Review E}\ }\textbf
  {\bibinfo {volume} {102}},\ \bibinfo {pages} {052129} (\bibinfo {year}
  {2020})}\BibitemShut {NoStop}%
\bibitem [{\citenamefont {Romanczuk}\ \emph {et~al.}(2012)\citenamefont
  {Romanczuk}, \citenamefont {B{\"a}r}, \citenamefont {Ebeling}, \citenamefont
  {Lindner},\ and\ \citenamefont {Schimansky-Geier}}]{romanczuk2012active}%
  \BibitemOpen
  \bibfield  {author} {\bibinfo {author} {\bibfnamefont {P.}~\bibnamefont
  {Romanczuk}}, \bibinfo {author} {\bibfnamefont {M.}~\bibnamefont {B{\"a}r}},
  \bibinfo {author} {\bibfnamefont {W.}~\bibnamefont {Ebeling}}, \bibinfo
  {author} {\bibfnamefont {B.}~\bibnamefont {Lindner}},\ and\ \bibinfo {author}
  {\bibfnamefont {L.}~\bibnamefont {Schimansky-Geier}},\ }\href@noop {}
  {\bibfield  {journal} {\bibinfo  {journal} {The European Physical Journal
  Special Topics}\ }\textbf {\bibinfo {volume} {202}} (\bibinfo {year}
  {2012})}\BibitemShut {NoStop}%
\bibitem [{\citenamefont {Romanczuk}\ and\ \citenamefont
  {Erdmann}(2010)}]{romanczuk2010collective}%
  \BibitemOpen
  \bibfield  {author} {\bibinfo {author} {\bibfnamefont {P.}~\bibnamefont
  {Romanczuk}}\ and\ \bibinfo {author} {\bibfnamefont {U.}~\bibnamefont
  {Erdmann}},\ }\href@noop {} {\bibfield  {journal} {\bibinfo  {journal} {The
  European Physical Journal Special Topics}\ }\textbf {\bibinfo {volume}
  {187}},\ \bibinfo {pages} {127} (\bibinfo {year} {2010})}\BibitemShut
  {NoStop}%
\bibitem [{\citenamefont {Santra}\ \emph {et~al.}(2020)\citenamefont {Santra},
  \citenamefont {Basu},\ and\ \citenamefont {Sabhapandit}}]{santra2020run}%
  \BibitemOpen
  \bibfield  {author} {\bibinfo {author} {\bibfnamefont {I.}~\bibnamefont
  {Santra}}, \bibinfo {author} {\bibfnamefont {U.}~\bibnamefont {Basu}},\ and\
  \bibinfo {author} {\bibfnamefont {S.}~\bibnamefont {Sabhapandit}},\
  }\href@noop {} {\bibfield  {journal} {\bibinfo  {journal} {Physical Review
  E}\ }\textbf {\bibinfo {volume} {101}},\ \bibinfo {pages} {062120} (\bibinfo
  {year} {2020})}\BibitemShut {NoStop}%
\bibitem [{\citenamefont {Solon}\ \emph {et~al.}(2015)\citenamefont {Solon},
  \citenamefont {Cates},\ and\ \citenamefont {Tailleur}}]{solon2015active}%
  \BibitemOpen
  \bibfield  {author} {\bibinfo {author} {\bibfnamefont {A.~P.}\ \bibnamefont
  {Solon}}, \bibinfo {author} {\bibfnamefont {M.~E.}\ \bibnamefont {Cates}},\
  and\ \bibinfo {author} {\bibfnamefont {J.}~\bibnamefont {Tailleur}},\
  }\href@noop {} {\bibfield  {journal} {\bibinfo  {journal} {The European
  Physical Journal Special Topics}\ }\textbf {\bibinfo {volume} {224}},\
  \bibinfo {pages} {1231} (\bibinfo {year} {2015})}\BibitemShut {NoStop}%
\bibitem [{\citenamefont {Montroll}\ and\ \citenamefont
  {Weiss}(1965)}]{montroll1965random}%
  \BibitemOpen
  \bibfield  {author} {\bibinfo {author} {\bibfnamefont {E.~W.}\ \bibnamefont
  {Montroll}}\ and\ \bibinfo {author} {\bibfnamefont {G.~H.}\ \bibnamefont
  {Weiss}},\ }\href@noop {} {\bibfield  {journal} {\bibinfo  {journal} {Journal
  of Mathematical Physics}\ }\textbf {\bibinfo {volume} {6}},\ \bibinfo {pages}
  {167} (\bibinfo {year} {1965})}\BibitemShut {NoStop}%
\bibitem [{\citenamefont {Montroll}\ and\ \citenamefont
  {West}(1979)}]{montroll1979enriched}%
  \BibitemOpen
  \bibfield  {author} {\bibinfo {author} {\bibfnamefont {E.~W.}\ \bibnamefont
  {Montroll}}\ and\ \bibinfo {author} {\bibfnamefont {B.~J.}\ \bibnamefont
  {West}},\ }\href@noop {} {\bibfield  {journal} {\bibinfo  {journal}
  {Fluctuation phenomena}\ }\textbf {\bibinfo {volume} {66}},\ \bibinfo {pages}
  {61} (\bibinfo {year} {1979})}\BibitemShut {NoStop}%
\bibitem [{\citenamefont {Kutner}\ and\ \citenamefont
  {Masoliver}(2017)}]{kutner2017continuous}%
  \BibitemOpen
  \bibfield  {author} {\bibinfo {author} {\bibfnamefont {R.}~\bibnamefont
  {Kutner}}\ and\ \bibinfo {author} {\bibfnamefont {J.}~\bibnamefont
  {Masoliver}},\ }\href@noop {} {\bibfield  {journal} {\bibinfo  {journal} {The
  European Physical Journal B}\ }\textbf {\bibinfo {volume} {90}},\ \bibinfo
  {pages} {1} (\bibinfo {year} {2017})}\BibitemShut {NoStop}%
\bibitem [{\citenamefont {Mainardi}(2020)}]{mainardi2020advent}%
  \BibitemOpen
  \bibfield  {author} {\bibinfo {author} {\bibfnamefont {F.}~\bibnamefont
  {Mainardi}},\ }\href@noop {} {\bibfield  {journal} {\bibinfo  {journal}
  {Mathematics}\ }\textbf {\bibinfo {volume} {8}},\ \bibinfo {pages} {641}
  (\bibinfo {year} {2020})}\BibitemShut {NoStop}%
\bibitem [{\citenamefont {Masoliver}\ and\ \citenamefont
  {Lindenberg}(2017)}]{masoliver2017continuous}%
  \BibitemOpen
  \bibfield  {author} {\bibinfo {author} {\bibfnamefont {J.}~\bibnamefont
  {Masoliver}}\ and\ \bibinfo {author} {\bibfnamefont {K.}~\bibnamefont
  {Lindenberg}},\ }\href@noop {} {\bibfield  {journal} {\bibinfo  {journal}
  {The European Physical Journal B}\ }\textbf {\bibinfo {volume} {90}},\
  \bibinfo {pages} {1} (\bibinfo {year} {2017})}\BibitemShut {NoStop}%
\bibitem [{\citenamefont {Shlesinger}(1979)}]{shlesinger1979correlation}%
  \BibitemOpen
  \bibfield  {author} {\bibinfo {author} {\bibfnamefont {M.~F.}\ \bibnamefont
  {Shlesinger}},\ }\href@noop {} {\bibfield  {journal} {\bibinfo  {journal}
  {Solid State Communications}\ }\textbf {\bibinfo {volume} {32}},\ \bibinfo
  {pages} {1207} (\bibinfo {year} {1979})}\BibitemShut {NoStop}%
\bibitem [{\citenamefont {Touchette}(2009)}]{touchette2009large}%
  \BibitemOpen
  \bibfield  {author} {\bibinfo {author} {\bibfnamefont {H.}~\bibnamefont
  {Touchette}},\ }\href@noop {} {\bibfield  {journal} {\bibinfo  {journal}
  {Physics Reports}\ }\textbf {\bibinfo {volume} {478}},\ \bibinfo {pages} {1}
  (\bibinfo {year} {2009})}\BibitemShut {NoStop}%
\bibitem [{\citenamefont {van Gisbergen}\ and\ \citenamefont
  {Redig}(2019)}]{van2019central}%
  \BibitemOpen
  \bibfield  {author} {\bibinfo {author} {\bibfnamefont {B.}~\bibnamefont {van
  Gisbergen}}\ and\ \bibinfo {author} {\bibfnamefont {F.}~\bibnamefont
  {Redig}},\ }\href@noop {} {\bibfield  {journal} {\bibinfo  {journal} {arXiv
  preprint arXiv:1910.03350}\ } (\bibinfo {year} {2019})}\BibitemShut {NoStop}%
\bibitem [{\citenamefont {Mori}\ \emph
  {et~al.}(2021{\natexlab{a}})\citenamefont {Mori}, \citenamefont {Le~Doussal},
  \citenamefont {Majumdar},\ and\ \citenamefont
  {Schehr}}]{mori2021condensation}%
  \BibitemOpen
  \bibfield  {author} {\bibinfo {author} {\bibfnamefont {F.}~\bibnamefont
  {Mori}}, \bibinfo {author} {\bibfnamefont {P.}~\bibnamefont {Le~Doussal}},
  \bibinfo {author} {\bibfnamefont {S.~N.}\ \bibnamefont {Majumdar}},\ and\
  \bibinfo {author} {\bibfnamefont {G.}~\bibnamefont {Schehr}},\ }\href@noop {}
  {\bibfield  {journal} {\bibinfo  {journal} {Physical Review E}\ }\textbf
  {\bibinfo {volume} {103}},\ \bibinfo {pages} {062134} (\bibinfo {year}
  {2021}{\natexlab{a}})}\BibitemShut {NoStop}%
\bibitem [{\citenamefont {Proesmans}\ \emph {et~al.}(2020)\citenamefont
  {Proesmans}, \citenamefont {Toral},\ and\ \citenamefont {Van~den
  Broeck}}]{proesmans2020phase}%
  \BibitemOpen
  \bibfield  {author} {\bibinfo {author} {\bibfnamefont {K.}~\bibnamefont
  {Proesmans}}, \bibinfo {author} {\bibfnamefont {R.}~\bibnamefont {Toral}},\
  and\ \bibinfo {author} {\bibfnamefont {C.}~\bibnamefont {Van~den Broeck}},\
  }\href@noop {} {\bibfield  {journal} {\bibinfo  {journal} {Physica A:
  Statistical Mechanics and its Applications}\ }\textbf {\bibinfo {volume}
  {552}},\ \bibinfo {pages} {121934} (\bibinfo {year} {2020})}\BibitemShut
  {NoStop}%
\bibitem [{\citenamefont {Mori}\ \emph
  {et~al.}(2021{\natexlab{b}})\citenamefont {Mori}, \citenamefont {Gradenigo},\
  and\ \citenamefont {Majumdar}}]{mori2021first}%
  \BibitemOpen
  \bibfield  {author} {\bibinfo {author} {\bibfnamefont {F.}~\bibnamefont
  {Mori}}, \bibinfo {author} {\bibfnamefont {G.}~\bibnamefont {Gradenigo}},\
  and\ \bibinfo {author} {\bibfnamefont {S.~N.}\ \bibnamefont {Majumdar}},\
  }\href@noop {} {\bibfield  {journal} {\bibinfo  {journal} {Journal of
  Statistical Mechanics: Theory and Experiment}\ }\textbf {\bibinfo {volume}
  {2021}},\ \bibinfo {pages} {103208} (\bibinfo {year}
  {2021}{\natexlab{b}})}\BibitemShut {NoStop}%
\bibitem [{\citenamefont {Mallmin}\ \emph {et~al.}(2019)\citenamefont
  {Mallmin}, \citenamefont {Blythe},\ and\ \citenamefont
  {Evans}}]{mallmin2019comparison}%
  \BibitemOpen
  \bibfield  {author} {\bibinfo {author} {\bibfnamefont {E.}~\bibnamefont
  {Mallmin}}, \bibinfo {author} {\bibfnamefont {R.~A.}\ \bibnamefont
  {Blythe}},\ and\ \bibinfo {author} {\bibfnamefont {M.~R.}\ \bibnamefont
  {Evans}},\ }\href@noop {} {\bibfield  {journal} {\bibinfo  {journal} {Journal
  of Physics A: Mathematical and Theoretical}\ }\textbf {\bibinfo {volume}
  {52}},\ \bibinfo {pages} {425002} (\bibinfo {year} {2019})}\BibitemShut
  {NoStop}%
\bibitem [{\citenamefont {Gradenigo}\ and\ \citenamefont
  {Majumdar}(2019)}]{gradenigo2019first}%
  \BibitemOpen
  \bibfield  {author} {\bibinfo {author} {\bibfnamefont {G.}~\bibnamefont
  {Gradenigo}}\ and\ \bibinfo {author} {\bibfnamefont {S.~N.}\ \bibnamefont
  {Majumdar}},\ }\href@noop {} {\bibfield  {journal} {\bibinfo  {journal}
  {Journal of Statistical Mechanics: Theory and Experiment}\ }\textbf {\bibinfo
  {volume} {2019}},\ \bibinfo {pages} {053206} (\bibinfo {year}
  {2019})}\BibitemShut {NoStop}%
\bibitem [{\citenamefont {Dean}\ \emph {et~al.}(2021)\citenamefont {Dean},
  \citenamefont {Majumdar},\ and\ \citenamefont {Schawe}}]{dean2021position}%
  \BibitemOpen
  \bibfield  {author} {\bibinfo {author} {\bibfnamefont {D.~S.}\ \bibnamefont
  {Dean}}, \bibinfo {author} {\bibfnamefont {S.~N.}\ \bibnamefont {Majumdar}},\
  and\ \bibinfo {author} {\bibfnamefont {H.}~\bibnamefont {Schawe}},\
  }\href@noop {} {\bibfield  {journal} {\bibinfo  {journal} {Physical Review
  E}\ }\textbf {\bibinfo {volume} {103}},\ \bibinfo {pages} {012130} (\bibinfo
  {year} {2021})}\BibitemShut {NoStop}%
\bibitem [{\citenamefont {Redner}(2001)}]{redner2001guide}%
  \BibitemOpen
  \bibfield  {author} {\bibinfo {author} {\bibfnamefont {S.}~\bibnamefont
  {Redner}},\ }\href@noop {} {\emph {\bibinfo {title} {A guide to first-passage
  processes}}}\ (\bibinfo  {publisher} {Cambridge university press},\ \bibinfo
  {year} {2001})\BibitemShut {NoStop}%
\bibitem [{\citenamefont {Hilfer}(1995)}]{hilfer1995exact}%
  \BibitemOpen
  \bibfield  {author} {\bibinfo {author} {\bibfnamefont {R.}~\bibnamefont
  {Hilfer}},\ }\href@noop {} {\bibfield  {journal} {\bibinfo  {journal}
  {Fractals}\ }\textbf {\bibinfo {volume} {3}},\ \bibinfo {pages} {211}
  (\bibinfo {year} {1995})}\BibitemShut {NoStop}%
\bibitem [{\citenamefont {Hilfer}(2003)}]{hilfer2003fractional}%
  \BibitemOpen
  \bibfield  {author} {\bibinfo {author} {\bibfnamefont {R.}~\bibnamefont
  {Hilfer}},\ }\href@noop {} {\bibfield  {journal} {\bibinfo  {journal}
  {Physica A: Statistical Mechanics and its Applications}\ }\textbf {\bibinfo
  {volume} {329}},\ \bibinfo {pages} {35} (\bibinfo {year} {2003})}\BibitemShut
  {NoStop}%
\bibitem [{\citenamefont {Prados}\ \emph {et~al.}(1997)\citenamefont {Prados},
  \citenamefont {Brey},\ and\ \citenamefont
  {S{\'a}nchez-Rey}}]{prados1997dynamical}%
  \BibitemOpen
  \bibfield  {author} {\bibinfo {author} {\bibfnamefont {A.}~\bibnamefont
  {Prados}}, \bibinfo {author} {\bibfnamefont {J.}~\bibnamefont {Brey}},\ and\
  \bibinfo {author} {\bibfnamefont {B.}~\bibnamefont {S{\'a}nchez-Rey}},\
  }\href@noop {} {\bibfield  {journal} {\bibinfo  {journal} {Journal of
  Statistical Physics}\ }\textbf {\bibinfo {volume} {89}},\ \bibinfo {pages}
  {709} (\bibinfo {year} {1997})}\BibitemShut {NoStop}%
\bibitem [{\citenamefont {Bortz}\ \emph {et~al.}(1975)\citenamefont {Bortz},
  \citenamefont {Kalos},\ and\ \citenamefont {Lebowitz}}]{bortz1975new}%
  \BibitemOpen
  \bibfield  {author} {\bibinfo {author} {\bibfnamefont {A.~B.}\ \bibnamefont
  {Bortz}}, \bibinfo {author} {\bibfnamefont {M.~H.}\ \bibnamefont {Kalos}},\
  and\ \bibinfo {author} {\bibfnamefont {J.~L.}\ \bibnamefont {Lebowitz}},\
  }\href@noop {} {\bibfield  {journal} {\bibinfo  {journal} {Journal of
  Computational Physics}\ }\textbf {\bibinfo {volume} {17}},\ \bibinfo {pages}
  {10} (\bibinfo {year} {1975})}\BibitemShut {NoStop}%
\bibitem [{\citenamefont {Voter}(2007)}]{voter2007introduction}%
  \BibitemOpen
  \bibfield  {author} {\bibinfo {author} {\bibfnamefont {A.~F.}\ \bibnamefont
  {Voter}},\ }in\ \href@noop {} {\emph {\bibinfo {booktitle} {Radiation effects
  in solids}}}\ (\bibinfo  {publisher} {Springer},\ \bibinfo {year} {2007})\
  pp.\ \bibinfo {pages} {1--23}\BibitemShut {NoStop}%
\bibitem [{SI()}]{SI}%
  \BibitemOpen
  \href@noop {} {}\bibinfo {note} {See Supplemental Material for
  details}\BibitemShut {NoStop}%
\bibitem [{\citenamefont {Ellis}(1984)}]{ellis1984large}%
  \BibitemOpen
  \bibfield  {author} {\bibinfo {author} {\bibfnamefont {R.~S.}\ \bibnamefont
  {Ellis}},\ }\href@noop {} {\bibfield  {journal} {\bibinfo  {journal} {The
  Annals of Probability}\ }\textbf {\bibinfo {volume} {12}},\ \bibinfo {pages}
  {1} (\bibinfo {year} {1984})}\BibitemShut {NoStop}%
\bibitem [{\citenamefont {Banerjee}\ \emph {et~al.}(2020)\citenamefont
  {Banerjee}, \citenamefont {Majumdar}, \citenamefont {Rosso},\ and\
  \citenamefont {Schehr}}]{banerjee2020current}%
  \BibitemOpen
  \bibfield  {author} {\bibinfo {author} {\bibfnamefont {T.}~\bibnamefont
  {Banerjee}}, \bibinfo {author} {\bibfnamefont {S.~N.}\ \bibnamefont
  {Majumdar}}, \bibinfo {author} {\bibfnamefont {A.}~\bibnamefont {Rosso}},\
  and\ \bibinfo {author} {\bibfnamefont {G.}~\bibnamefont {Schehr}},\
  }\href@noop {} {\bibfield  {journal} {\bibinfo  {journal} {Physical Review
  E}\ }\textbf {\bibinfo {volume} {101}},\ \bibinfo {pages} {052101} (\bibinfo
  {year} {2020})}\BibitemShut {NoStop}%
\bibitem [{\citenamefont {Lacroix-A-Chez-Toine}\ and\ \citenamefont
  {Mori}(2020)}]{lacroix2020universal}%
  \BibitemOpen
  \bibfield  {author} {\bibinfo {author} {\bibfnamefont {B.}~\bibnamefont
  {Lacroix-A-Chez-Toine}}\ and\ \bibinfo {author} {\bibfnamefont
  {F.}~\bibnamefont {Mori}},\ }\href@noop {} {\bibfield  {journal} {\bibinfo
  {journal} {Journal of Physics A: Mathematical and Theoretical}\ }\textbf
  {\bibinfo {volume} {53}},\ \bibinfo {pages} {495002} (\bibinfo {year}
  {2020})}\BibitemShut {NoStop}%
\bibitem [{\citenamefont {De~Bruyne}\ \emph {et~al.}(2021)\citenamefont
  {De~Bruyne}, \citenamefont {Majumdar},\ and\ \citenamefont
  {Schehr}}]{de2021survival}%
  \BibitemOpen
  \bibfield  {author} {\bibinfo {author} {\bibfnamefont {B.}~\bibnamefont
  {De~Bruyne}}, \bibinfo {author} {\bibfnamefont {S.~N.}\ \bibnamefont
  {Majumdar}},\ and\ \bibinfo {author} {\bibfnamefont {G.}~\bibnamefont
  {Schehr}},\ }\href@noop {} {\bibfield  {journal} {\bibinfo  {journal}
  {Journal of Statistical Mechanics: Theory and Experiment}\ }\textbf {\bibinfo
  {volume} {2021}},\ \bibinfo {pages} {043211} (\bibinfo {year}
  {2021})}\BibitemShut {NoStop}%
\bibitem [{\citenamefont {De~Masi}\ \emph {et~al.}(1985)\citenamefont
  {De~Masi}, \citenamefont {Ferrari},\ and\ \citenamefont
  {Lebowitz}}]{de1985rigorous}%
  \BibitemOpen
  \bibfield  {author} {\bibinfo {author} {\bibfnamefont {A.}~\bibnamefont
  {De~Masi}}, \bibinfo {author} {\bibfnamefont {P.}~\bibnamefont {Ferrari}},\
  and\ \bibinfo {author} {\bibfnamefont {J.}~\bibnamefont {Lebowitz}},\
  }\href@noop {} {\bibfield  {journal} {\bibinfo  {journal} {Physical Review
  Letters}\ }\textbf {\bibinfo {volume} {55}},\ \bibinfo {pages} {1947}
  (\bibinfo {year} {1985})}\BibitemShut {NoStop}%
\bibitem [{\citenamefont {De~Masi}\ \emph {et~al.}(1986)\citenamefont
  {De~Masi}, \citenamefont {Ferrari},\ and\ \citenamefont
  {Lebowitz}}]{de1986reaction}%
  \BibitemOpen
  \bibfield  {author} {\bibinfo {author} {\bibfnamefont {A.}~\bibnamefont
  {De~Masi}}, \bibinfo {author} {\bibfnamefont {P.~A.}\ \bibnamefont
  {Ferrari}},\ and\ \bibinfo {author} {\bibfnamefont {J.~L.}\ \bibnamefont
  {Lebowitz}},\ }\href@noop {} {\bibfield  {journal} {\bibinfo  {journal}
  {Journal of Statistical Physics}\ }\textbf {\bibinfo {volume} {44}},\
  \bibinfo {pages} {589} (\bibinfo {year} {1986})}\BibitemShut {NoStop}%
\bibitem [{\citenamefont {Agranov}\ \emph {et~al.}(2021)\citenamefont
  {Agranov}, \citenamefont {Ro}, \citenamefont {Kafri},\ and\ \citenamefont
  {Lecomte}}]{agranov2021exact}%
  \BibitemOpen
  \bibfield  {author} {\bibinfo {author} {\bibfnamefont {T.}~\bibnamefont
  {Agranov}}, \bibinfo {author} {\bibfnamefont {S.}~\bibnamefont {Ro}},
  \bibinfo {author} {\bibfnamefont {Y.}~\bibnamefont {Kafri}},\ and\ \bibinfo
  {author} {\bibfnamefont {V.}~\bibnamefont {Lecomte}},\ }\href@noop {}
  {\bibfield  {journal} {\bibinfo  {journal} {Journal of Statistical Mechanics:
  Theory and Experiment}\ }\textbf {\bibinfo {volume} {2021}},\ \bibinfo
  {pages} {083208} (\bibinfo {year} {2021})}\BibitemShut {NoStop}%
\end{thebibliography}%
 
\begin{widetext}

\section*{\large Supplemental Material for\\ ''Active Random Walks in One and Two Dimensions''}

In this document, we provide details related to the results presented in the main text. Throughout the calculations, we have set the intrinsic diffusion constants $D_{1d}=\frac{1}{2}$ in one dimension and $D_{2d}=\frac{1}{4}$ for convenience.
\section{ List of Moments and Cumulants - One Dimension}
\label{A_supp}
\subsubsection{Discrete Space Domain}
Listed below are the first few non-zero even moments computed using Eq.~(23) in the main text;
\begin{equation} 
\label{msd1d}
\langle x^{2}(t) \rangle = \frac{2 \left(-1+e^{-2 t \gamma }\right) \epsilon ^2}{\gamma ^2}+t \left(1+\frac{4 \epsilon ^2}{\gamma }\right),
\end{equation}

\begin{eqnarray} 
\label{mfd1d}
\langle x^{4}(t) \rangle &=&-\frac{8 e^{-2 t \gamma } \left(-1+e^{2 t \gamma }\right) \epsilon ^2 \left(\gamma ^2-9 \epsilon ^2\right)}{\gamma ^4}+\frac{t
   \left(\gamma ^3+12 \left(-1+e^{-2 t \gamma }\right) \gamma  \epsilon ^2+16 \gamma ^2 \epsilon ^2+48 \left(-2-e^{-2 t \gamma
   }\right) \epsilon ^4\right)}{\gamma ^3}\nonumber\\&&+\frac{3 t^2 \left(\gamma +4 \epsilon ^2\right)^2}{\gamma ^2},
\end{eqnarray}

\begin{eqnarray}
\label{msixd1d}
\langle x^{6}(t) \rangle &=&-\frac{64 e^{-t \gamma } \epsilon ^2 \left(\gamma ^4-45 \gamma ^2 \epsilon ^2+225 \epsilon ^4\right) \sinh (t \gamma )}{\gamma
   ^6}\nonumber\\&&+\frac{t \left(\gamma ^5+2 \gamma ^3 (-75+32 \gamma ) \epsilon ^2+120 (9-16 \gamma ) \gamma  \epsilon ^4+8640 \epsilon ^6+30
   e^{-2 t \gamma } \epsilon ^2 \left(5 \gamma ^3-4 \gamma  (9+8 \gamma ) \epsilon ^2+192 \epsilon ^4\right)\right)}{\gamma
   ^5}\nonumber\\&&+\frac{15 e^{-2 t \gamma } t^2 \left(6 \epsilon ^2 \left(\gamma -4 \epsilon ^2\right)^2+e^{2 t \gamma } \left(\gamma +4
   \epsilon ^2\right) \left(\gamma ^3+2 \gamma  (-3+8 \gamma ) \epsilon ^2-72 \epsilon ^4\right)\right)}{\gamma ^4}+\frac{15 t^3
   \left(\gamma +4 \epsilon ^2\right)^3}{\gamma ^3}.
\end{eqnarray}

\begin{figure}[t]
\centering
\subfigure[]{\includegraphics[width=2.5in]{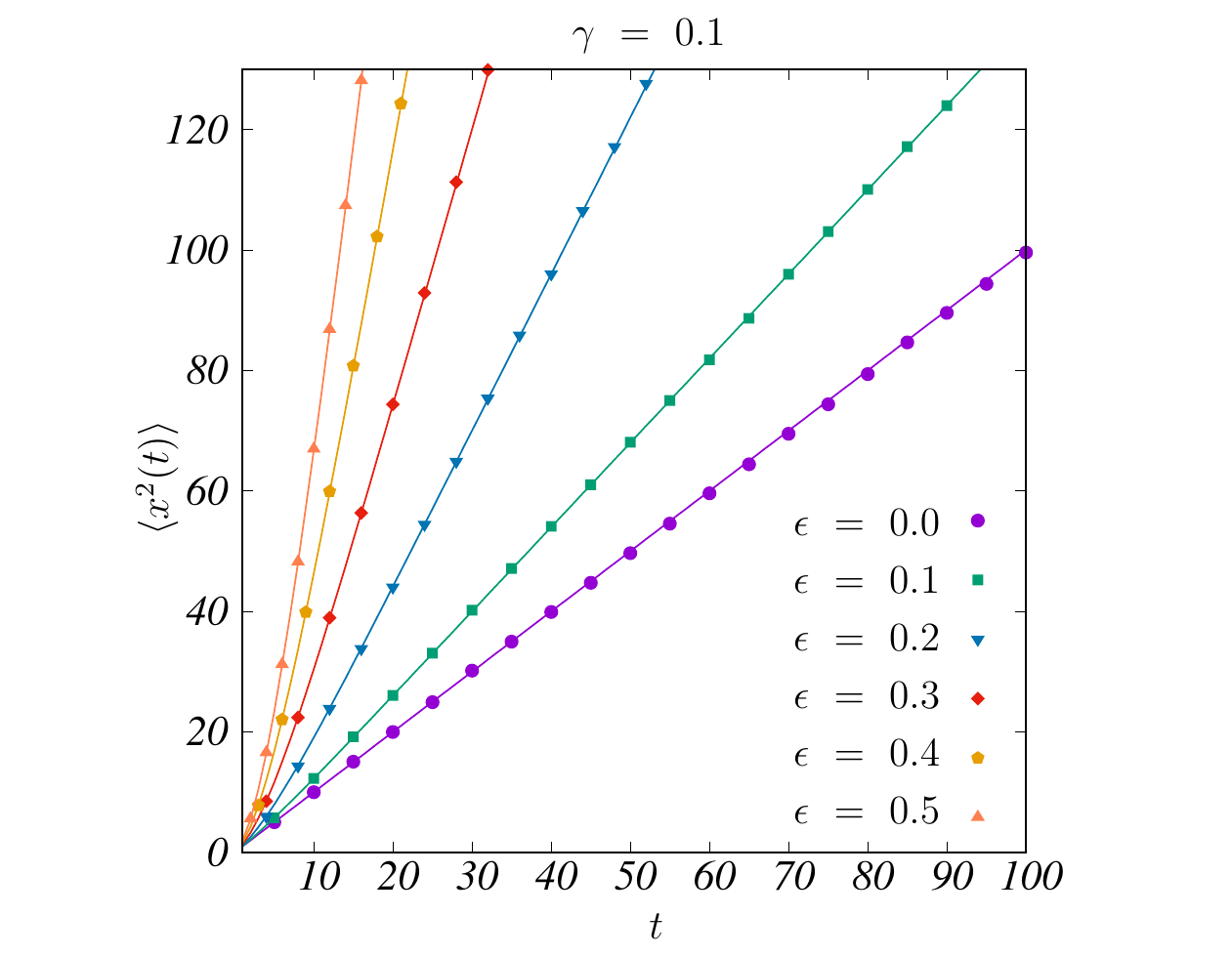}}\hspace{-1 cm}
\subfigure[]{\includegraphics[width=2.5in]{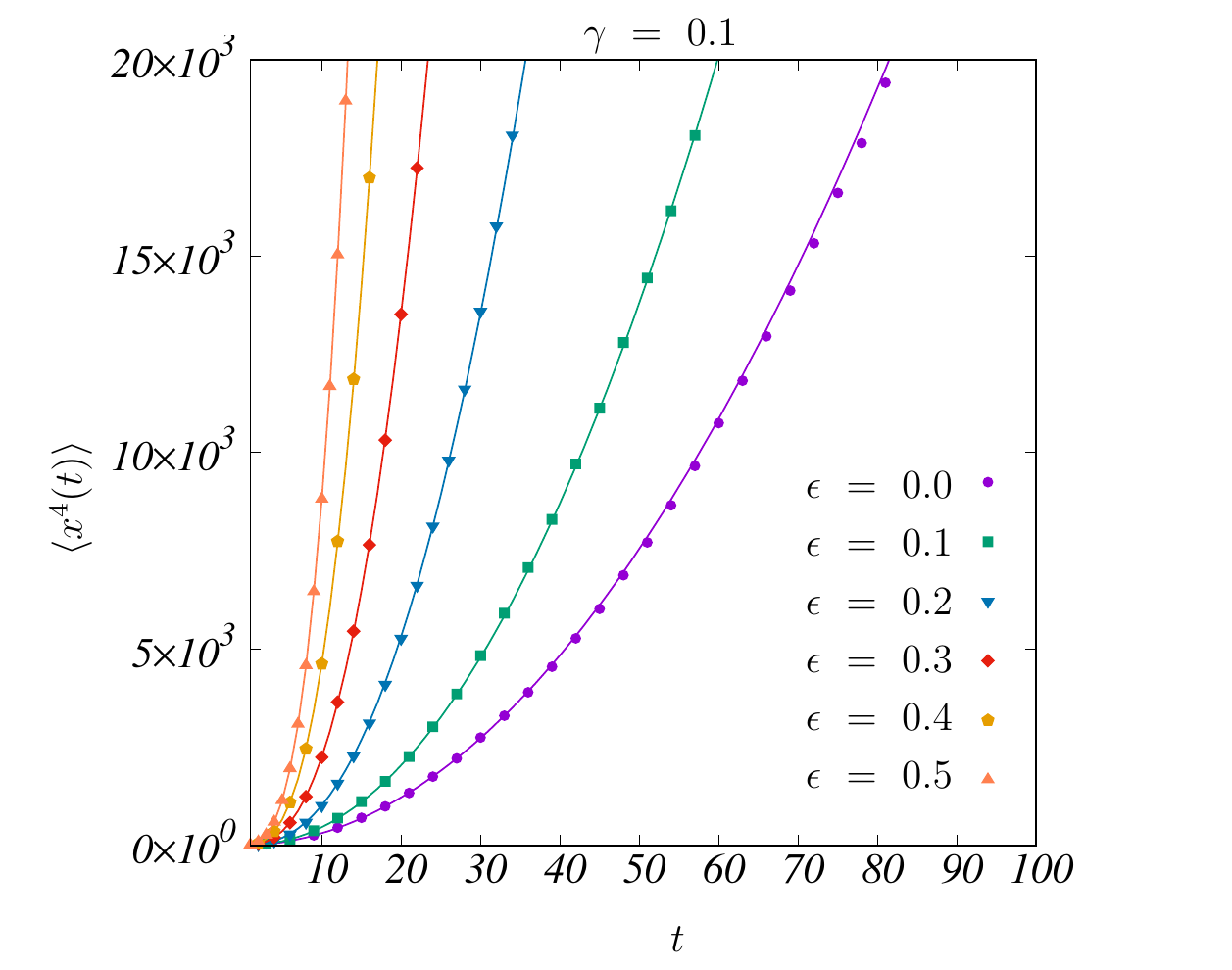} } \hspace{-1 cm}
\subfigure[]{\includegraphics[width=2.5in]{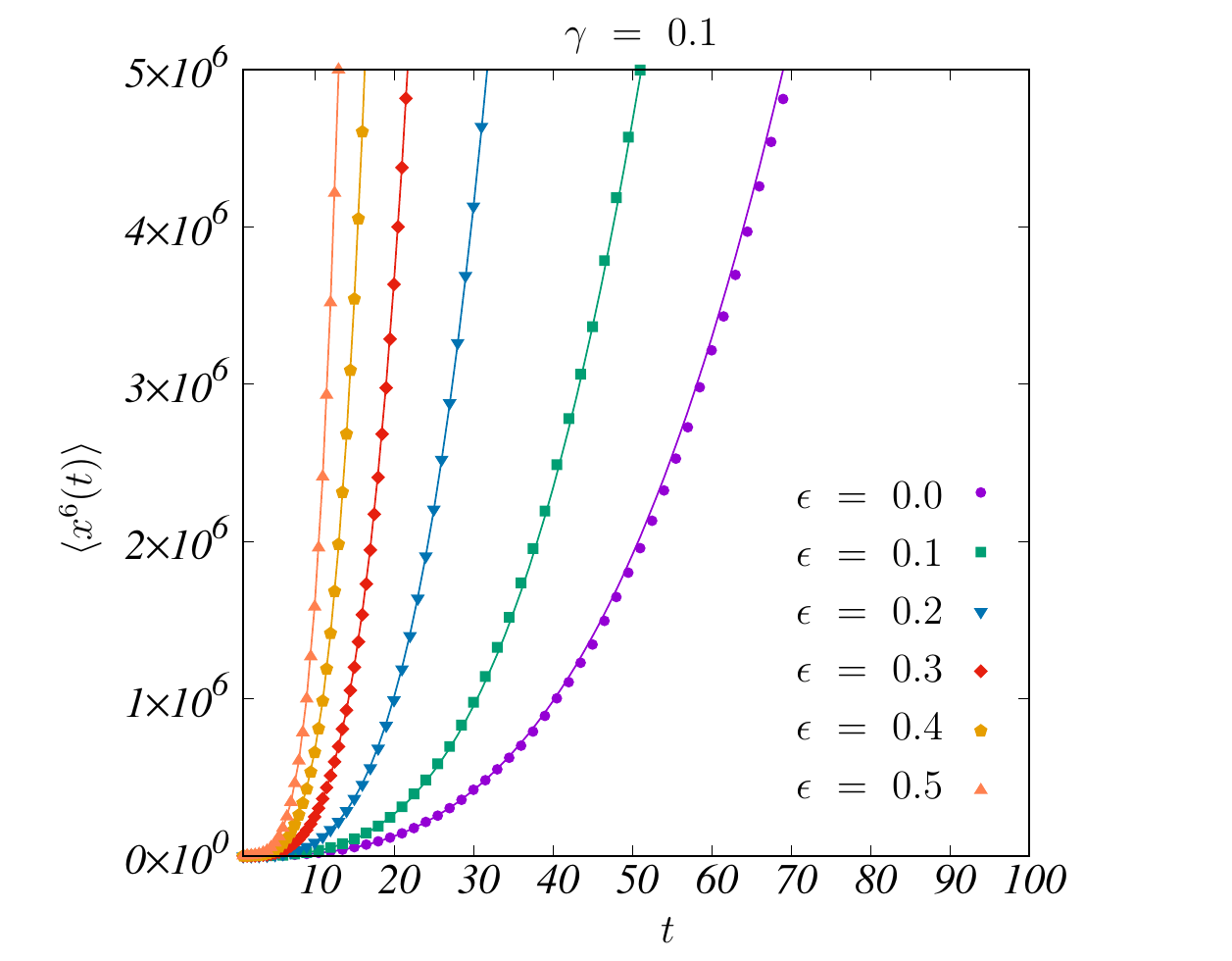} }
\subfigure[]{\includegraphics[width=2.5in]{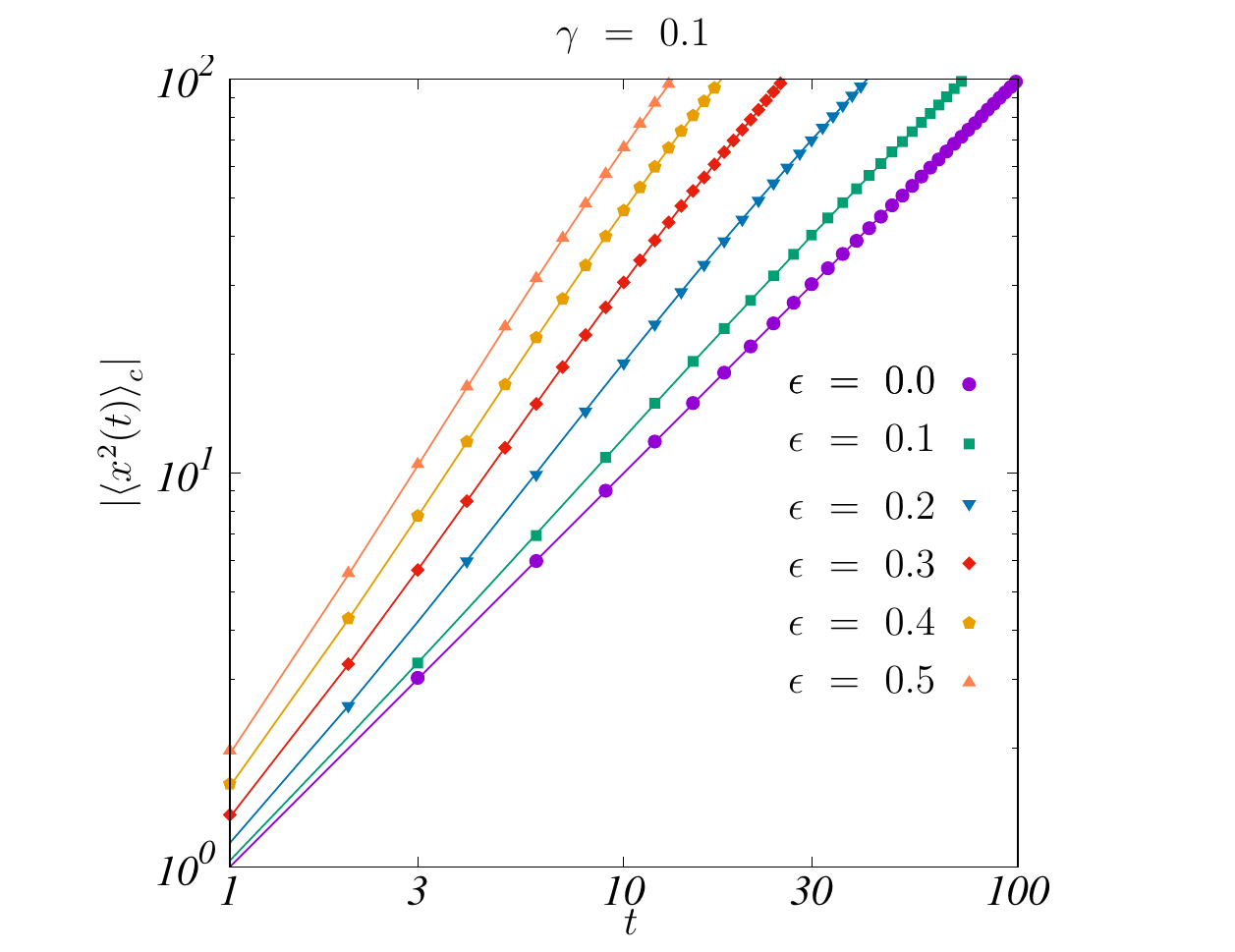}}\hspace{-1 cm}
\subfigure[]{\includegraphics[width=2.5in]{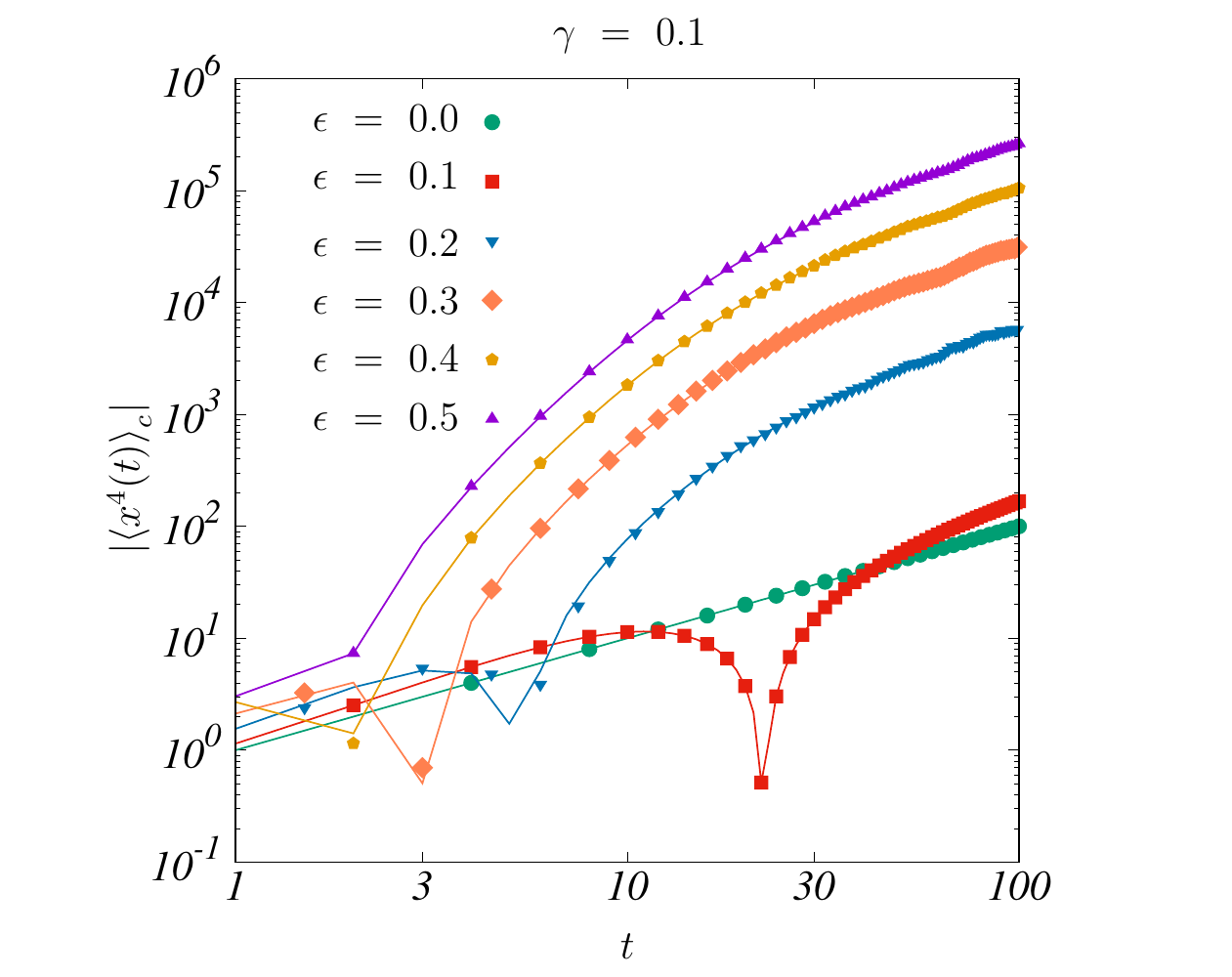} } \hspace{-1 cm}
\subfigure[]{\includegraphics[width=2.5in]{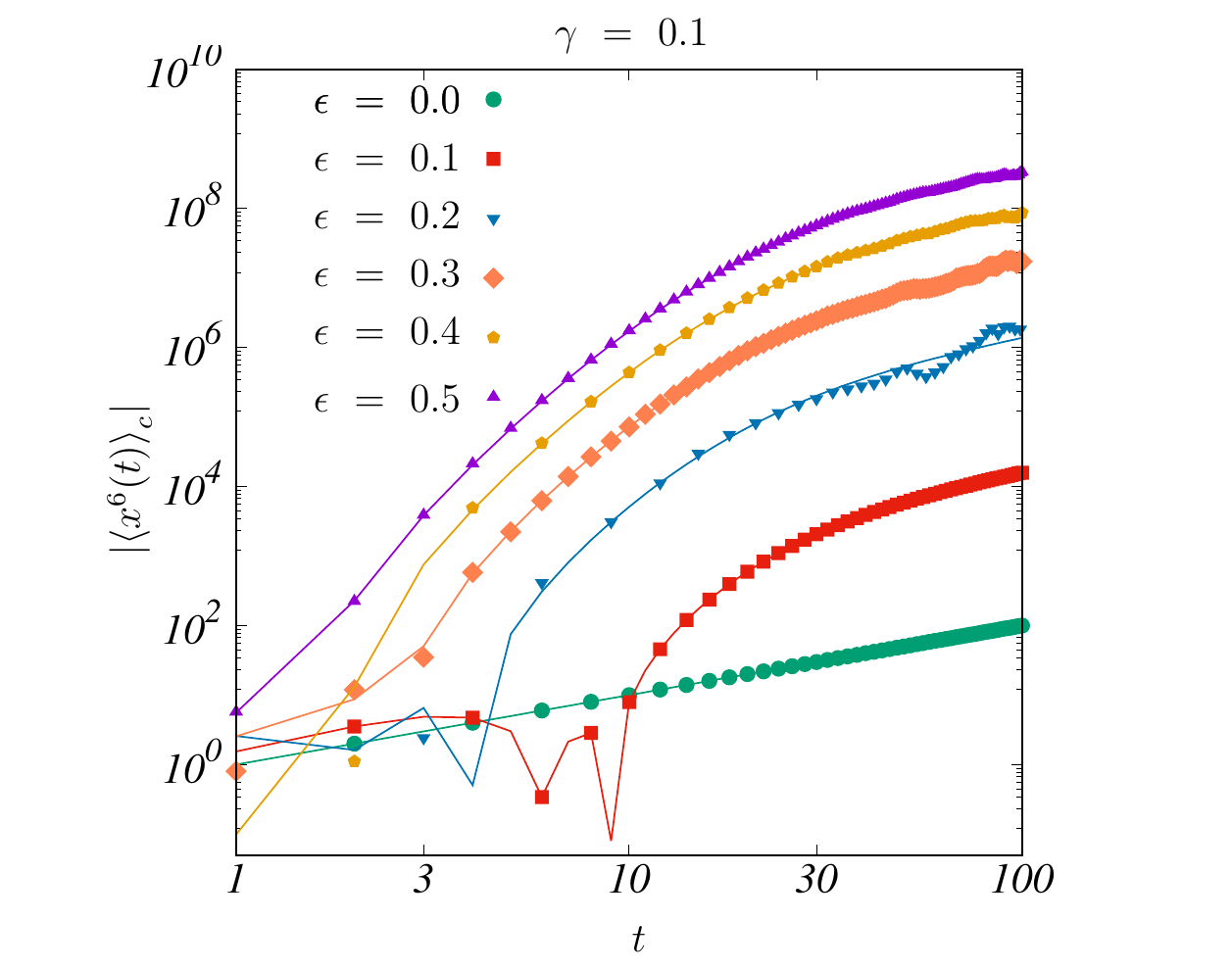} }
\caption{The moments and cumulants of the position of a run and tumble particle on a one dimensional infinite lattice plotted as a function of time for different values of $\epsilon$. The fixed parameter used is $\gamma=0.1$. (a) $\langle x^2(t) \rangle$ plotted as a function of $t$. The solid lines correspond to the theoretical result in Eq.~(\ref{msd1d}) and the points are from simulations. (b) $\langle x^4(t) \rangle$ plotted as a function of $t$. The solid lines correspond to the theoretical result in Eq.~(\ref{mfd1d}) and the points are from simulations. (c) $\langle x^6(t) \rangle$ plotted as a function of $t$. The solid lines correspond to the theoretical result in Eq.~(\ref{msixd1d}) and the points are from simulations. (d) ${\langle x^2(t) \rangle}_c$ plotted (in logscale) as a function of $t$. The solid lines correspond to the theoretical result in Eq.~(\ref{msdc1d}) and the points are from simulations. (e) ${\langle x^4(t) \rangle}_c$ plotted (in log scale) as a function of $t$. The solid lines correspond to the theoretical result in Eq.~(\ref{mfdc1d}) and the points are from simulations. (f) ${\langle x^6(t) \rangle}_c$ plotted (log scale) as a function of $t$. The solid lines correspond to the theoretical result in Eq.~(\ref{msixdc1d}) and the points are from simulations. 
} \label{fig:ap2}
\end{figure}

Listed below are the first few non-zero cumulants computed using Eq.~(29) in the main text;
\begin{equation} 
\label{msdc1d}
{\langle x^{2}(t) \rangle}_c = \frac{2 \left(-1+e^{-2 t \gamma }\right) \epsilon ^2}{\gamma ^2}+t \left(1+\frac{4 \epsilon ^2}{\gamma }\right),
\end{equation}

\begin{eqnarray} 
\label{mfdc1d}
{\langle x^{4}(t) \rangle}_c &=&\frac{4 e^{-4 t \gamma } \left(-1+e^{2 t \gamma }\right) \epsilon ^2 \left(3 \epsilon ^2+e^{2 t \gamma } \left(-2 \gamma ^2+15
   \epsilon ^2\right)\right)}{\gamma ^4}+t \left(1+\frac{16 \epsilon ^2 \left(\gamma ^2+3 \left(-1-2 e^{-2 t \gamma }\right)
   \epsilon ^2\right)}{\gamma ^3}\right),
\end{eqnarray}

\begin{eqnarray}
\label{msixdc1d}
{\langle x^{6}(t) \rangle}_c &=&-\frac{16 e^{-6 t \gamma } \left(-1+e^{2 t \gamma }\right) \epsilon ^2 \left(15 \epsilon ^4-15 e^{2 t \gamma } \epsilon ^2
   \left(\gamma ^2-7 \epsilon ^2\right)+e^{4 t \gamma } \left(2 \gamma ^4-75 \gamma ^2 \epsilon ^2+330 \epsilon
   ^4\right)\right)}{\gamma ^6}\nonumber\\&&+t \left(1+\frac{64 \epsilon ^2}{\gamma
   }-\frac{960 e^{-2 t \gamma } \left(2+e^{2 t \gamma }\right) \epsilon ^4}{\gamma ^3}+\frac{2880 e^{-4 t \gamma } \left(1+3 e^{2 t
   \gamma }+e^{4 t \gamma }\right) \epsilon ^6}{\gamma ^5}\right)+\frac{5760 e^{-2 t \gamma } t^2 \epsilon ^6}{\gamma ^4}.
\end{eqnarray}
In Fig.~\ref{fig:ap2}, the moments and cumulants listed above are compared with the results from direct numerical simulations of a RTP on a one dimensional infinite lattice and are found to be in good agreement.\\\\

\textit{Asymptotic limit of cumulants from the free energy function:}\\

Listed below are the asymptotic limits of the first few cumulants computed using the expression for large deviation free energy function given in Eq.~(32) along with Eq.~(31) in the main text.
\begin{equation} 
\label{msdc1dasym}
\lim_{t \rightarrow \infty}{\langle x^{2}(t) \rangle}_c = t \left(1+\frac{4 \epsilon ^2}{\gamma }\right),
\end{equation}

\begin{eqnarray} 
\label{mfdc1dasym}
\lim_{t \rightarrow \infty}{\langle x^{4}(t) \rangle}_c &=&t \left(1+\frac{16 \epsilon ^2 \left(\gamma ^2-3 \epsilon ^2\right)}{\gamma ^3}\right),
\end{eqnarray}

\begin{eqnarray}
\label{msixdc1dasym}
\lim_{t \rightarrow \infty}{\langle x^{6}(t) \rangle}_c &=&t \left(1+\frac{64 \epsilon ^2 \left(\gamma ^4-15 \gamma ^2 \epsilon ^2+45 \epsilon
   ^4\right)}{\gamma ^5}\right).
\end{eqnarray}

The expressions for the asymptotic limits of the cumulants given in Eq.~(\ref{msdc1dasym}), Eq.~(\ref{mfdc1dasym}) and Eq.~(\ref{msixdc1dasym}) can also be derived by taking a $t \rightarrow \infty$ limit of Eq.~(\ref{msdc1d}), Eq.~(\ref{mfdc1d}) and Eq.~(\ref{msixdc1d}) respectively.
\subsubsection{Continuous Space Domain}

Listed below are a few non-zero even moments computed using the continuum limit of Eq.~(23) in the main text:

\begin{equation}
\langle x^{2}(t) \rangle = \frac{2 \left(-1+e^{-2 t \gamma }\right) \epsilon ^2}{\gamma ^2}+t \left(1+\frac{4 \epsilon ^2}{\gamma }\right),
\end{equation}

\begin{equation} 
\langle x^{4}(t) \rangle =\frac{72 \epsilon ^4 \left(1-e^{-2 t \gamma }\right)}{\gamma ^4}-\frac{t \left(12 e^{-2 t \gamma } \epsilon ^2 \left(-\gamma +4
   \epsilon ^2+e^{2 t \gamma } \left(\gamma +8 \epsilon ^2\right)\right)\right)}{\gamma ^3}+\frac{3 t^2 \left(\gamma +4 \epsilon
   ^2\right)^2}{\gamma ^2},
\end{equation}

\begin{eqnarray}
\langle x^{6}(t) \rangle &=&-\frac{7200 e^{-2 t \gamma } \left(-1+e^{2 t \gamma }\right) \epsilon ^6}{\gamma ^6}+\frac{t \left(360 e^{-2 t \gamma } \epsilon ^4
   \left(-3 \gamma +16 \epsilon ^2+3 e^{2 t \gamma } \left(\gamma +8 \epsilon ^2\right)\right)\right)}{\gamma ^5}\nonumber\\&&+\frac{t^2
   \left(90 e^{-2 t \gamma } \epsilon ^2 \left(\left(\gamma -4 \epsilon ^2\right)^2-e^{2 t \gamma } \left(\gamma +4 \epsilon
   ^2\right) \left(\gamma +12 \epsilon ^2\right)\right)\right)}{\gamma ^4}+\frac{15 t^3 \left(\gamma +4 \epsilon
   ^2\right)^3}{\gamma ^3}.
\end{eqnarray}

Listed below are a few non-zero cumulants computed using the continuum limit of Eq.~(29) in the main text;
\begin{equation} 
\label{cmsdc1d}
{\langle x^{2}(t) \rangle}_c = \frac{2 \left(-1+e^{-2 t \gamma }\right) \epsilon ^2}{\gamma ^2}+t \left(1+\frac{4 \epsilon ^2}{\gamma }\right),
\end{equation}

\begin{eqnarray} 
\label{cmfdc1d}
{\langle x^{4}(t) \rangle}_c &=&\frac{12 e^{-4 t \gamma } \left(-1-4 e^{2 t \gamma }+5 e^{4 t \gamma }\right) \epsilon ^4}{\gamma ^4}-\frac{48 t e^{-2 t \gamma }
   \left(2+e^{2 t \gamma }\right) \epsilon ^4}{\gamma ^3},
\end{eqnarray}

\begin{eqnarray}
\label{cmsixdc1d}
{\langle x^{6}(t) \rangle}_c &=&-\frac{240 e^{-6 t \gamma } \left(-1-6 e^{2 t \gamma }-15 e^{4 t \gamma }+22 e^{6 t \gamma }\right) \epsilon ^6}{\gamma
   ^6}+\frac{2880 t e^{-4 t \gamma } \left(1+3 e^{2 t \gamma }+e^{4 t \gamma }\right) \epsilon ^6}{\gamma ^5}+\frac{5760 e^{-2 t
   \gamma } t^2 \epsilon ^6}{\gamma ^4}
\end{eqnarray}

\section{ List of Moments and Cumulants - Two Dimensions}
\label{B_supp}

\subsubsection{Discrete Space Domain}

Listed are the first few non-zero moments and correlations computed using Eq.~(54) in the main text;
\begin{equation} 
\label{msd2d}
\langle x^{2}(t) \rangle=\langle y^{2}(t) \rangle  = -\frac{4 \left(1-e^{-t \gamma }\right) \epsilon ^2}{\gamma ^2}+t \left(\frac{1}{2}+\frac{4 \epsilon ^2}{\gamma }\right),
\end{equation}

\begin{eqnarray}
\label{mfd2d}
\langle x^{4}(t) \rangle=\langle y^{4}(t) \rangle  &=&
\frac{8 e^{-2 t \gamma }
   \left(-1+e^{t \gamma }\right) \epsilon ^2 \left(-3 \epsilon ^2+e^{t \gamma } \left(-2 \gamma ^2+21 \epsilon
   ^2\right)\right)}{\gamma ^4}\nonumber\\&&+\frac{t \left(\gamma ^3+24 \left(-1+e^{-t \gamma
   }\right) \gamma  \epsilon ^2+32 \gamma ^2 \epsilon ^2-288 \epsilon ^4\right)}{2 \gamma ^3}+\frac{t^2 \left(3 \left(\gamma +8 \epsilon ^2\right)^2\right)}{4 \gamma ^2},
   \end{eqnarray}

\begin{eqnarray}
\label{msixd2d}
\langle x^{6}(t) \rangle=\langle y^{6}(t) \rangle  &=&
  -\frac{32 e^{-2 t \gamma
   } \left(-1+e^{t \gamma }\right) \epsilon ^2 \left(15 \epsilon ^2 \left(\gamma ^2-9 \epsilon ^2\right)+e^{t \gamma } \left(2
   \gamma ^4-105 \gamma ^2 \epsilon ^2+585 \epsilon ^4\right)\right)}{\gamma ^6} \nonumber\\&&+\frac{t}{2 \gamma ^5} \Bigg(e^{-2 t \gamma } \Big(60 e^{t \gamma } \gamma  \epsilon ^2 \left(5 \gamma ^2-48 \epsilon
   ^2\right)+360 \epsilon ^4 \left(\gamma -8 \epsilon ^2\right)\nonumber\\&&+e^{2 t \gamma } \left(\gamma ^5+4 \gamma ^3 (-75+32 \gamma )
   \epsilon ^2+360 (7-16 \gamma ) \gamma  \epsilon ^4+31680 \epsilon ^6\right)\Big)\Bigg)\nonumber\\&&+t^2 \left(\frac{45 e^{-t \gamma } \epsilon ^2}{\gamma
   ^2}+\frac{15 \left(\gamma +8 \epsilon ^2\right) \left(\gamma ^3+4 \gamma  (-3+8 \gamma ) \epsilon ^2-192 \epsilon ^4\right)}{4
   \gamma ^4}\right)+\frac{t^3 \left(15 \left(\gamma +8 \epsilon ^2\right)^3\right)}{8 \gamma ^3} .
   \end{eqnarray}

\textit{Cross Correlations}
   \begin{equation} 
\langle {(x^{2n+1}(t)y^m(t))} \rangle = \langle {(y^{2n+1}(t)x^m(t))} \rangle= 0, n=m=0,1,2,3,...
\end{equation}

   \begin{eqnarray}
   \label{aq}
\langle {(x(t)y(t))}^{2} \rangle &=&\frac{8 e^{-2 t \gamma } \left(-1+e^{t \gamma }\right) \left(1+17 e^{t \gamma }\right)
   \epsilon ^4}{\gamma ^4}-\frac{4 t e^{-t \gamma } \epsilon ^2 \left(\left(-1+e^{t \gamma
   }\right) \gamma +4 \left(4+5 e^{t \gamma }\right) \epsilon ^2\right)}{\gamma
   ^3}+\frac{t^2 \left(\gamma +8 \epsilon ^2\right)^2}{4 \gamma ^2},\nonumber\\
\end{eqnarray}

   \begin{eqnarray}
\langle {x(t)}^{2} \rangle\langle {y(t)}^{2} \rangle &=&\frac{16 e^{-2 t \gamma } \left(-1+e^{t \gamma }\right)^2 \epsilon ^4}{\gamma ^4}-\frac{4 t
   e^{-t \gamma } \left(-1+e^{t \gamma }\right) \epsilon ^2 \left(\gamma +8 \epsilon
   ^2\right)}{\gamma ^3}+\frac{t^2 \left(\gamma +8 \epsilon ^2\right)^2}{4 \gamma ^2},
\end{eqnarray}

   \begin{eqnarray}
   \label{aq1}
\langle {(x^4(t)y^2(t))} \rangle &=&=\langle {(y^4(t)x^2(t))} \rangle \nonumber\\&&=\frac{32 e^{-2 t \gamma } \left(-1+e^{t \gamma }\right) \epsilon ^4 \left(\left(1+17 e^{t
   \gamma }\right) \gamma ^2-9 \left(1+49 e^{t \gamma }\right) \epsilon ^2\right)}{\gamma
   ^6}+\frac{t}{\gamma ^5} \Big[2 e^{-2 t \gamma } \epsilon ^2 (-6 \epsilon ^2 (\gamma -8
   \epsilon ^2)\nonumber\\&&+e^{t \gamma } (5 \gamma ^3-16 \gamma  (15+8 \gamma ) \epsilon
   ^2+3072 \epsilon ^4)+e^{2 t \gamma } (-5 \gamma ^3+2 (123-80 \gamma ) \gamma 
   \epsilon ^2+4080 \epsilon ^4))\Big]\nonumber\\&&+\frac{t^2 \left(12 e^{-t
   \gamma } \epsilon ^2 \left(\gamma -16 \epsilon ^2\right) \left(3 \gamma -16 \epsilon
   ^2\right)+\left(\gamma +8 \epsilon ^2\right) \left(\gamma ^3+4 \gamma  (-9+8 \gamma )
   \epsilon ^2-960 \epsilon ^4\right)\right)}{4 \gamma ^4}+\frac{3 t^3 \left(\gamma +8
   \epsilon ^2\right)^3}{8 \gamma ^3},\nonumber\\
\end{eqnarray}

   \begin{eqnarray}
\langle {x^{4}(t)} \rangle\langle {y^{2}(t)} \rangle &=&\langle {y^{4}(t)} \rangle\langle {x^{2}(t)} \rangle \nonumber\\&&=\frac{32 e^{-3 t \gamma } \left(-1+e^{t \gamma }\right)^2 \epsilon ^4 \left(3 \epsilon
   ^2+e^{t \gamma } \left(2 \gamma ^2-21 \epsilon ^2\right)\right)}{\gamma ^6}+\frac{t}{\gamma
   ^5}
   \Big[-(2 e^{-2 t \gamma } (-1+e^{t \gamma }) \epsilon ^2 (6
   \epsilon ^2 (5 \gamma +8 \epsilon ^2)\nonumber\\&&+e^{t \gamma } (5 \gamma ^3+2 \gamma
    (-33+32 \gamma ) \epsilon ^2-624 \epsilon ^4)))\Big]+\frac{t^2}{4 \gamma ^4} \Big[e^{-t \gamma } (\gamma +8 \epsilon ^2) (36 \gamma 
   \epsilon ^2+96 \epsilon ^4\nonumber\\&&+e^{t \gamma } (\gamma ^3+4 \gamma  (-9+8 \gamma )
   \epsilon ^2-384 \epsilon ^4))\Big]+\frac{3 t^3 \left(\gamma
   +8 \epsilon ^2\right)^3}{8 \gamma ^3},
\end{eqnarray}
\begin{figure}[t]
\centering
\subfigure[]{\includegraphics[width=2.4in]{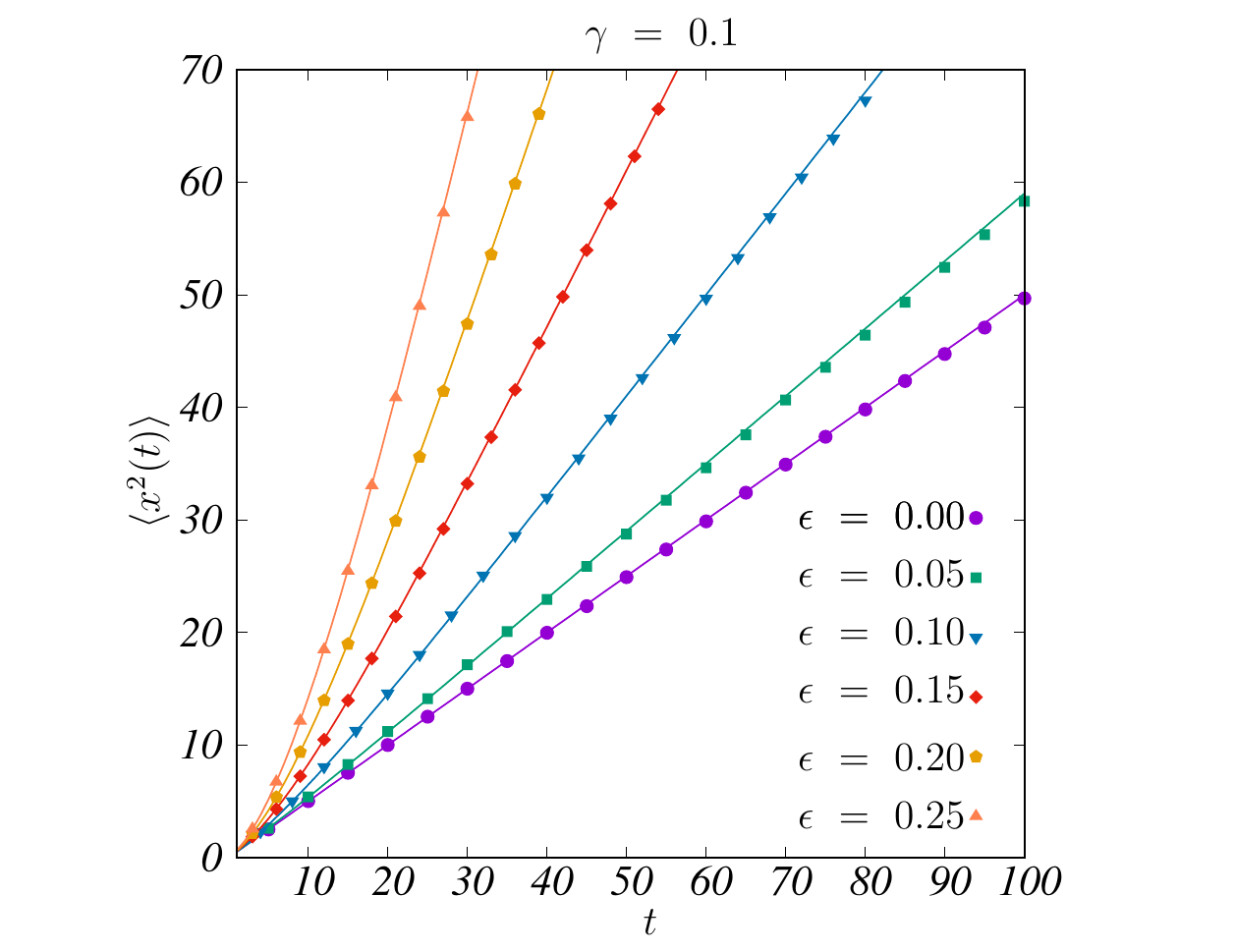}}\hspace{-1 cm}
\subfigure[]{\includegraphics[width=2.4in]{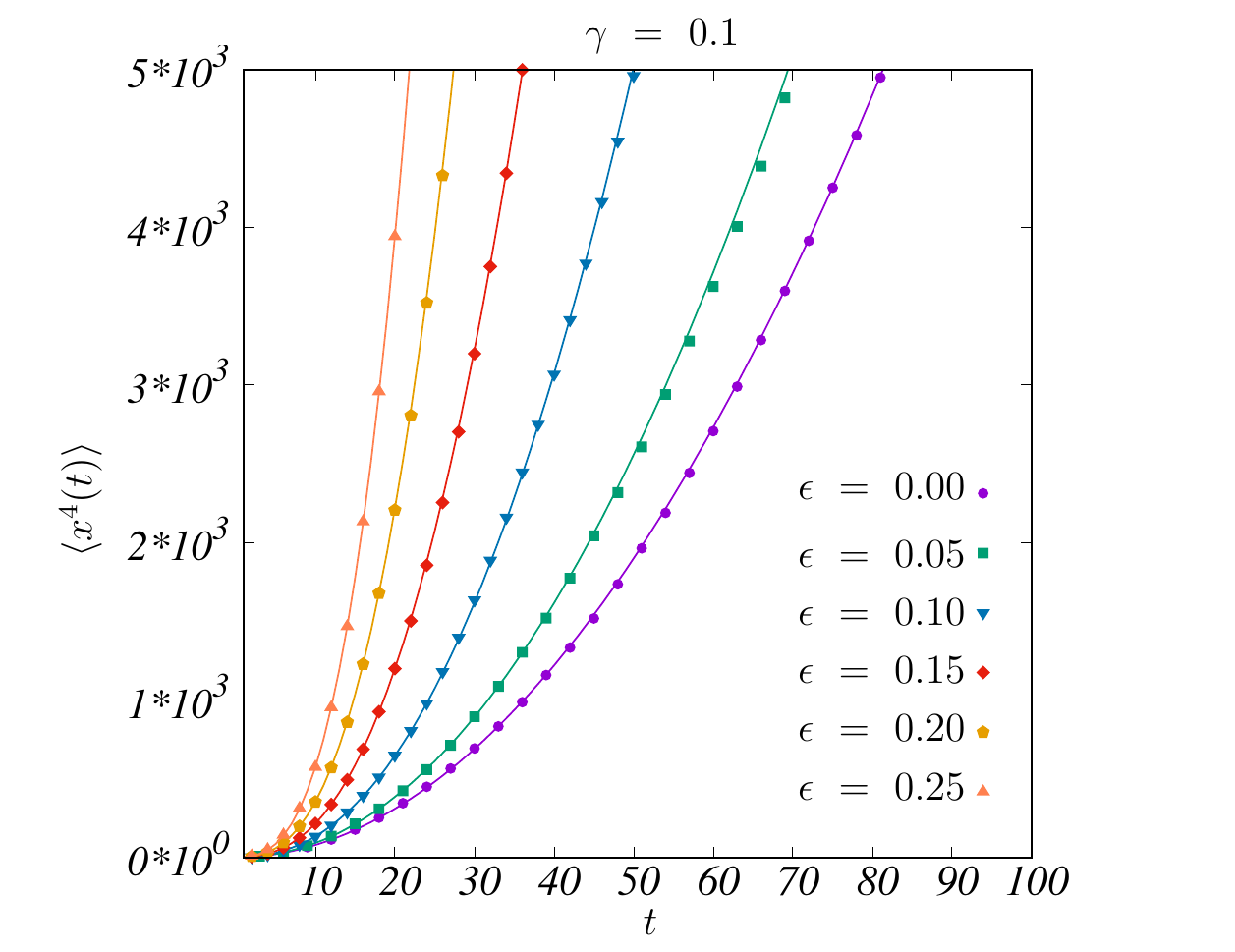} }\hspace{-1 cm}
\subfigure[]{\includegraphics[width=2.4in]{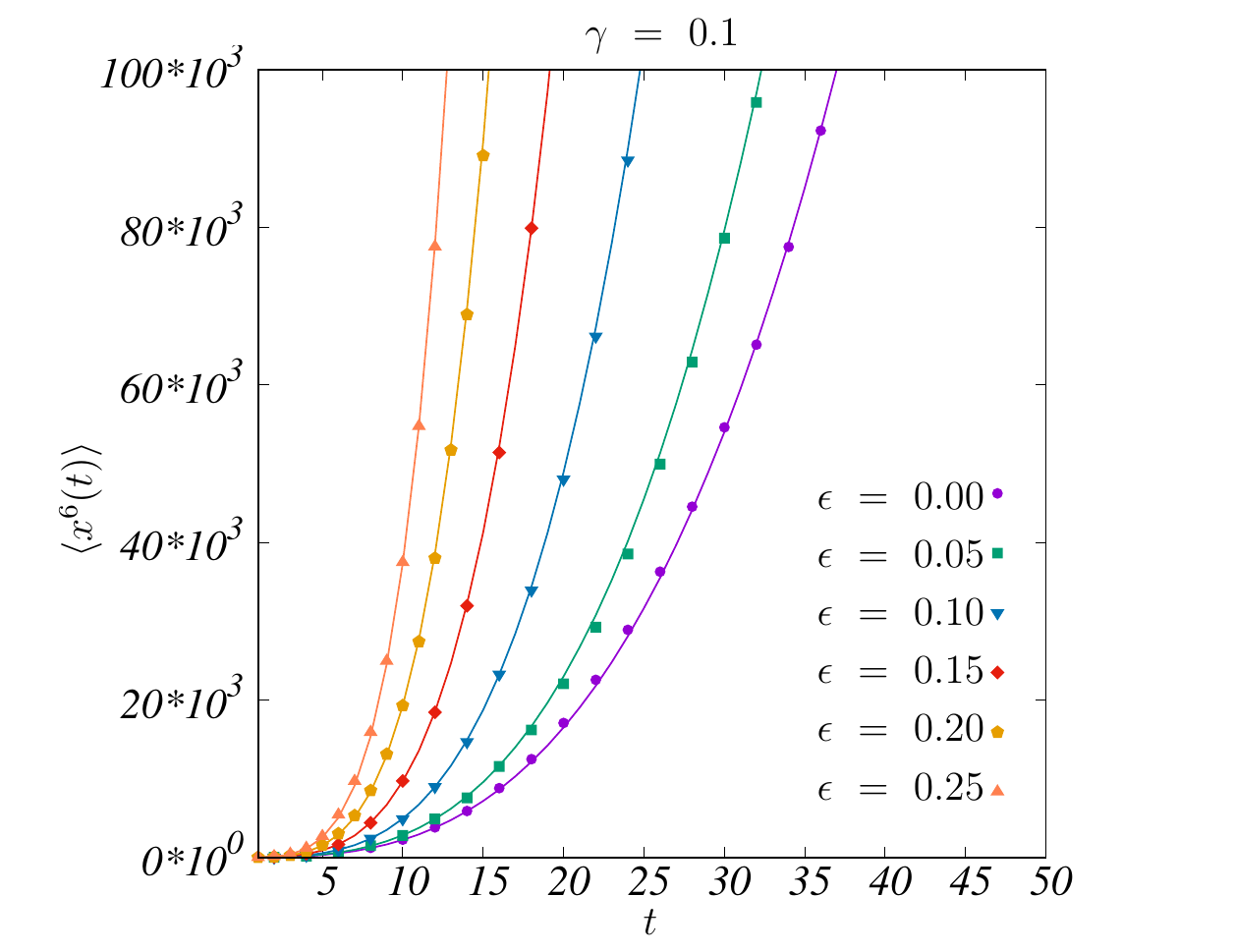} }\hspace{-1 cm}
\subfigure[]{\includegraphics[width=2.4in]{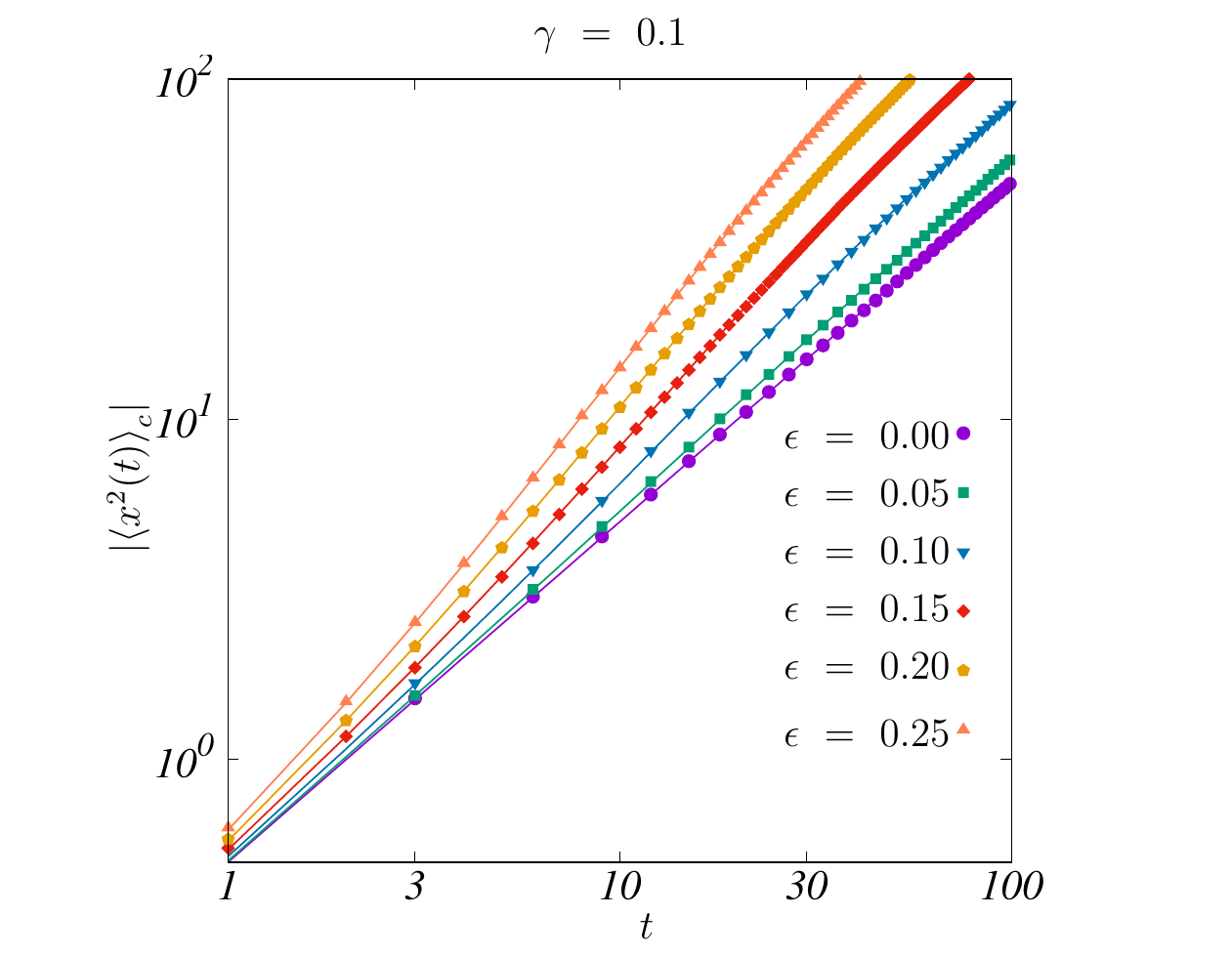} }\hspace{-1 cm}
\subfigure[]{\includegraphics[width=2.4in]{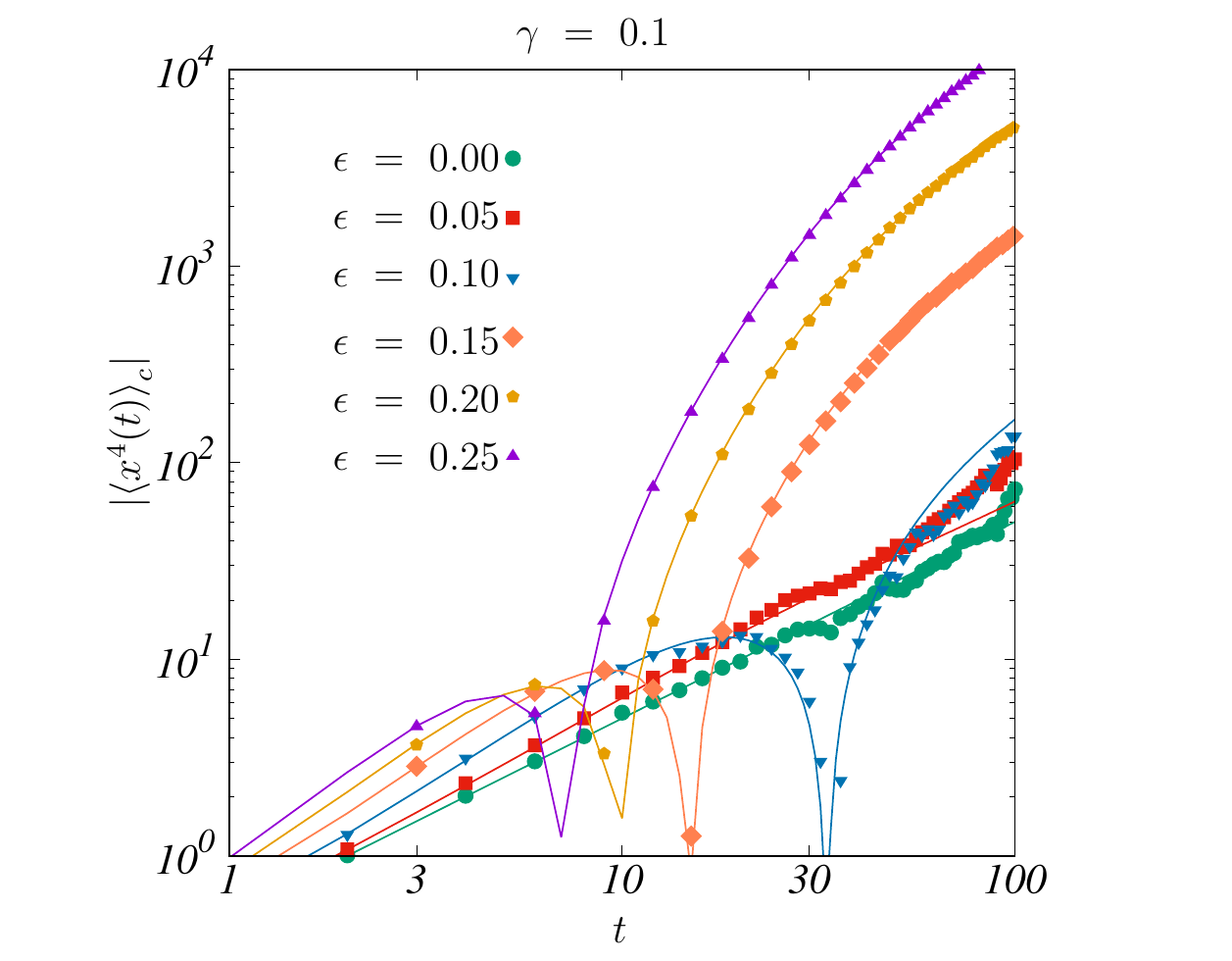} }
\caption{The moments and cumulants of the probability distribution of a run and tumble particle on a two dimensional infinite lattice plotted as a function of time for different values of $\epsilon$. The fixed parameter used is $\gamma=0.1$. (a) $\langle x^2(t) \rangle$ plotted as a function of $t$. The solid lines correspond to the theoretical result in Eq.~(\ref{msd2d}) and the points are from simulations. (b) $\langle x^4(t) \rangle$ plotted as a function of $t$. The solid lines correspond to the theoretical result in Eq.~(\ref{mfd2d}) and the points are from simulations. (c) $\langle x^6(t) \rangle$ plotted as a function of $t$. The solid lines correspond to the theoretical result in Eq.~(\ref{msixd2d}) and the points are from simulations. (d) ${\langle x^2(t) \rangle}_c$ plotted (in logscale) as a function of $t$. The solid lines correspond to the theoretical result in Eq.~(\ref{msdc2d}) and the points are from simulations. (d) ${\langle x^4(t) \rangle}_c$ plotted (in logscale) as a function of $t$. The solid lines correspond to the theoretical result in Eq.~(\ref{mfdc2d}) and the points are from simulations.} \label{fig:ap3}
\end{figure}

Listed below are a few cumulants that can be found from the moments listed above;\\
\begin{eqnarray}
\label{msdc2d}
{\langle x^{2}(t) \rangle}_c &=&{\langle y^{2}(t) \rangle}_c=\langle x^{2}(t) \rangle \nonumber\\&& = -\frac{4 \left(1-e^{-t \gamma }\right) \epsilon ^2}{\gamma ^2}+t \left(\frac{1}{2}+\frac{4 \epsilon ^2}{\gamma }\right),
\end{eqnarray}

\begin{eqnarray}
\label{mfdc2d}
{\langle x^{4}(t) \rangle}_c &=&{\langle y^{4}(t) \rangle}_c =\langle x^{4}(t) \rangle-3 {\langle x^{2}(t) \rangle}^2\nonumber\\&&=
\frac{8 e^{-2 t \gamma } \left(-1+e^{t \gamma }\right) \epsilon ^2 \left(3 \epsilon ^2+e^{t
   \gamma } \left(-2 \gamma ^2+15 \epsilon ^2\right)\right)}{\gamma ^4}+t
   \left(\frac{1}{2}+\frac{16 \epsilon ^2 \left(\gamma ^2+3 \left(-1-2 e^{-t \gamma }\right)
   \epsilon ^2\right)}{\gamma ^3}\right),
   \end{eqnarray}
   
\begin{eqnarray}
\label{msixdc2d}
{\langle x^{6}(t) \rangle}_c &=&{\langle y^{6}(t) \rangle}_c  =\langle x^{6}(t) \rangle -15 \langle x^{4}(t) \rangle \langle x^{2}(t) \rangle +30 {\langle x^{2}(t) \rangle}^3 \nonumber\\&&=
  -\frac{32 e^{-3 t \gamma } \left(-1+e^{t \gamma }\right) \epsilon ^2 \left(15 \epsilon ^4-15
   e^{t \gamma } \epsilon ^2 \left(\gamma ^2-7 \epsilon ^2\right)+e^{2 t \gamma } \left(2
   \gamma ^4-75 \gamma ^2 \epsilon ^2+330 \epsilon ^4\right)\right)}{\gamma ^6}\nonumber\\&&+t
   \left(\frac{1}{2}+\frac{64 e^{-2 t \gamma } \epsilon ^2 \left(45 \epsilon ^4+15 e^{t
   \gamma } \epsilon ^2 \left(-2 \gamma ^2+9 \epsilon ^2\right)+e^{2 t \gamma } \left(\gamma
   ^4-15 \gamma ^2 \epsilon ^2+45 \epsilon ^4\right)\right)}{\gamma ^5}\right)+\frac{2880
   e^{-t \gamma } t^2 \epsilon ^6}{\gamma ^4},\nonumber\\
   \end{eqnarray}

\begin{eqnarray}
\label{mx2y2c2d}
{\langle x^{2}(t)y^{2}(t)\rangle}_c &=& \langle x^{2}(t)y^{2}(t) \rangle- {\langle x^{2}(t) \rangle}^2\nonumber\\&&=
\frac{24 e^{-2 t \gamma } \left(-1+e^{t \gamma }\right) \left(1+5 e^{t \gamma }\right)
   \epsilon ^4}{\gamma ^4}-\frac{48 t e^{-t \gamma } \left(2+e^{t \gamma }\right) \epsilon
   ^4}{\gamma ^3},
   \end{eqnarray}
   
\begin{eqnarray}
\label{mx2y4c2d}
{\langle x^{2}(t)y^{4}(t)\rangle}_c &=& 6{\langle x^{2}(t) \rangle}^3-\langle x^{2}(t) \rangle \langle x^{4}(t) \rangle -6 \langle x^{2}(t) \rangle \langle x^{2}(t) y^{2}(t) \rangle+\langle x^{2}(t)y^{4}(t) \rangle\nonumber\\&&=
\frac{96 e^{-3 t \gamma } \left(-1+e^{t \gamma }\right) \epsilon ^4 \left(-5 \epsilon
   ^2+e^{t \gamma } \left(\gamma ^2-35 \epsilon ^2+5 e^{t \gamma } \left(\gamma ^2-22
   \epsilon ^2\right)\right)\right)}{\gamma ^6}+\frac{2880 e^{-t \gamma } t^2 \epsilon
   ^6}{\gamma ^4}\nonumber\\&&-\frac{t \left(192 e^{-t \gamma } \epsilon ^4 \left(2 \gamma ^2-45 \epsilon
   ^2+\left(\gamma ^2-30 \epsilon ^2\right) \cosh (t \gamma )+\gamma ^2 \sinh (t \gamma
   )\right)\right)}{\gamma ^5}.
   \end{eqnarray}
In Fig.~\ref{fig:ap3}, the first few non zero moments and cumulants in two dimensions are compared with the results from direct numerical simulations of a RTP on a two dimensional infinite lattice and are found to be in good agreement.\\\\

\textit{Asymptotic limit of cumulants from the free energy function:}\\

Listed below are the asymptotic limits of the first few cumulants computed using the expression for large deviation free energy function given in Eq.~(59) along with Eq.~(63) in the main text.

\begin{eqnarray}
\label{msdc2dasym}
\lim_{t \rightarrow \infty}{\langle x^{2}(t) \rangle}_c &=&{\langle y^{2}(t) \rangle}_c=t \left(\frac{1}{2}+\frac{4 \epsilon ^2}{\gamma }\right),
\end{eqnarray}

\begin{eqnarray}
\label{mfdc2dasym}
\lim_{t \rightarrow \infty}{\langle x^{4}(t) \rangle}_c &=&{\langle y^{4}(t) \rangle}_c =t \left(\frac{1}{2}+\frac{16 \epsilon ^2 \left(\gamma ^2-3 \epsilon ^2\right)}{\gamma
   ^3}\right),
   \end{eqnarray}
   
\begin{eqnarray}
\label{msixdc2dasym}
\lim_{t \rightarrow \infty}{\langle x^{6}(t) \rangle}_c &=&{\langle y^{6}(t) \rangle}_c  =t \left(\frac{1}{2}+\frac{64 \epsilon ^2 \left(\gamma ^4-15 \gamma ^2 \epsilon ^2+45
   \epsilon ^4\right)}{\gamma ^5}\right),
   \end{eqnarray}

\begin{eqnarray}
\label{mx2y2c2dasym}
\lim_{t \rightarrow \infty}{\langle x^{2}(t)y^{2}(t)\rangle}_c =
-t\frac{ 48 \epsilon ^4}{\gamma ^3},
   \end{eqnarray}
   
\begin{eqnarray}
\label{mx2y4c2dasym}
\lim_{t \rightarrow \infty}{\langle x^{2}(t)y^{4}(t)\rangle}_c =-t\frac{192 \epsilon ^4 \left(\gamma ^2-15 \epsilon ^2\right)}{\gamma ^5}.
   \end{eqnarray}
The expressions for the asymptotic limits of the cumulants given in Eqs.~(\ref{msdc2dasym})-(\ref{mx2y4c2dasym}) can alternatively be derived by taking a $t \rightarrow \infty$ limit of Eqs.~(\ref{msdc2d})-(\ref{mx2y4c2d}) respectively.

\section{ Comparison of Discrete and Continuous Time Simulations of Active Random Walks}
\begin{figure} [ht]
 \includegraphics[width=0.3\linewidth]{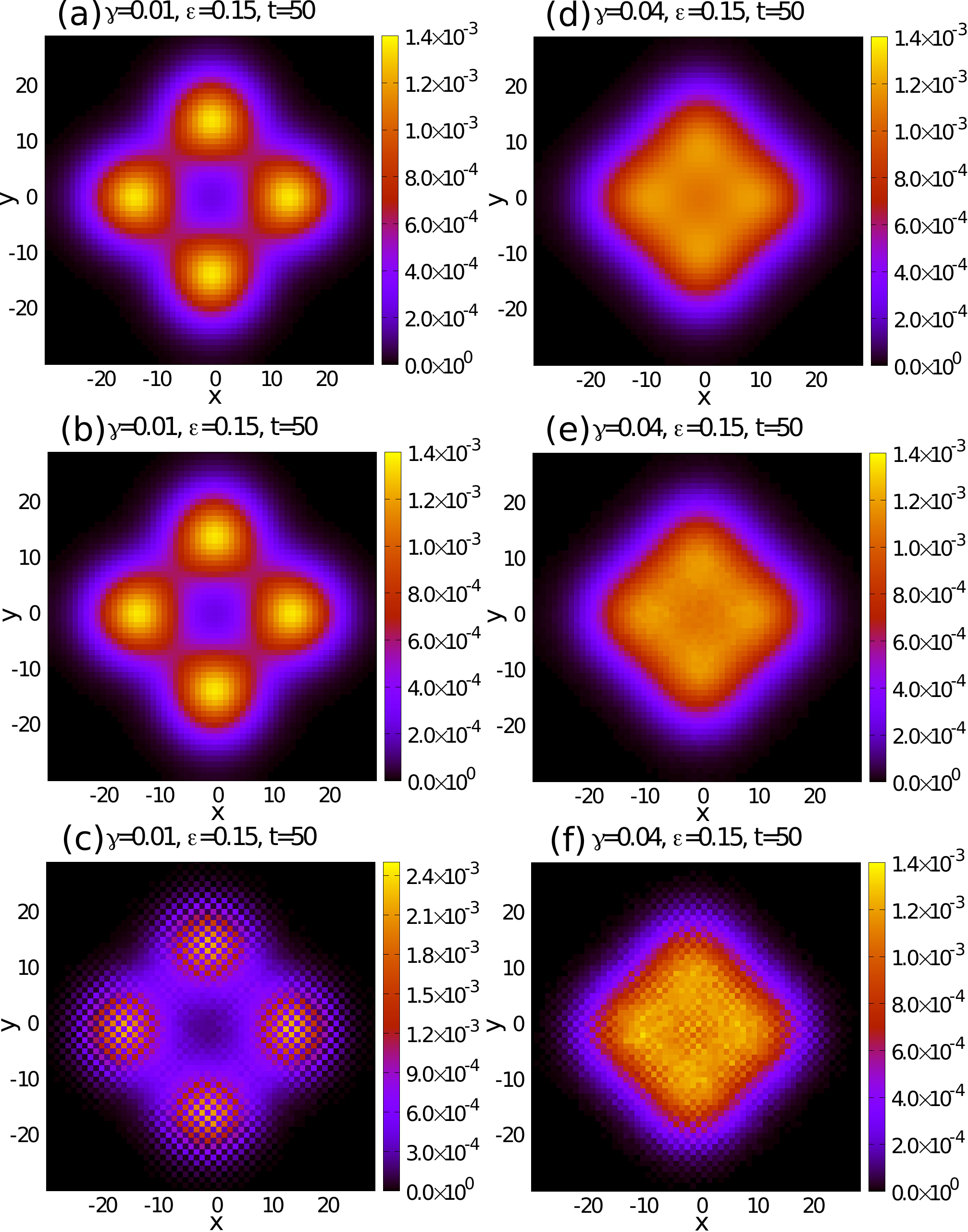}
\caption{Plot of occupation probabilities of a single active random walker on an infinite square lattice obtained from (a,d) continuous time theory (b,e) kinetic Monte Carlo simulations, and (c,f) discrete time simulations for parameter values $\gamma=0.01$ and $\gamma=0.04$ with fixed $\epsilon=0.15$, $D_{2d}=0.25$ and $t=50$. The discrete time simulations exhibit pronounced even-odd oscillations that are not present in the continuous time case.}
\label{compare_analytic_kmc_discrete}
\end{figure} 
\newpage
\section{CONTINUUM SPACE LIMIT OF ACTIVE RANDOM WALKS IN CONTINUOUS TIME}
\label{continuum_limit_CTRW}
\begin{figure} [ht]
 \includegraphics[width=0.6\linewidth]{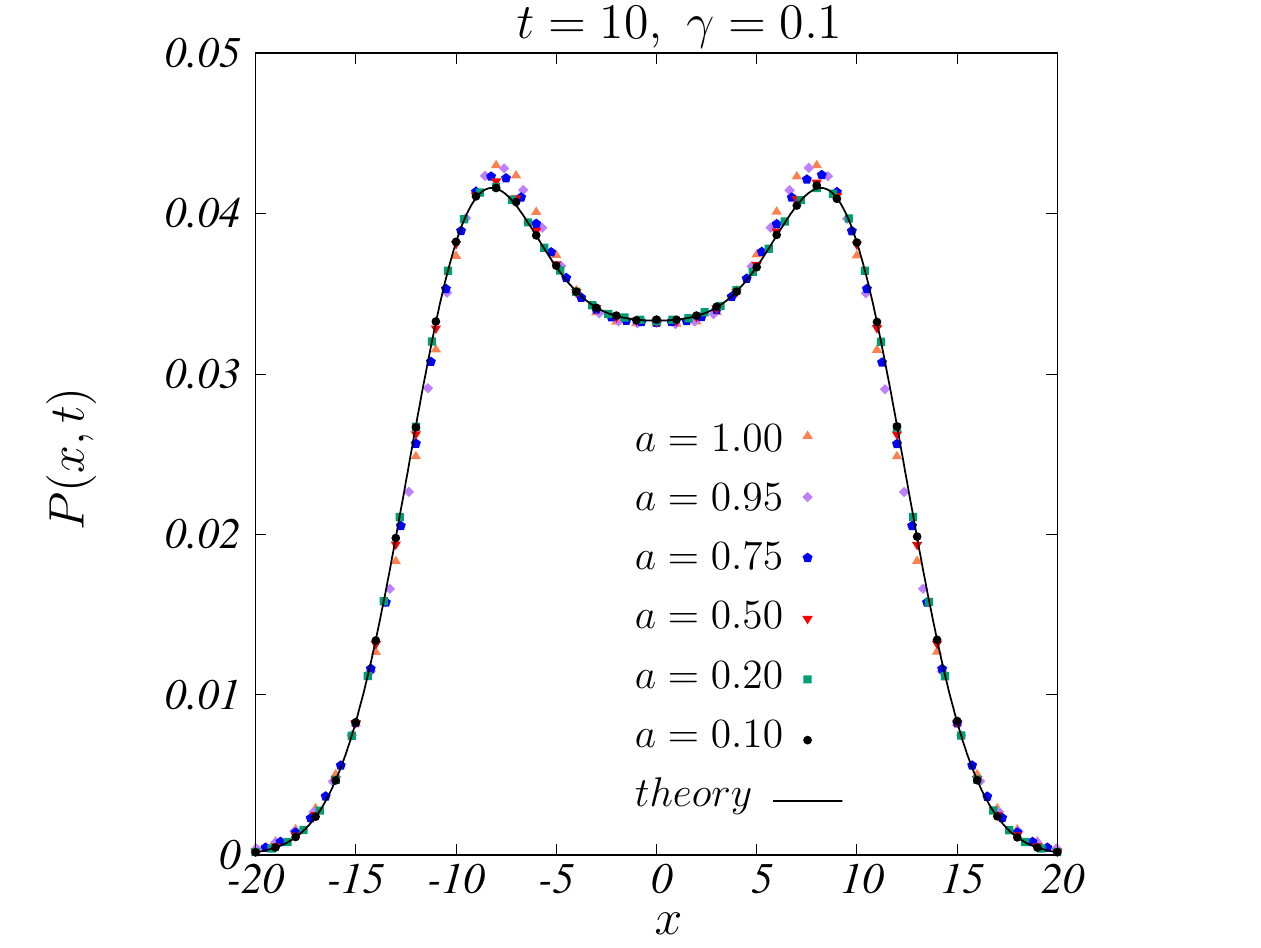}
\caption{ Occupation probability of a single Run and Tumble Particle on an infinite one dimensional lattice plotted at time $t=10$. The points are obtained from kinetic Monte Carlo simulations of an active random walker on a one dimensional lattice with different lattice constants $a$. The bias and diffusion rates are rescaled as $\epsilon \to \frac{\epsilon}{a}$ and $D_{1d} \to \frac{D_{1d}}{a^2}$. 
The solid curve corresponds to the analytic expression in Eq.~(14) of the main text for the occupation probability of a RTP in continuous space for fixed parameter values $D_{1d}=0.5, \epsilon=0.5, \gamma=0.1$. The simulation results converge to the continuum limit expression as the lattice spacing $a$ is reduced.}
\end{figure}


\clearpage
\end{widetext}


\end{document}